\documentclass[11pt]{article}
\usepackage{graphicx}
\usepackage{amssymb}
\usepackage{amsmath}
\usepackage{epsfig} 
\usepackage[]{caption}
\usepackage{verbatim}
\usepackage{amsfonts}
\usepackage{amsthm}
\usepackage{enumerate}
\usepackage{float}
\usepackage{morefloats}
\usepackage{authblk}
\usepackage{algorithm}
\usepackage{algpseudocode}

\title{Some considerations on  crowd Congestion Level}
\author{Francesco Zanlungo$^1$, Claudio Feliciani$^2$, Zeynep Y\"ucel$^1$, Katsuhiro Nishinari$^2$ and Takayuki Kanda$^3$}
\date{%
    $^1$Okayama University\\%
  $^2$Tokyo University\\
  $^3$ Kyoto University \\ 
    \today
}
\begin{document}
\maketitle
\section{Introduction}
The concept of (crowd) Congestion Level ($CL$) was introduced in \cite{CL1}, and presented at the PED 2018 conference by C. Feliciani \cite{CL2}. Following the PED presentation,
along with appreciation for the novel contribution, a few interesting questions were raised, concerning the integral/differential nature of the definition of $CL$,
and the possibility of defining a related pure number. In these short notes we are going, although with no attempt at rigour or formality, to present some
considerations suggesting that the two problems are related, and providing a possible solution. Furthermore, using both theoretical arguments and analysis of simulated data in complex scenarios,
we will try to shed further light on the meaning and applications of this concept. Finally, we analyse some results of an experiment performed with human participants in a ``crossing-flows'' scenario.

In the following, we are going to:
\begin{itemize}
\item Recall the $CL$ original definition and possible theoretical issues
\item Propose a related but possibly theoretically more appealing concept, Congestion Number ($CN$)
\item Analyse the values obtained by $CL$ and $CN$ in some simple scenarios, using respectively:
 \begin{itemize}
 \item fully discrete models,
   \item fully continuous models,
 \end{itemize}
 and use these models to better clarify the original numerical details in the $CL$ definition
\item Perform, using simulations, the $CN$ analysis of a complex scenario (two high density crossing pedestrian flows) under the following assumptions:
  \begin{itemize}
 \item centralised motion, in which the velocity and positions of pedestrians are decided in advance in order to avoid any collision
 \item zero body size limit, in which pedestrians move independently but their bodies are so small that no collision is possible (and no collision avoidance is performed)
   \item ``realistic'' physical dynamics and collision avoidance behaviour
  \end{itemize}
  and verify the ability of the $CL$/$CN$ concept to reflect the increase in congestion levels as the model becomes more realistic.
  \item Analyse  $CL$/$CN$ data from an experiment with human participants in a scenario similar to the simulation setting (two crossing pedestrian flows).
\end{itemize}

This work will not, nevertheless, try to justify the relevance of $CN$ (or $CL$) in the assessment of crowd safety. These notes are intended  as a clarification of some mathematical and numerical
issues, and as a, hopefully significant and more coherent, re-definition of some quantities. Through the study of simulated and controlled crowds we will suggest that the proposed metric possess
some of the minimal properties expected to be relevant in crowds assessment, but for the actual justification of the relevance of the congestion metrics we refer to the original works \cite{CL1,CL2}.
\section{Definition and issues}
\subsection{Definition}
Let us first recall the definition: assuming a field of pedestrian (crowd) velocity $\mathbf{v}$ is given, we may define in a point $\mathbf{x}=(x,y)$ of planar space
\begin{equation}
  \label{defcl}
CL(\mathbf{x})=\frac{\max_{R(\mathbf{x})}(\nabla \wedge \mathbf{v})_z(\mathbf{x})-\min_{R(\mathbf{x})}(\nabla \wedge \mathbf{v})_z(\mathbf{x})}{<v>_{R(\mathbf{x})}}.
\end{equation}
In this equation, $v$ stands for the magnitude of the velocity field, extrema and averages are computed over a Region Of Interest (ROI) centred around $\mathbf{x}$, and denoted as $R(\mathbf{x})$, while $(\nabla \wedge \mathbf{v})_z$ denotes the $z$ component of the rotor of the velocity field (i.e., the only non-zero component). We remind the readers that the numerator is defined as ``Rotation Range'', while ``Crowd Danger'' is defined as the product of $CL$ and crowd density \cite{CL1}, although we will not consider these important metrics in this work.

Obviously, defining a velocity field for a crowd is not a trivial issue, as it involves problems related to the characteristic scales of crowd dynamics. Indeed, any
``fluid dynamics'' approach to crowd dynamics involves non trivial problems, since often the interesting scale of the problem (e.g., corridor width) is
very similar to that of the basic component of the crowd (humans). Although these problems will be extremely relevant to the following discussion, for technical details
such as choosing the correct time and space smoothing tools to define the velocity field from tracking trajectories we refer the reader to the original work \cite{CL1}.
\subsection{Issues}
In this work we are interested in two possible problems with this definition, and namely:
\begin{enumerate}
\item $CL$ is not a pure number, and thus does not have an easy to interpret scale (its dimension is $[L]^{-1}$, i.e. the inverse of a length),
\item the definition mixes local (differential) quantities such as the rotor with global ones (extrema, averages), which, along with the original problems in definition of a
  velocity field (or, equivalently, of a density field), makes again its interpretation difficult.
\end{enumerate}

Concerning the physical dimension aspect, it should be noticed that, being the ratio between the variation of the velocity rotor and the average velocity, $CL$ is left unchanged by a re-scaling of the velocity field.
As a result, changing the time units will not change the $CL$ value (since both the the numerator and denominator will be scaled of the same factor, or, in an equivalent way, since
$CL$ has no dimensional dependence on time). This also means that a high velocity field and a low velocity field will have the same $CL$ value if the are just obtained through a constant (over space) scaling. This may seem counter-intuitive, but it is actually a nice property of the $CL$ concept, as we will see in our analysis\footnote{Obviously, crowd velocity may be relevant
  to crowd risk. It has to be expected that relatively simple measures may not fully take in consideration every aspect of crowd risk, and $CL$ should be always used along other measures that may take velocity in consideration.}.

On the other hand, $CL$ will change if we change the spatial units, or, in an analogous way, if the spatial properties of the field are scaled. As the original definition introduces a spatial scale,
the grid size $R$, the adimensional product $R \cdot CL$ should be expected to be related to a meaningful definition of a crowd metric. Let us see how to formalise this idea.
\section{Definition of a Congestion Number}
\subsection{Relation to the rotor gradient (Differential congestion)}
As stated above, \cite{CL1} provides a detailed discussion on the scales at which the grid for the velocity field and the ROI should be computed to obtain results that
are significant and useful from a crowd management point of view. It seems reasonable that the ROI should be large enough to be able to assess
the change in the $z$ component of the rotor, but not too large, as this would lead to an under estimate of the growth rate. Namely, assuming $L$ to be the
linear scale of the ROI, we may use a linear approximation
\begin{equation}
  \label{linap}
\max_{R(\mathbf{x})}(\nabla \wedge \mathbf{v})_z(\mathbf{x})-\min_{R(\mathbf{x})}(\nabla \wedge \mathbf{v})_z(\mathbf{x})\approx L ||\mathbf{\nabla} [(\nabla \wedge \mathbf{v})_z(\mathbf{x})]||.
\end{equation}
We may now see that the numerator is proportional, according to the proposed approximation and interpretation, to the magnitude of differential operator applied on the velocity field,
namely the gradient of
the only non-zero component of the rotor of the velocity field. We may call the latter  Differential Congestion ($DC$), namely
\begin{equation}
DC(\mathbf{x})=||\mathbf{\nabla} [(\nabla \wedge \mathbf{v})_z(\mathbf{x})]||,
\end{equation}
\subsection{Congestion Number as a ratio with an Extreme Differential Congestion}
We still need to explain the remaining terms. The proposed explanation is that Congestion is indeed related to $DC$,
but the remaining terms arise by taking a ratio between the measured $DC$ and a reference value, that we may call Extreme Differential Congestion, or $EDC$.
Such a ratio will be obviously a pure number, which we may call Congestion Number ($CN$)
\begin{equation}
  \label{defcn}
CN(\mathbf{x})=\frac{DC(\mathbf{x})}{EDC(\mathbf{x})}.
\end{equation}
In the following, we are going to define $EDC$ (which still will depend on $\mathbf{x}$, as it is related to the local magnitude of the velocity) and derive the relation between $CL$ and $CN$.
\subsection{Definition of Extreme Differential Congestion}
To introduce a ``maximally rotating field'', let us first recall Stokes theorem
\begin{equation}
\int_{S} (\nabla \wedge \mathbf{v}) \cdot \mathbf{dn} = \int_{C} \mathbf{v} \cdot \mathbf{dl},
\end{equation}
where the 1-D path $C$ is the boundary of the 2-D region $S$. For a circular path of radius $R$, denoted as $C_R$, over which the magnitude of the velocity field is fixed to a
constant value, $v_R$, the maximum value for the integrals defined above will be given by a field exactly directed along the tangent to the path
\begin{equation}
\mathbf{v}(R,\varphi)=v_R \mathbf{e}_\varphi,
\end{equation}
where $\mathbf{e}_\varphi$ is the angular versor whose Cartesian component are $(-\sin \varphi,\cos \varphi)$ or $(-y/r,x/r)$. For this field, we have
\begin{equation}
\int_{D_R} (\nabla \wedge \mathbf{v}) \cdot \mathbf{dn} = \int_{C_R} \mathbf{v} \cdot \mathbf{dl}=2 \pi v_R R,
\end{equation}
regardless of the value attained inside the disc $D_R$. Obviously, the reversed field $-v_R \mathbf{e}_\varphi$ assumes the minimum value $-2 \pi v_R R$.

We may choose the values assumed inside $D_R$ by requiring the field to go continuously and linearly to 0 towards the centre
\begin{equation}
  \label{mr}
\mathbf{v}(r,\varphi)=v_R \frac{r}{R} \mathbf{e}_\varphi.
\end{equation}
For this field, the non zero component of the rotor is given by the cylindrical coordinate formula\footnote{Or equivalently by $\partial_x v_y - \partial_y v_x= (\partial_x x -\partial_y (-y)) v_R/R  = 2 v_R/R$.}
\begin{equation}
  \label{maxrot}
(\nabla \wedge \mathbf{v})_z=\frac{1}{r} \partial_r\left( r v_R \frac{r}{R}\right)=\frac{2 v_R}{R}
\end{equation}

The original formula for $CL$ includes the average value of $v$, and for this reason it may be useful to compute the average value of $v$ over $D_R$, which is
\begin{equation}
  <v>_{D_R}=\frac{\int_{D_R} v}{\pi R^2}=\frac{\int_{D_R} v_R \frac{r}{R}}{\pi R^2}=\frac{\int_0^{2\pi}d\varphi \int_0^{R} r v_R \frac{r}{R} dr}{\pi R^2}
  =\frac{2\pi \frac{v_R}{R} \int_0^{R} r^2 dr}{\pi R^2} = \frac{2 v_R}{R^3} \frac{R^3}{3}= \frac{2}{3} v_R,
\end{equation}
or
\begin{equation}
  v_R=\frac{3}{2} <v>_{D_R}.
\end{equation}
Substituting in eq. (\ref{maxrot}) we obtain
\begin{equation}
(\nabla \wedge \mathbf{v})_z=\frac{2 v_R}{R}=\frac{3 <v>_{D_R}}{R}.
\end{equation}

These considerations on vectorial analysis are obviously independent of pedestrian dynamics. Let us now assume that the maximum value for the rotor in a pedestrian velocity field is given when a
maximally rotating field, as defined above, occurs on a scale comparable to the pedestrian body size, $R\approx 0.2$ m, which is incidentally the value chosen in \cite{CL1} for the size of the cell grid\footnote{Readers may notice that for defining the maximally rotating vector field we used a continuous field defined obviously at a scale much smaller than $R$; we will come back on this issue in section
\ref{ca}.}.

We now define the $EDC$ as a gradient corresponding to a change from two different but opposed maximally rotating fields located at a distance $L$, which corresponds to the linear size of the ROI.
Namely\footnote{We will again consider later the relation between this ``approximated macroscopic gradient'' and the actual local gradient defined on a continuous field, see in particular section \ref{tm}.}
\begin{equation}
EDC=\frac{\frac{3 <v>_{D_R}}{R}-(-\frac{3 <v>_{D_R}}{R})}{L}=\frac{6 <v>_{D_R}}{R L}.
\end{equation}

Furthermore, we are going to assume that the average of the magnitude of vector field on the ROI is equivalent to that of the average over the maximally rotating area\footnote{Again, we will
  discuss this approximation in section \ref{tm}.}, which leads to
\begin{equation}
EDC(\mathbf{x}) \approx\frac{6 <v>_{R(\mathbf{x})}}{R L}.
\end{equation}
\subsection{Operational definition of $CN$}
Recalling our definition of $CN$, eq. (\ref{defcn}), and the approximation eq. (\ref{linap}), we have
\begin{equation}
  CN(\mathbf{x}) \approx \frac{(\max_{R(\mathbf{x})}(\nabla \wedge \mathbf{v})_z(\mathbf{x})-\min_{R(\mathbf{x})}(\nabla \wedge \mathbf{v})_z(\mathbf{x})) R L}{6 <v>_{R(\mathbf{x})} L},
\end{equation}
or
\begin{equation}
  \label{cnapprox}
  CN(\mathbf{x}) \approx CL(\mathbf{x}) \frac{R}{6},
\end{equation}
which may be also considered as the operational definition of $CN$.

According to this derivation, in the experiments reported by \cite{CL1}, maximum registered values of $CN$ would be around 0.5. It looks clear that values of $CN \ll 1$ should correspond to non-congested
crowds, while as $CN$ gets comparable to 1, the crowd should be extremely congested. Nevertheless, it should be noticed, as we will show in detail later (section \ref{tm}), that values of $CN>1$ are possible, as it is possible to define fields for which the $DC$ is larger than the proposed $EDC$. Anyway, as the maximum possible $DC$ depends on the discretisation and numerical scheme, and corresponds to highly artificial setting, we will not try to define $CN$ in such a way to have always $0 \leq CN \leq1$.
\section{Toy models of high $CN$ settings}
\label{tm}
\subsection{Discrete approach}
\label{da}
Let us study some settings that correspond to very high $CN$ values, first using a derivation
more closely related to the numerical nature of the computation. The following ``discrete derivation'' has the following interesting properties:
\begin{enumerate}
\item does not rely on a fictitious continuous field defined at a scale much smaller than $R$,
\item clearly explains the relation between $L$ and $R$,
\item clearly explains the role of the approximation  $<v>_{R(\mathbf{x})}\approx <v>_{D_R}$ in our derivation,  
\item explains the role of numerical approximations (e.g. choice of integration/differentiation schemes).
\end{enumerate}

In the subsection \ref{ca}, we will consider the opposite approach, i.e., perform all computations on a microscopic, continuous velocity field.
\subsubsection{Separated, random constant velocity}
The discrete equivalent of two opposing maximally rotating fields located at a close distance can be realised on the grid shown in Fig. \ref{f1}. The maximum rotor value occurs in the cell 2, while
the minimum occurs in cell 3. We are going to compute $CN$ in cell 1, which is in the middle of the two flows (and separated from them). If the cells have size $R$, the distance between cell 2 and 3 is $L=4 R$. The vector field has
magnitude $v$ in the direction given
by the arrows, and we will assume it to have constant magnitude $v$ but random direction on all other cells. In such a way, regardless of the choice of the ROI, the
value of $<v>_{ROI}$ is going to be $v$.

We recall that
\begin{equation}
(\nabla \wedge \mathbf{v})_z= \partial_x v_y -\partial_y v_x.
\end{equation}
Using the most trivial numerical differentiation scheme, in cell 2 we have
\begin{equation}
(\nabla \wedge \mathbf{v})_z= \frac{2 v}{2 R}-(-\frac{2 v}{2 R})= \frac{2 v}{R},
\end{equation}
while obviously in 3 we will have the opposite value, so that  
\begin{equation}
\max (\nabla \wedge \mathbf{v})_z - \min (\nabla \wedge \mathbf{v})_z = \frac{4 v}{R},
\end{equation}
and
\begin{equation}
CN= \frac{4 v}{R} \frac{R}{6 v} =\frac{2}{3}
\end{equation}

The reason we obtained a value different from 1 is due to the difference between the values of the average velocity in the continuous model used to define $CN$ and in the discrete model. Anyway, higher values of $CN$ are possible, as shown next.
\subsubsection{Separated, negligible velocity outside flows}
\begin{figure}[t]
\begin{center}
\includegraphics[width=.45\textwidth]{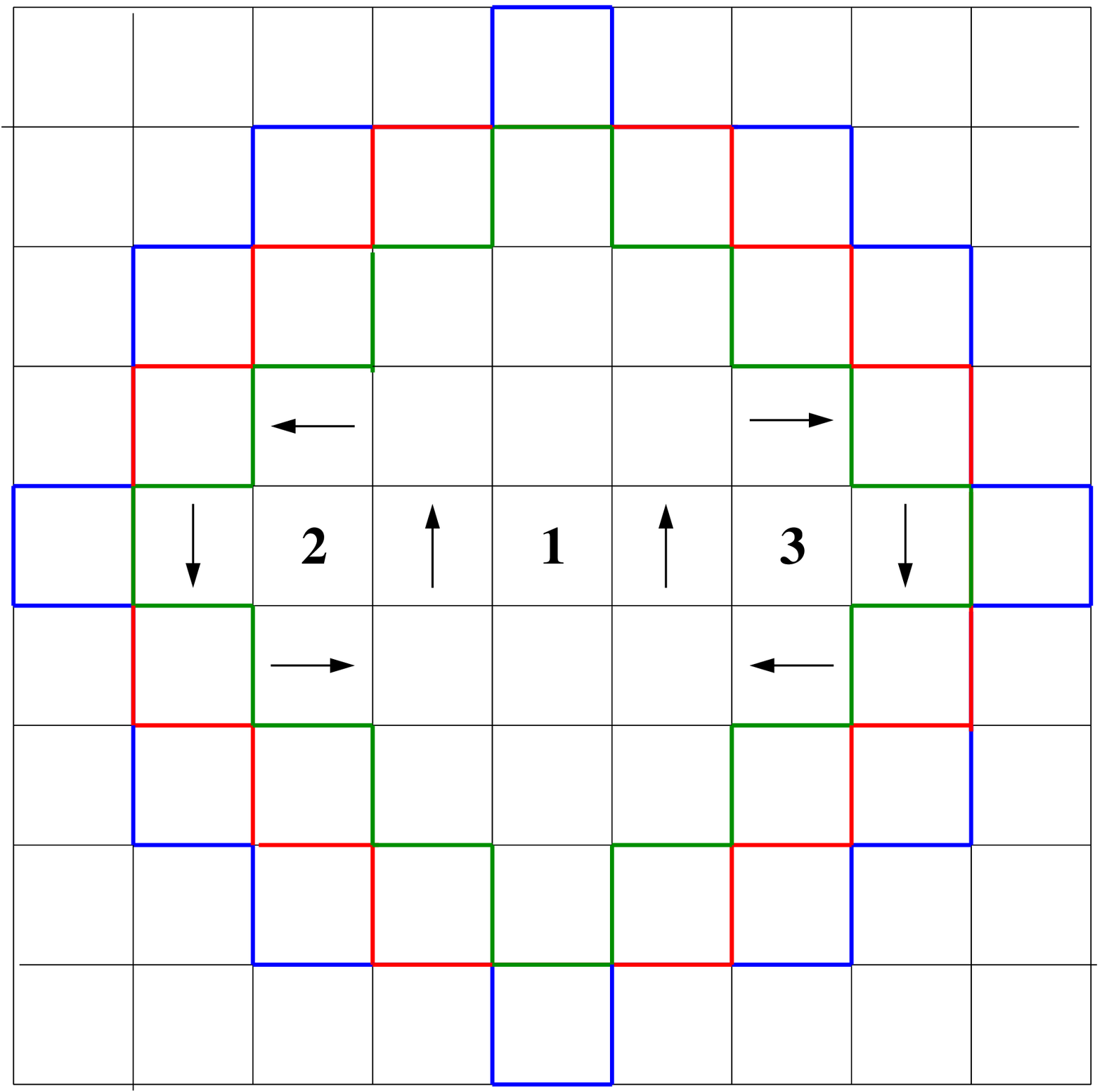} \hspace{0.07\textwidth}\includegraphics[width=.45\textwidth]{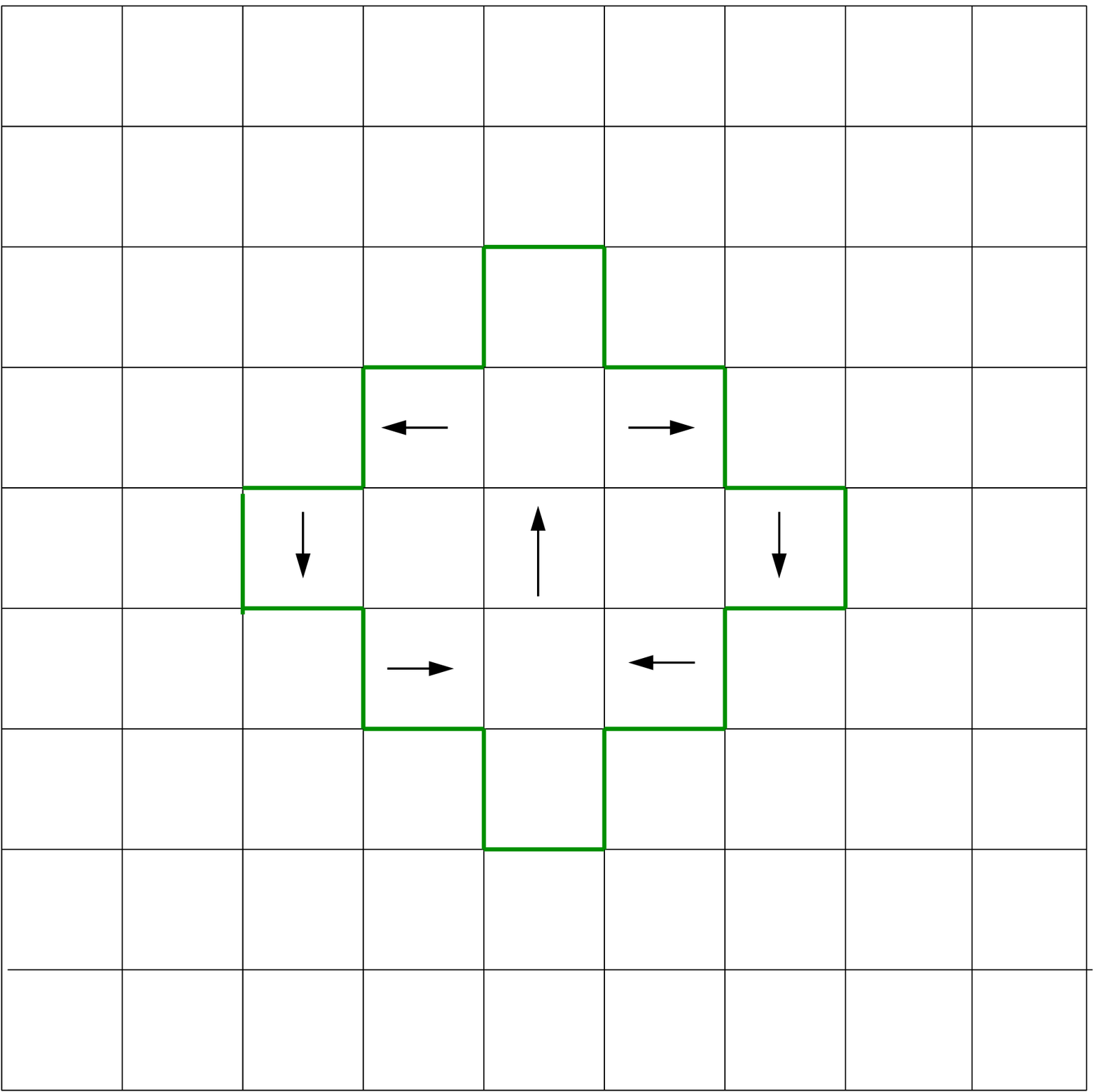}
\caption{Left: separated flows. MM scheme boundary in green, OE scheme boundary in red, ME scheme boundary in blue. Right: overlapping flows.}
\label{f1}
\end{center}
\end{figure}

Let us now go back to Fig. \ref{f1}, but this time we may assume that velocity is  {\it almost} 0 (i.e., $\ll v$)
where arrows are lacking (including cells 1, 2 and 3). This choice of having a lot of cells with very low velocity is due to the
attempt of keeping $<v>$ as small as possible, and thus $CN$ as large as possible. It should be noticed that an empty cell is, from the crowd dynamics viewpoint,
extremely different from a non-empty cell with a low velocity, and indeed \cite{CL1} reminds us to compute average velocities only using occupied cells. Anyway, in the computations below, the cells without arrows may be considered with such a small velocity (e.g., $10^{-3} v$) that their contribution to averages and rotors is simply ignored
(from an actual crowd dynamics viewpoint, obviously, this is quite unrealistic).

The value of $<v>_{ROI}$ depends now strongly on the definition of the ROI. It seems obvious that the ROI should include all non-zero cells, so that its ``diameter''
(maximum distance between two included cells) has to be at least $7R$, which is exactly the value empirically proposed in \cite{CL1}. The actual choice of the ROI depends then on the definition of the
distance on the grid. We propose here 3 schemes:
\begin{enumerate}
\item Manhattan distance: only cells at a Manhattan distance $d_M=|dx|+|dy|<3$ are included. This includes exactly $N_{ROI}=25$ cells, as shown in Fig. \ref{f1} left (Minimum Manhattan scheme, or MM).
\item The original scheme proposes by \cite{CL1}, i.e., including, since the diameter of the ROI is 7, all cells that have an Euclidean distance between their centres $d_E\leq 7 R/2$. This includes 37 cells
  (Original Euclidean scheme, or OE).
\item As above, using Euclidean distance but, since the distance between 2 and 3 is $L=4 R$, requiring $d_E\leq 4 R$. This includes 49 cells (Maximum Euclidean scheme, or ME).  
\end{enumerate}

Since only 8 cells have non-zero $v$, the value of $<v>$ would be 8 $v/N_{ROI}$, and thus
\begin{equation}
CN=\frac{N_{ROI}}{8 v} \frac{4 v}{R} \frac{R}{6} =\frac{N_{ROI}}{12},
  \end{equation}
which gives a value higher than 4 in the ME scheme. It seems anyway reasonable that these extremely artificial conditions may not occur in the real world and
values of $CN \approx 1$ should already be considered as extremely high\footnote{In such a setting, the same proposed values are found
  if we use a single maximally rotating field (e.g. in the centre of the cell), since the
numerator is decreased by a factor two, but also the denominator is (the number of non-empty cells would be 4). These results may be puzzling, but, again, we should
remember that $CL$ has been proposed in \cite{CL1} to deal with actual crowds, and not with our handpicked situations. Furthermore, ``empty cells'' in our settings should actually represent occupied cells with very low velocity (packed crowd), and it is clear that having a strongly rotational movement inside a packed, almost non-moving
crowd should be
an hint of a potentially very dangerous situation in an actual crowd.}.
\subsubsection{Overlapping, random constant velocity}
If in Fig.  \ref{f1} we displace both maximally rotating fields of one cell towards the centre (Fig. \ref{f1} right)
we get a non-zero field in 1 ($\mathbf{v}=(0, 2v)$). In this case,
\begin{equation}
(\nabla \wedge \mathbf{v})_z= \frac{3 v}{2 R}-(-\frac{2 v}{2 R})= \frac{5 v}{2R}.
\end{equation}
If the velocity in the other cells has constant magnitude we get
\begin{equation}
CN= \frac{5 v}{R} \frac{R}{6 v} =\frac{5}{6}
\end{equation}
\subsubsection{Overlapping, negligible velocity outside flows}
Assuming negligible velocity outside the flows, we now get, by having just 7 ``non-zero'' cells, one of them contributing to $2v$
\begin{equation}
CN=\frac{N_{ROI}}{8 v} \frac{5 v}{R} \frac{R}{6} =\frac{5 N_{ROI}}{48}.
\end{equation}
Interestingly, anyway, in this situation is possible to include all non-zero cells in a $L=2R$ radius ball, which would have $N_{ROI}=13$  (Fig. \ref{f1} right) and thus $CN$ close to 1 ($65/48$).
\subsection{Continuous approach}
\label{ca}
Models using a velocity field defined on a continuous scale smaller than $R$ have little physical or computational value, but allowing for analytical computations are helpful in better clarifying basic concepts. Let us generalise eq. \ref{mr} to
\begin{equation}
\mathbf{v}(r,\varphi)=\frac{v_R}{R} f(r) \mathbf{e}_\varphi.
\end{equation}
so that
\begin{equation}
(\nabla \wedge \mathbf{v})_z=\frac{v_R}{r R} \partial_r\left( r f(r) \right) \equiv \frac{v_R}{r R} \partial_r\left( g(r) \right),
\end{equation}
where we defined
\begin{equation}
g(r)\equiv r f(r).
\end{equation}
Now, if $f(r)=r$, we have (eq. \ref{maxrot}) $(\nabla \wedge \mathbf{v})_z=2 v_R/R$, so we may define
\begin{equation}
Rot^{MAX}=\frac{2 v_R}{R}.
\end{equation}
Furthermore, we define $l(r)$ such that
\begin{equation}
(\nabla \wedge \mathbf{v})_z\equiv Rot^{MAX} l(r)
\end{equation}
so that
\begin{equation}
h(r)\equiv 2r l(r) = \partial_r\left( g(r) \right),
\end{equation}
which gives us a differential equation for the value of the velocity field given the value of the rotor one. Furthermore, we have
\begin{equation}
<v>_{D_r}= \frac{(v_r/R) \int_0^{2\pi} d\varphi \int_0^rf(\rho) \rho d\rho}{\pi r^2}=\frac{(v_r/R) 2 \pi \int_0^rf(\rho) \rho d\rho}{\pi r^2}=\frac{Rot^{MAX}}{r^2} \int_0^rf(\rho) \rho d\rho
\end{equation}
\subsubsection{Continuous rotor field}
Let us assume now that between two maximally rotating fields the value of the rotor passes gradually from $2 v_R$ to $-2 v_R$ in a $2 R$ distance. This may be attained by using the following function\footnote{The function
  is clearly not differentiable in $R$ and $2R$, but a differentiable version in which the radial derivative of $h$ transits from 0 to $-1/R$ (and vice versa in $2 R$) in a $\lambda \ll R$ scale
  can be obtained using a ``smoothed step function'', adapting the detailed description on bump functions found in \cite{Tu}.} for $l(r)$
\begin{equation}
l(r)=
\begin{cases}
1, \text{ if } 0 \leq r<R\\
2-\frac{r}{R} \text{ if } R \leq r< 2R\\
0  \text{ if } r \leq  2R
\end{cases}
\end{equation}
or
\begin{equation}
h(r)=
\begin{cases}
2 r, \text{ if } 0 \leq r<R\\
4 r-\frac{2 r^2}{R} \text{ if } R \leq r< 2R\\
0  \text{ if } r \leq  2R
\end{cases}
\end{equation}
Using
\begin{equation}
g(r)=\int_0^r h(\rho) d \rho
\end{equation}
we obtain
\begin{equation}
g(r)=
\begin{cases}
r^2, \text{ if } 0 \leq r<R\\
2 r^2-\frac{2 r^3}{3 R} -\frac{R^2}{3} \text{ if } R \leq r< 2R\\
\frac{7 R^2}{3}  \text{ if } r \geq  2R
\end{cases}
\end{equation}
or
\begin{equation}
f(r)=
\begin{cases}
r, \text{ if } 0 \leq r<R\\
2 r-\frac{2 r^2}{3 R} -\frac{R^2}{3 r} \text{ if } R \leq r< 2R\\
\frac{7 R^2}{3 r}  \text{ if } r \geq  2R
\end{cases}
\end{equation}
and
\begin{equation}
<v>_{D_r}=
\begin{cases}
  Rot^{MAX} \frac{r}{3} , \text{ if } 0 \leq r<R\\
  Rot^{MAX}(\frac{2 r}{3}-\frac{r^2}{6 R}-\frac{R^2}{3 r}-\frac{R^3}{6 r^2}) \text{ if } R \leq r< 2R\\
Rot^{MAX} \frac{R^2(14 r -15 R)}{6 r^2}  \text{ if } r \geq  2R
\end{cases}
\end{equation}
We introduced this model because it reproduces our ``linear approximation'' for the gradient, so we may use it to check the validity of the approximation according to which
$<v>_{ROI}=2/3 v_R$. Using the above results, for $2R$ we obtain $<v>_{D_{2R}}=13/12 v_R$. In the case of two different opposing rotating fields, we may use as ROI a disc of radius
$4R$ located in the middle point. A numerical integration gives for this case $<v>_{D_{4R}} \approx 1.0095 v_R$. This values suggest again that the maximal possible $CN$ value should not depart strongly from 1.
\subsubsection{Discontinuous rotor field}
Defining the velocity field starting from the rotor may seem counter intuitive. We could have started from the velocity field, e.g., asking the velocity to be zero outside a disc of radius $2R$
\begin{equation}
f(r)=
\begin{cases}
r \text{ if } 0 \leq r<R\\
2R-r \text{ if } R \leq r< 2R\\
0  \text{ if } r \geq  2R
\end{cases}
\end{equation}
By straightforward differentiation we have now
\begin{equation}
l(r)=
\begin{cases}
1, \text{ if } 0 \leq r<R\\
\frac{R}{r}-1 \text{ if } R < r< 2R\\
0  \text{ if } r >  2R
\end{cases}
\end{equation}
The rotor field is now discontinuous. The continuity can be regained by using bump function \cite{Tu} in such a way that $f(r)$ is re-defined as a differentiable function in $R$,
i.e. by having $l(r)$ to make a continuous transition from 1 to zero over a scale $\lambda \ll R$. Anyway, in this way the gradient of the rotor would be increased of a scale $R/\lambda$ with respect to
the ``macroscopic variation'' scale $v_R/R^2$. This toy model serves thus to remind us that trying to define the differential congestion ($DC$ and $EDC$) on a scale smaller than $R$ is meaningless,
and justifies our use of a linear approximation, i.e. of the comparison between the maximum and minimum value on the ROI.

By integrating $f(r)$ we obtain for this model
\begin{equation}
<v>_{D_r}=
\begin{cases}
  Rot^{MAX} \frac{r}{3} , \text{ if } 0 \leq r<R\\
  Rot^{MAX}(R -\frac{r}{3}-\frac{R^3}{3 r^2}) \text{ if } R \leq r< 2R\\
Rot^{MAX} \frac{R^3}{r^2}  \text{ if } r \geq  2R
\end{cases}
\end{equation}
The result for $r>2R$ is the mathematical average, but since the velocity for $r>2R$ is zero, the average according to the $CL$ definition is the $2R$ value, or
$<v>=v_R/2$. If the velocity for $r>2R$ is negligible but non-zero, for two opposite flows included in in a $D_{4R}$ ROI we have $<v>=v_R/4$ and $CN=8/3$, showing that also for continuous models we
may have $CN$ quite large in these seemingly artificial settings.
\section{Simulations}
Let us now study some interesting properties of this crowd metric when applied to a crowd system. In this theoretical work, we are using artificial data from simulations, and we are on purpose comparing a more realistic model with some very unrealistic ones, to show the ability of the metric to distinguish the former from the latter\footnote{Here we decided to compute $CN$ in each cell, including those that do not present non-zero velocity and/or do not have a defined rotor field; obviously such cells have a non-zero $CN$ only if in their ROI there is a non-zero rotor cell. The ROI was chosen as an Euclidean $D_{7/2 R}$.}.

\begin{figure}[t]
\begin{center}
\includegraphics[width=0.28\textwidth]{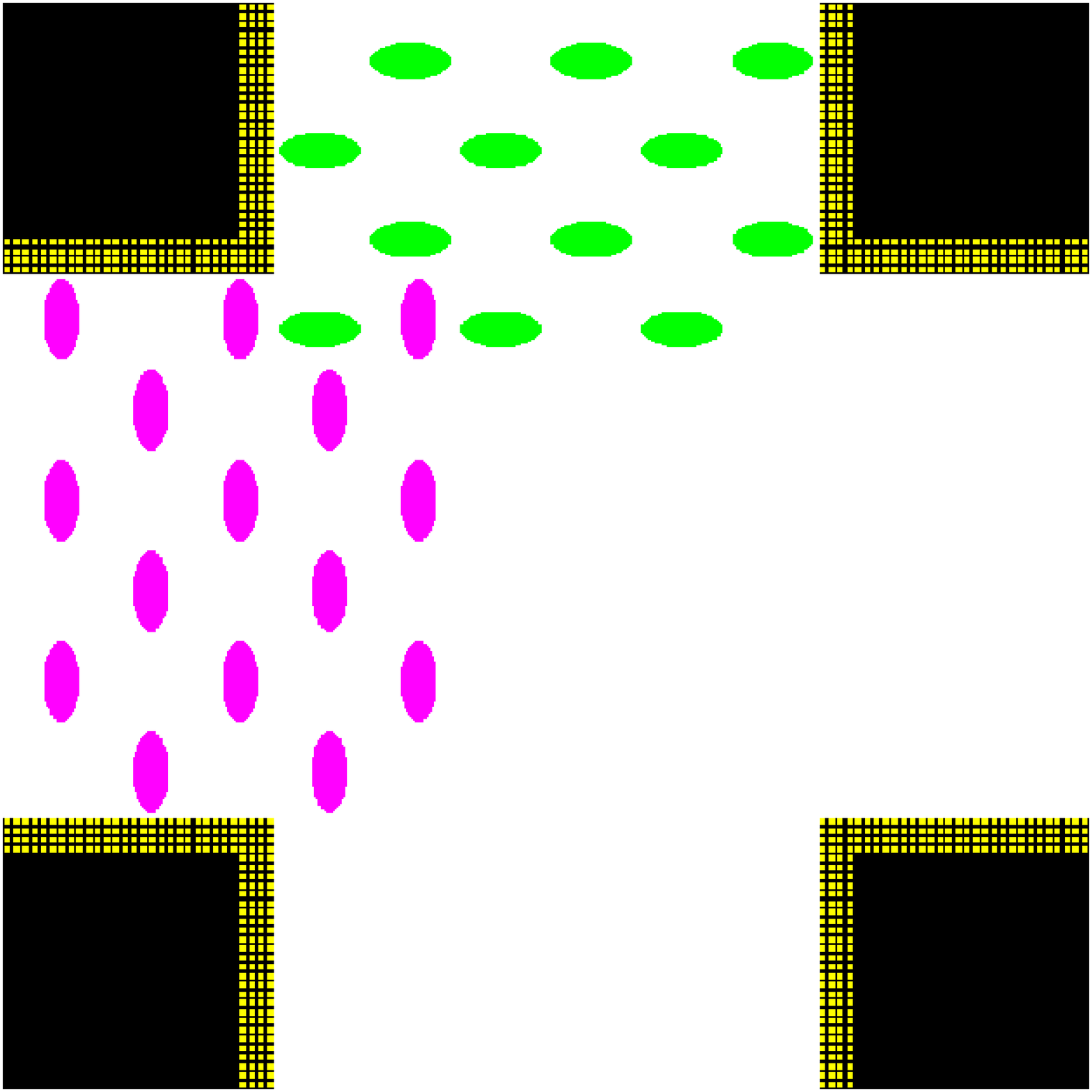}\hspace{.05\textwidth}\includegraphics[width=0.28\textwidth]{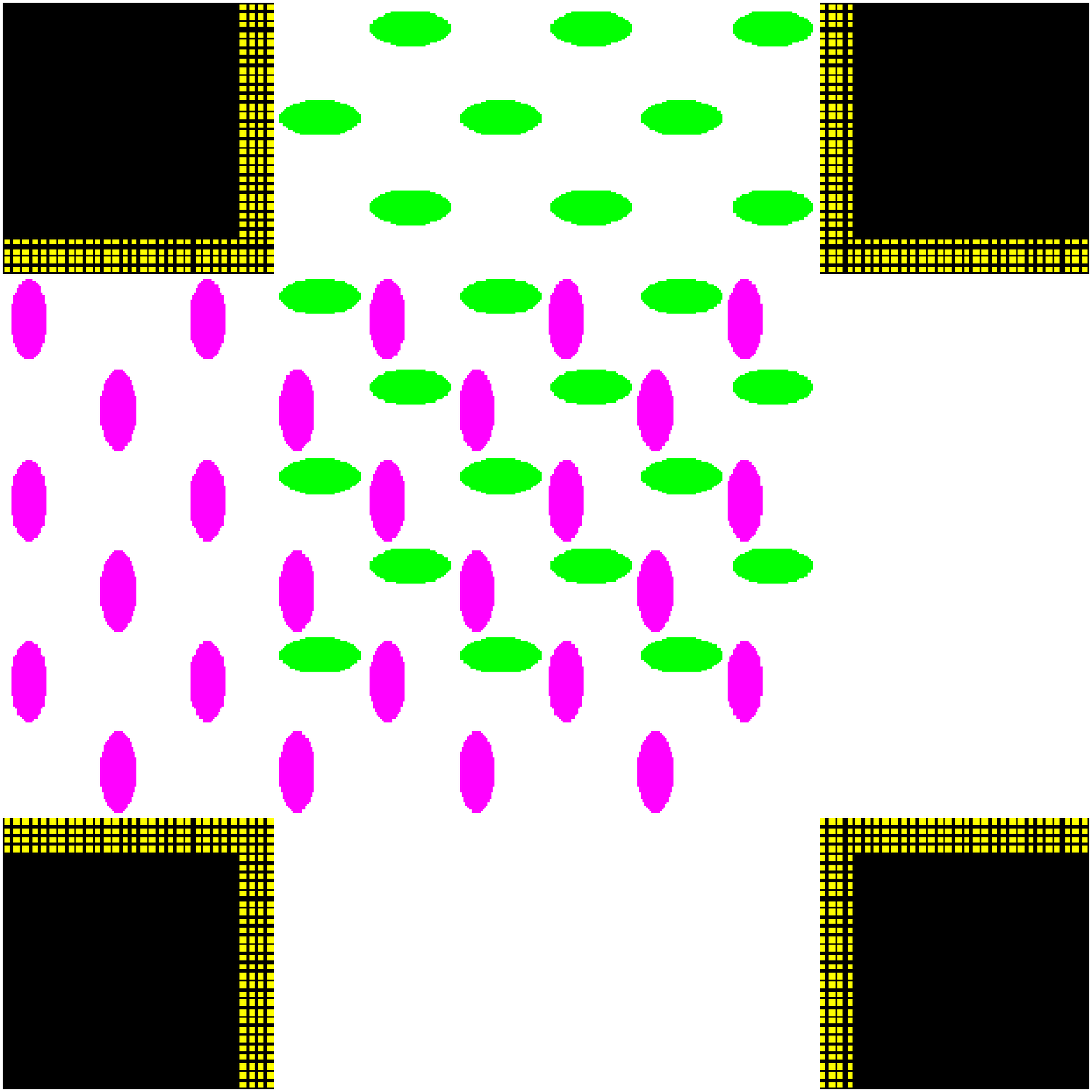}\hspace{.05\textwidth}\includegraphics[width=0.28\textwidth]{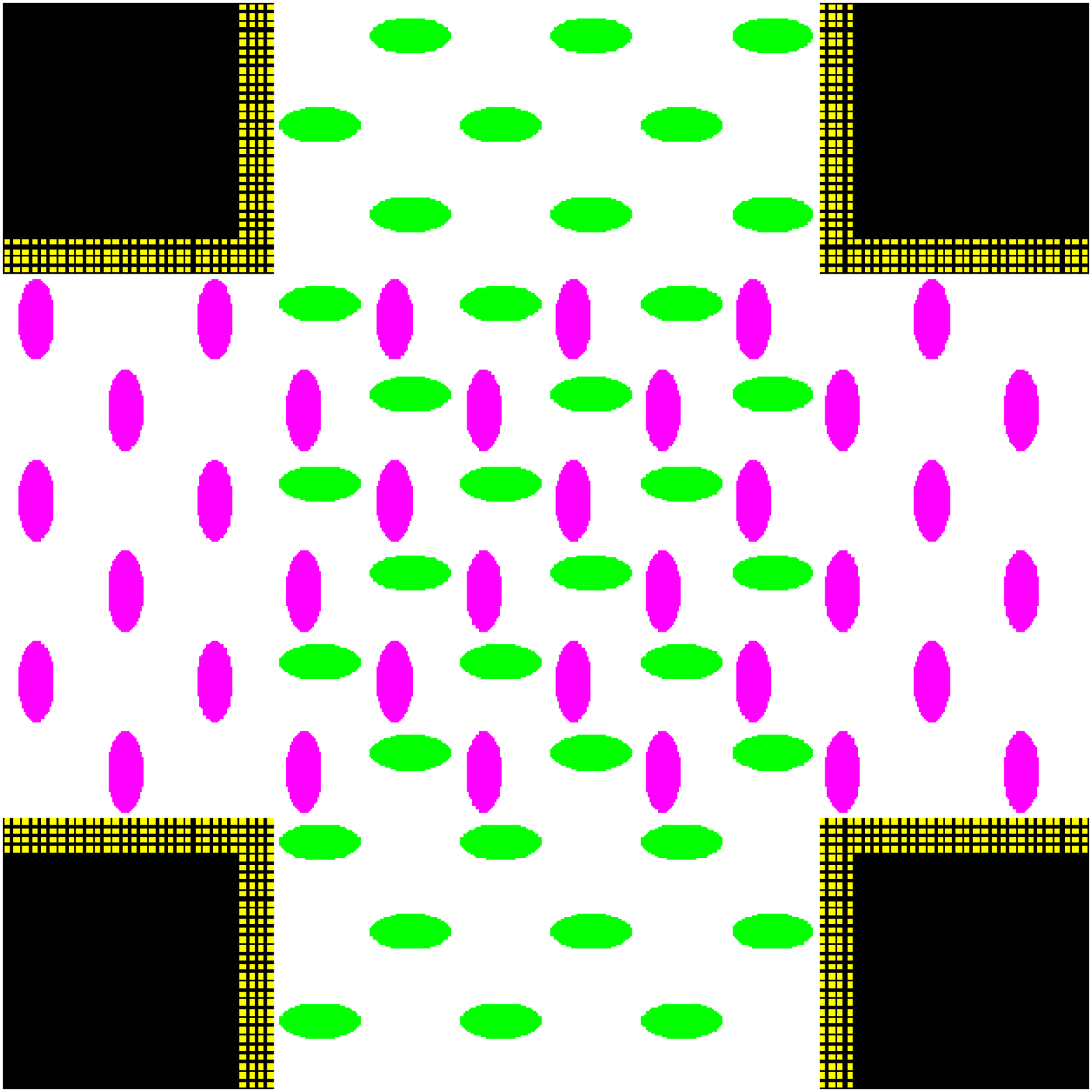}
\caption{Snapshots for ``marching'' pedestrians, unidirectional flows with 2 ped/m$^2$ density.}
\label{fv1}
\end{center}
\end{figure}

\begin{figure}[t]
\begin{center}
\includegraphics[width=0.28\textwidth]{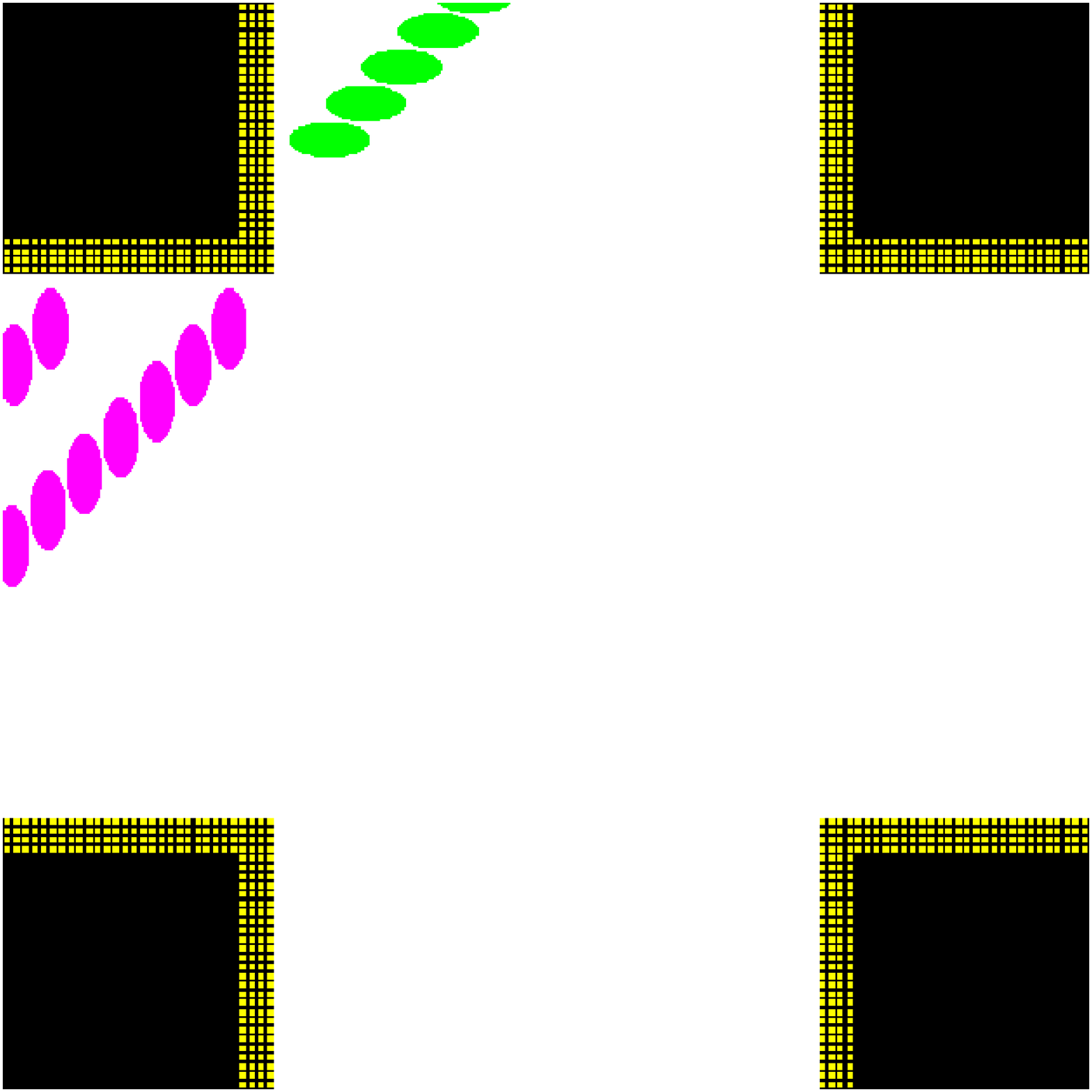}\hspace{.05\textwidth}\includegraphics[width=0.28\textwidth]{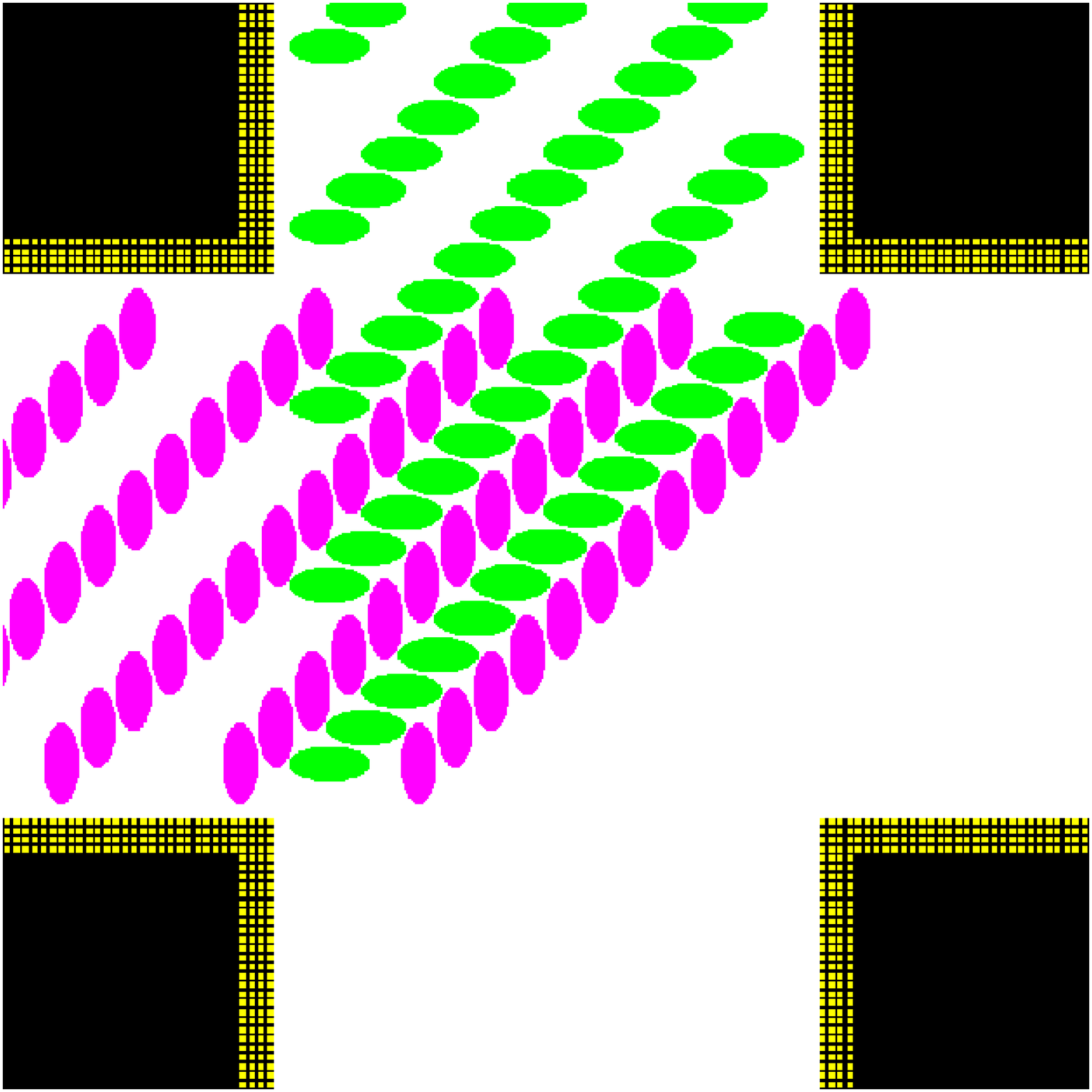}\hspace{.05\textwidth}\includegraphics[width=0.28\textwidth]{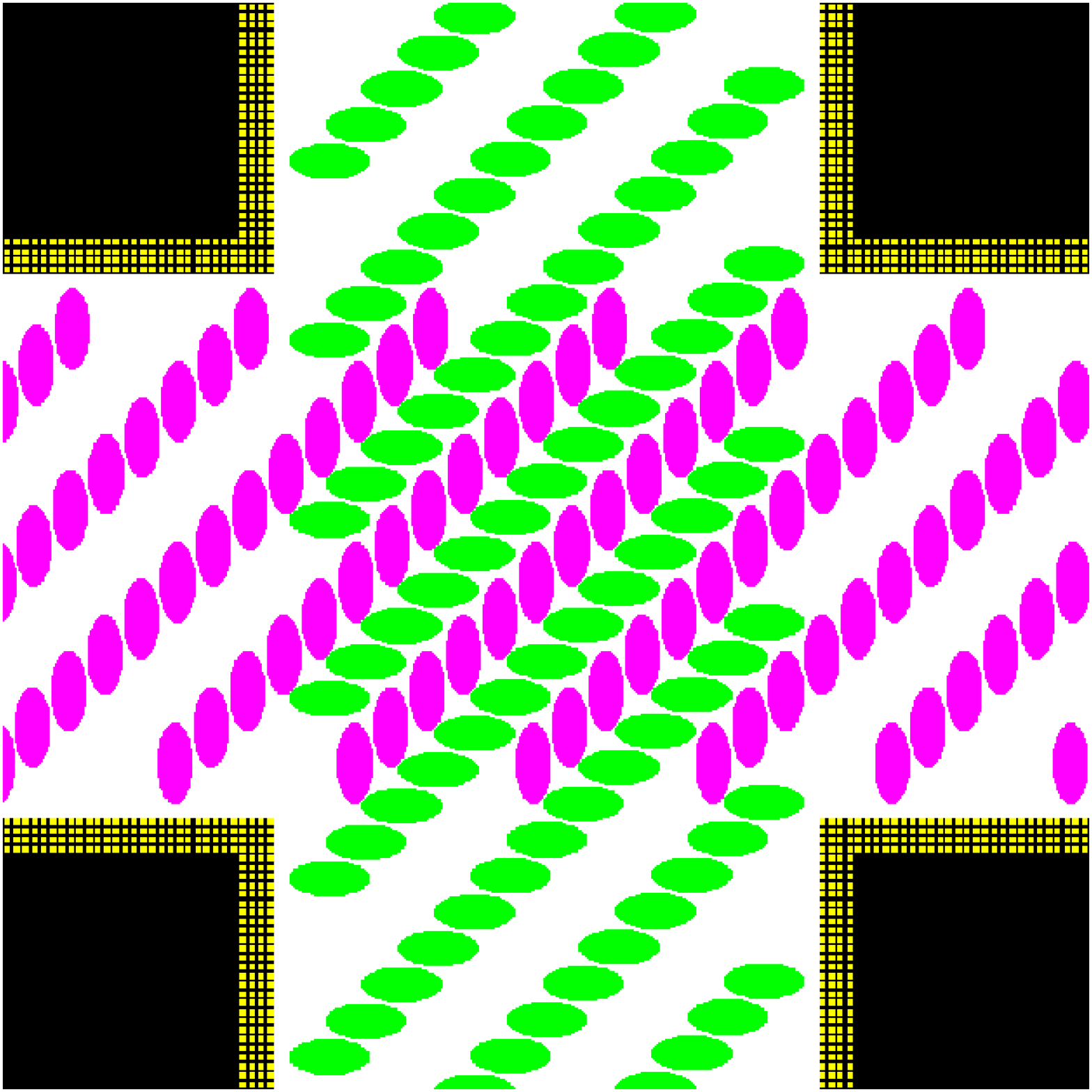}
\caption{Snapshots for ``marching'' pedestrians, unidirectional flows with 4.5 ped/m$^2$ density.}
\label{fv2}
\end{center}
\end{figure}

\begin{figure}[t]
\begin{center}
\includegraphics[width=0.28\textwidth]{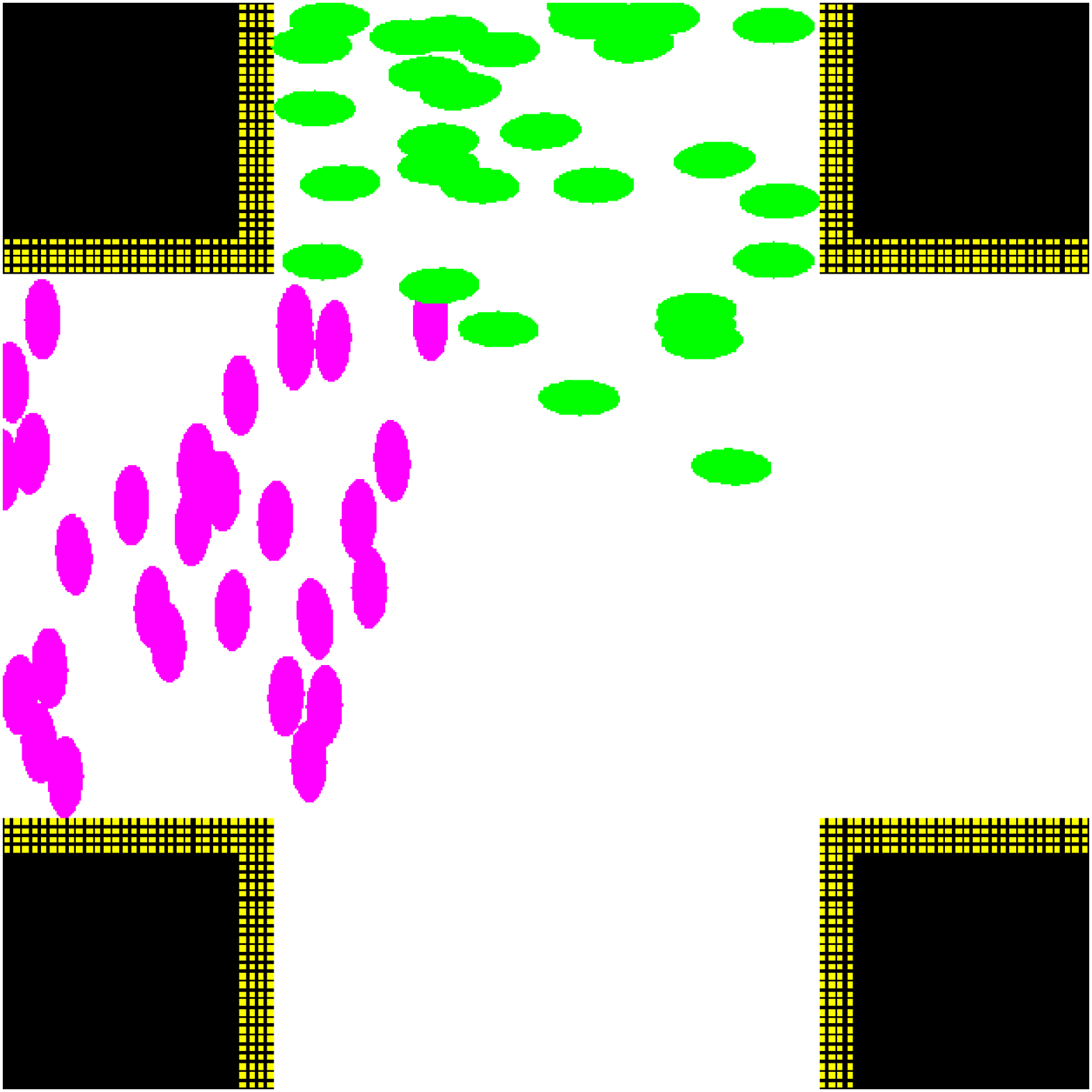}\hspace{.05\textwidth}\includegraphics[width=0.28\textwidth]{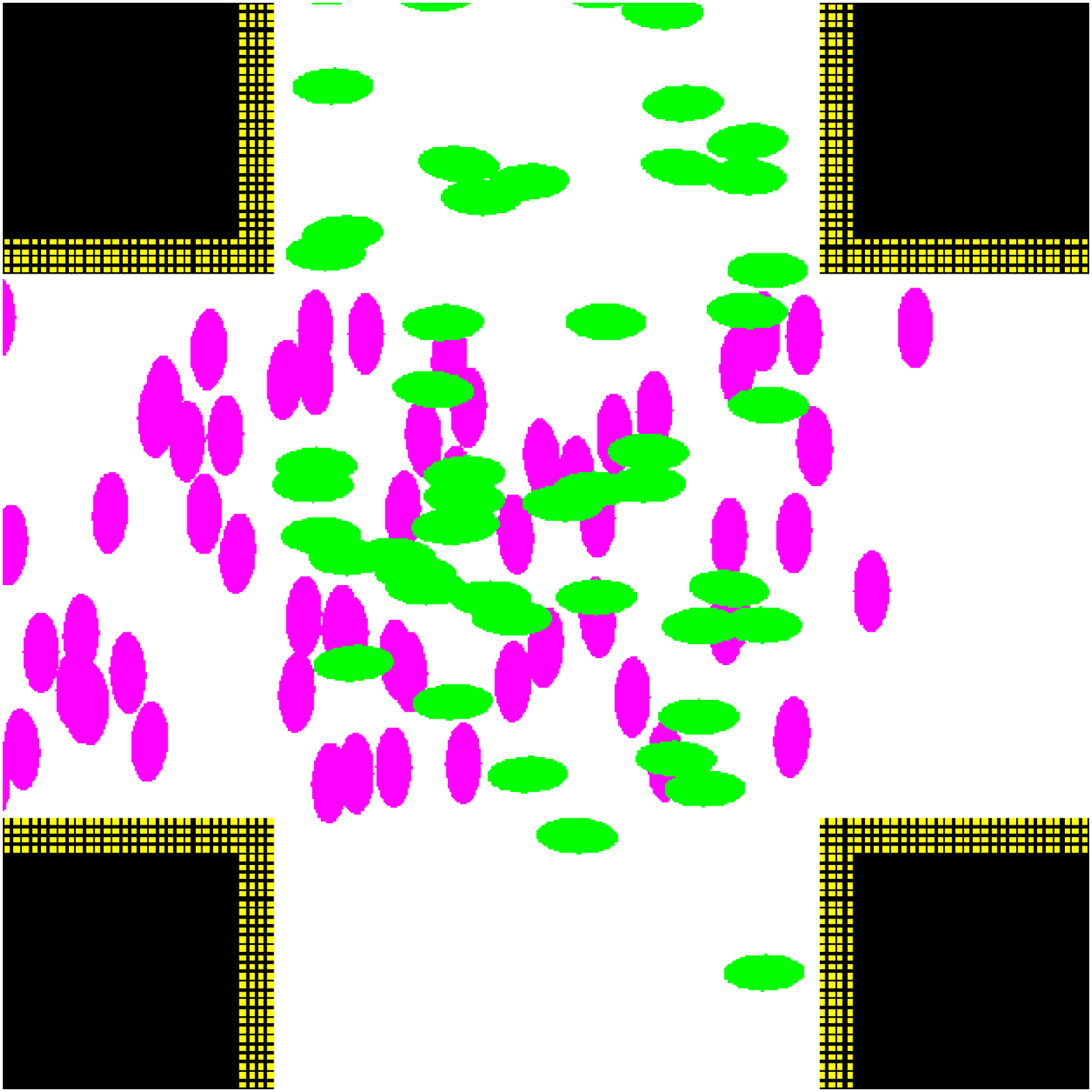}\hspace{.05\textwidth}\includegraphics[width=0.28\textwidth]{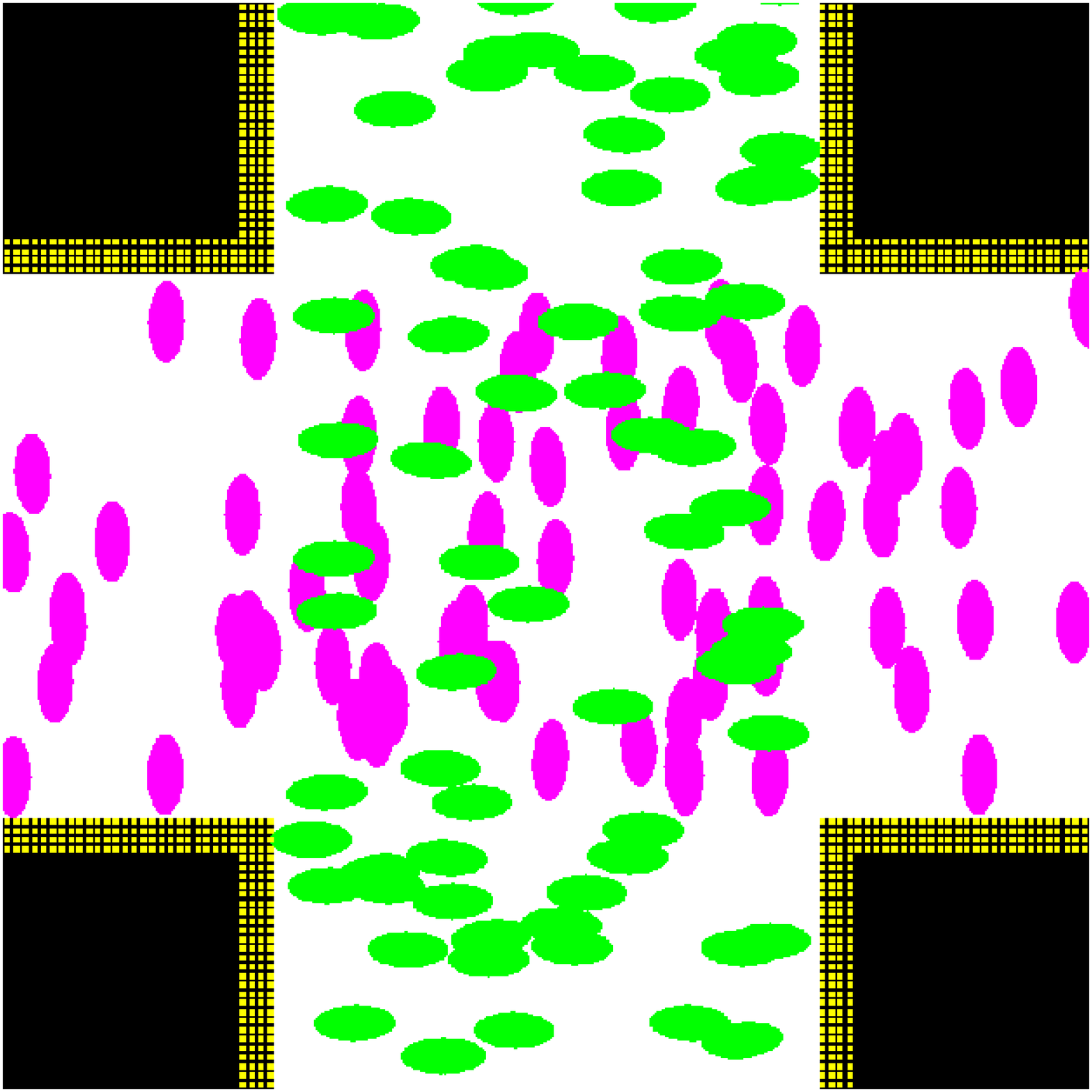}
\caption{Snapshots for ``bodiless'' pedestrians, unidirectional flows with 4 ped/m$^2$ density.}
\label{fv3}
\end{center}
\end{figure}

\begin{figure}[t]
\begin{center}
\includegraphics[width=0.28\textwidth]{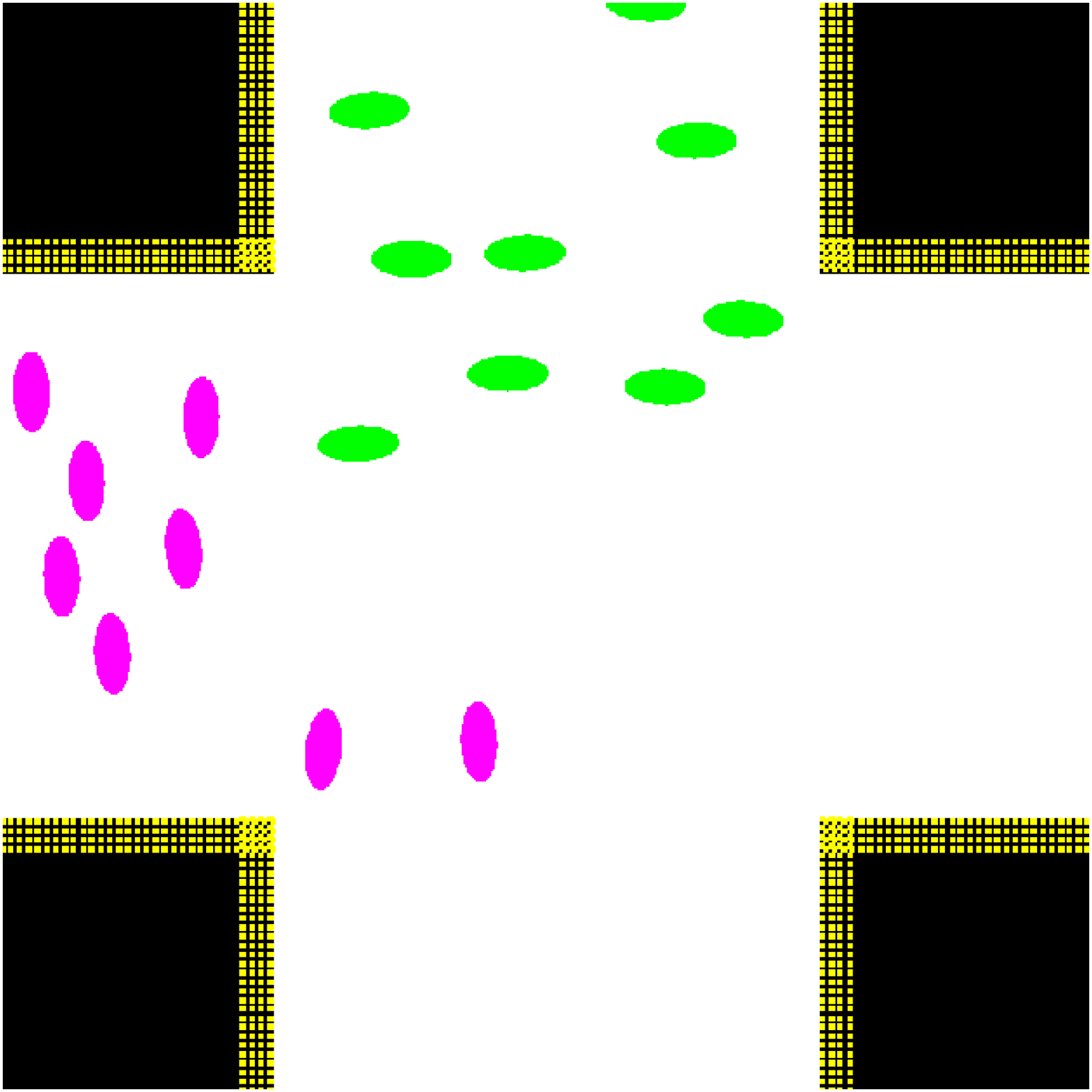}\hspace{.05\textwidth}\includegraphics[width=0.28\textwidth]{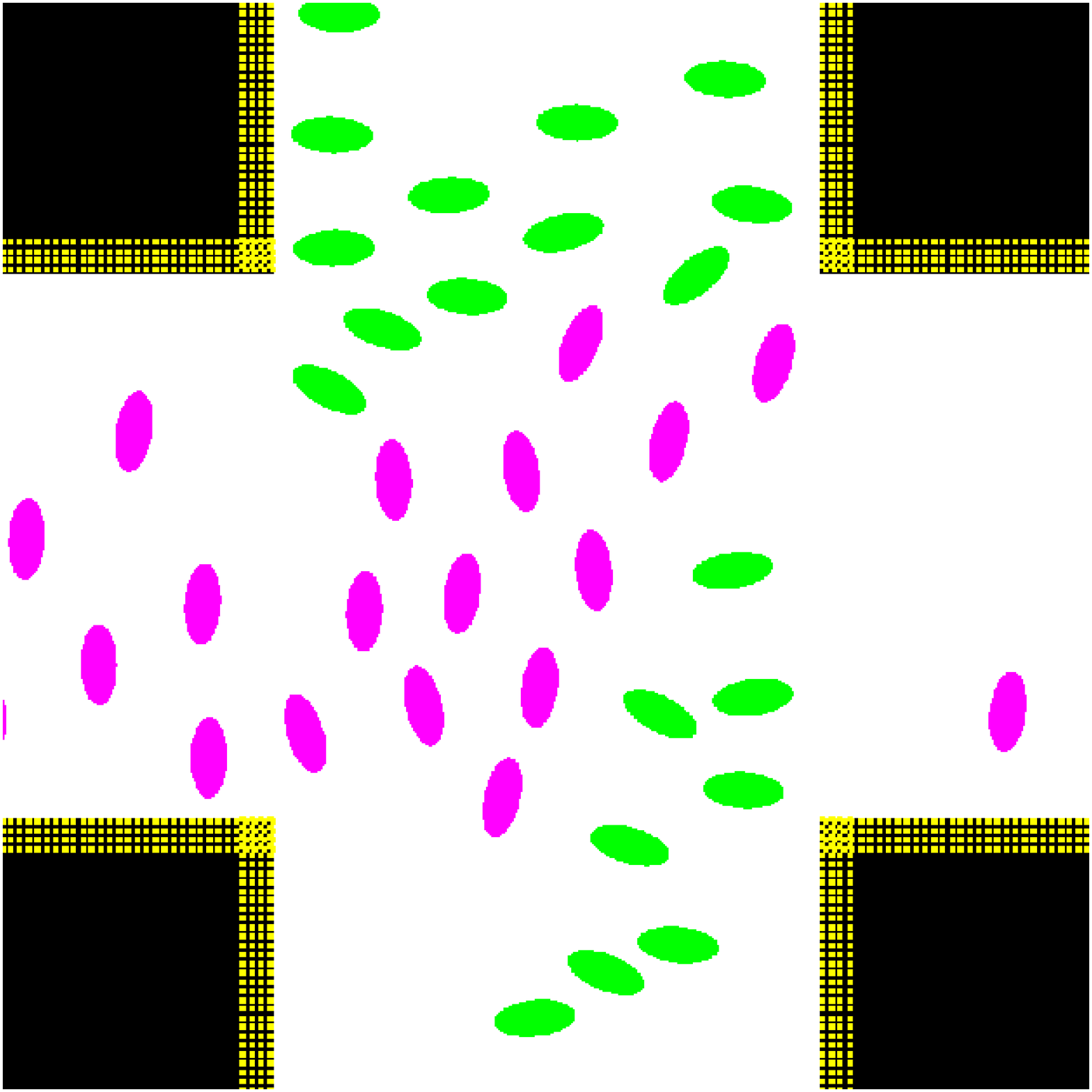}\hspace{.05\textwidth}\includegraphics[width=0.28\textwidth]{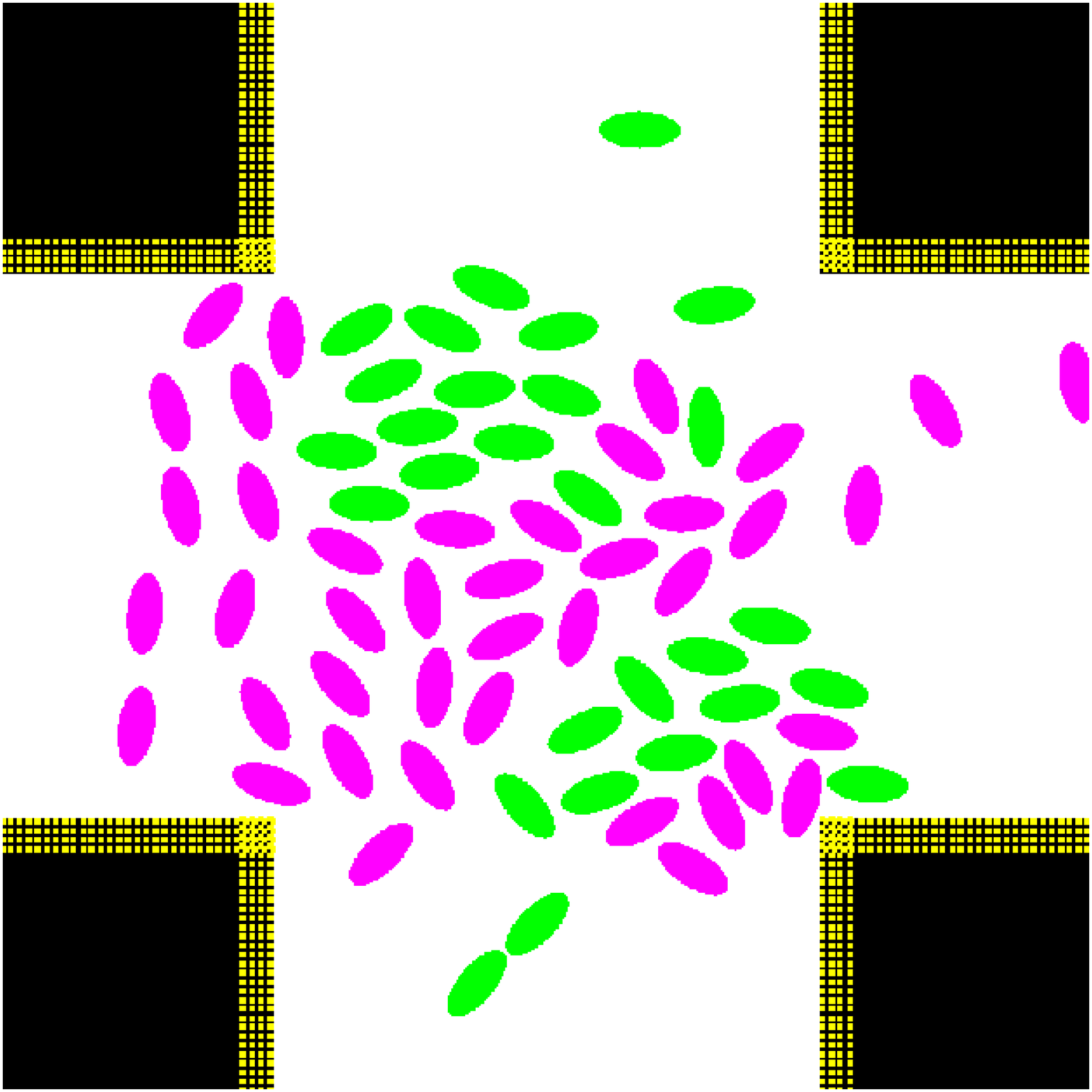}
\caption{Snapshots for ``realistic'' pedestrians, unidirectional flows with 1 ped/m$^2$ density.}
\label{fv4}
\end{center}
\end{figure}

\begin{figure}[t]
\begin{center}
\includegraphics[width=0.28\textwidth]{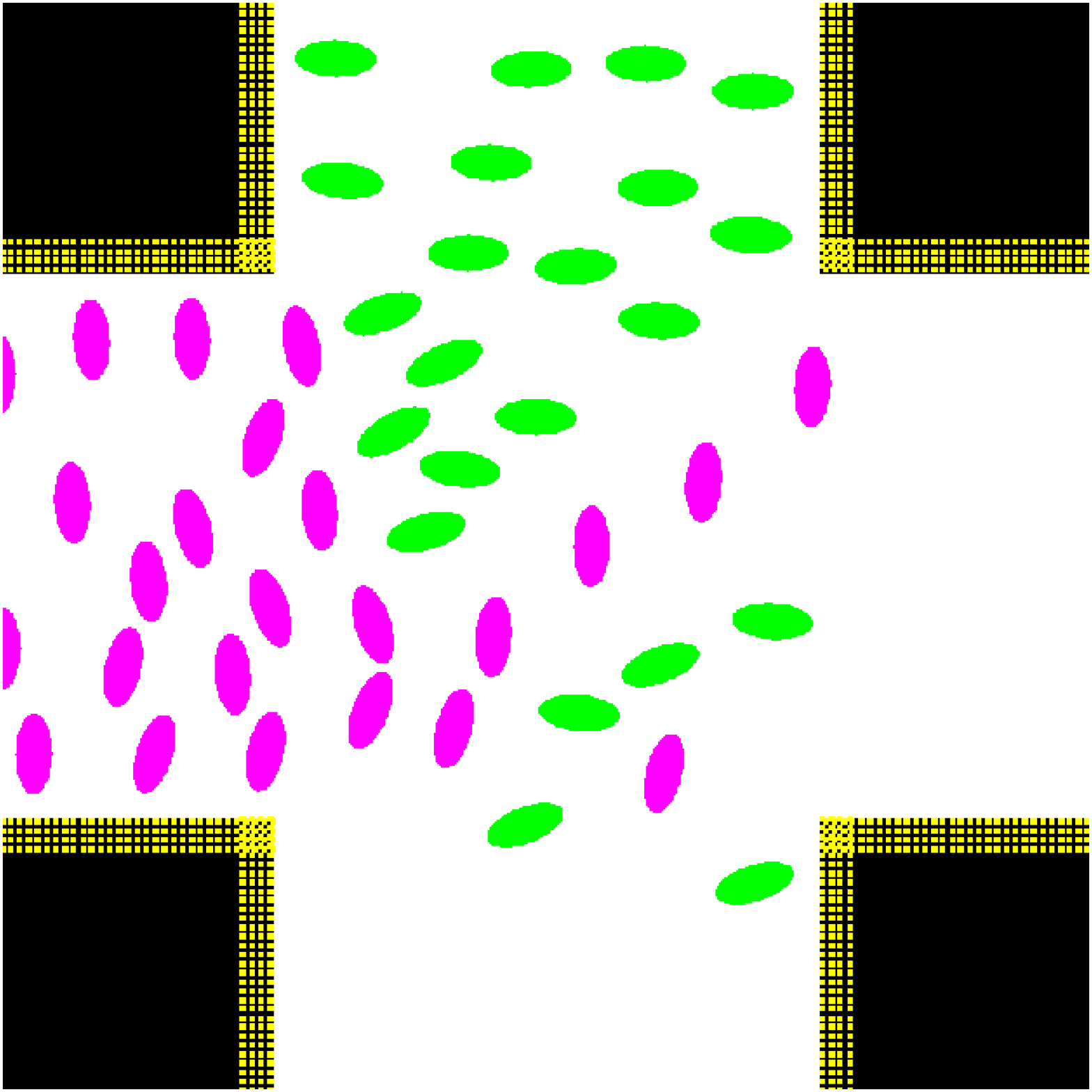}\hspace{.05\textwidth}\includegraphics[width=0.28\textwidth]{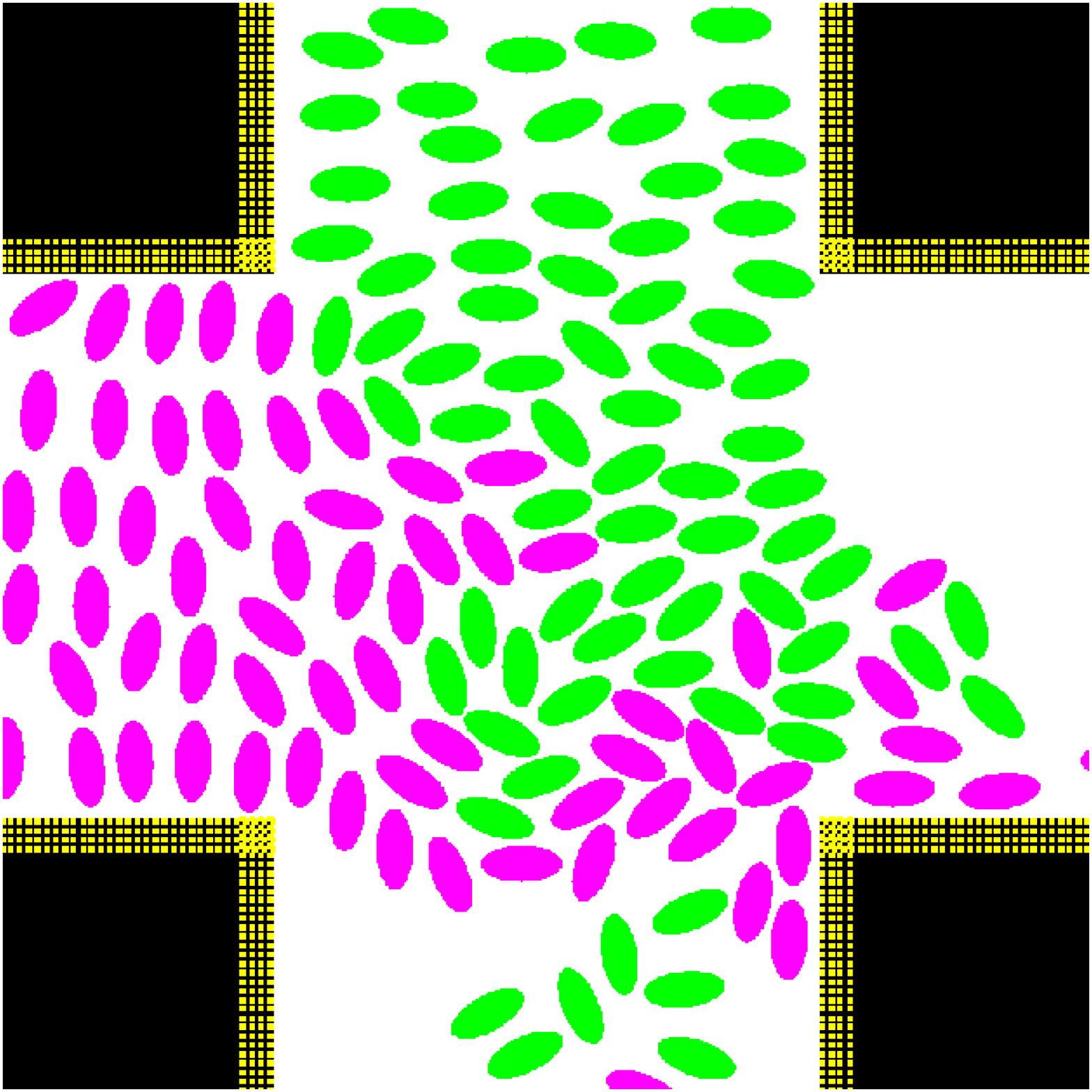}\hspace{.05\textwidth}\includegraphics[width=0.28\textwidth]{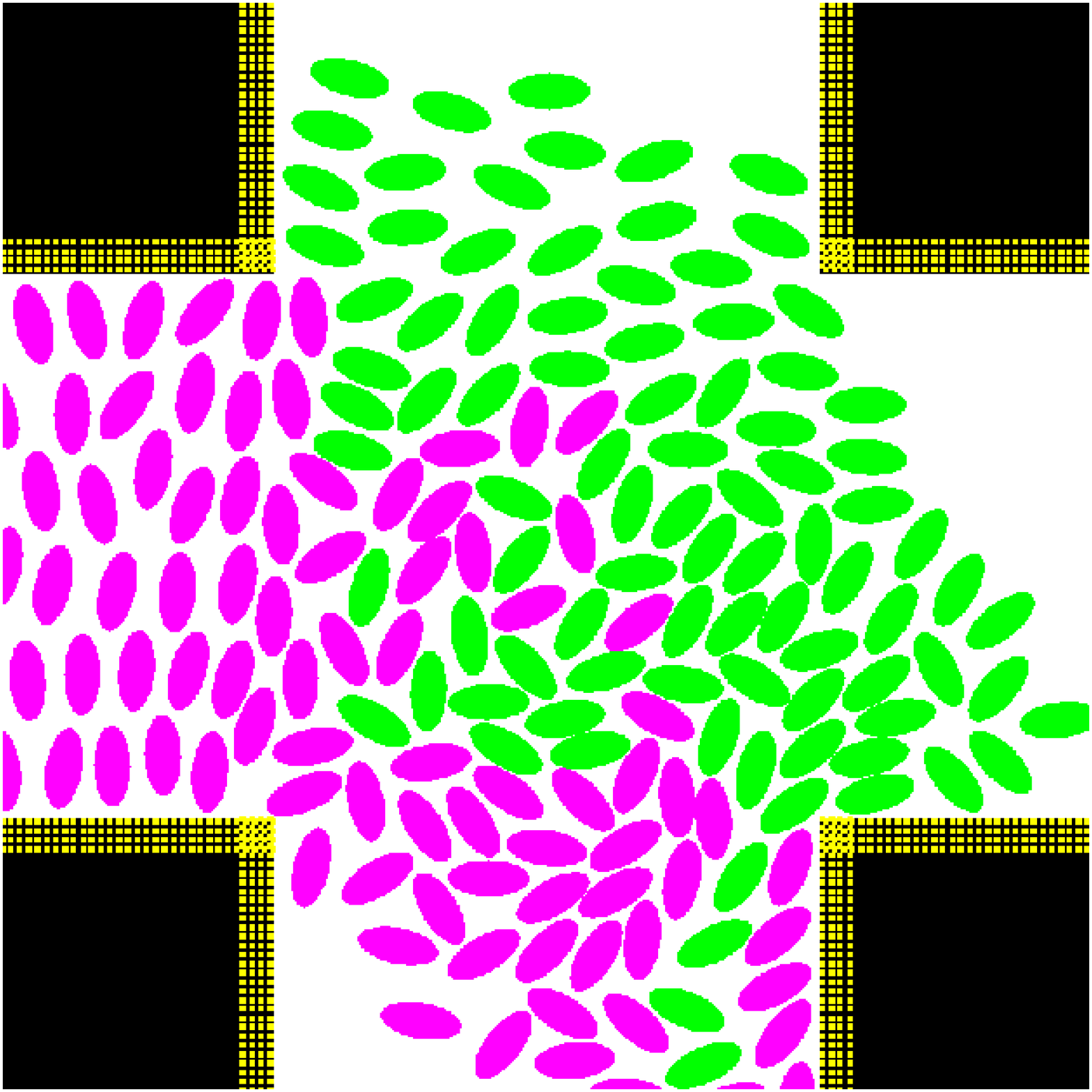}
\caption{Snapshots for ``realistic'' pedestrians, unidirectional flows with 2 ped/m$^2$ density.}
\label{fv5}
\end{center}
\end{figure}

\begin{figure}[t]
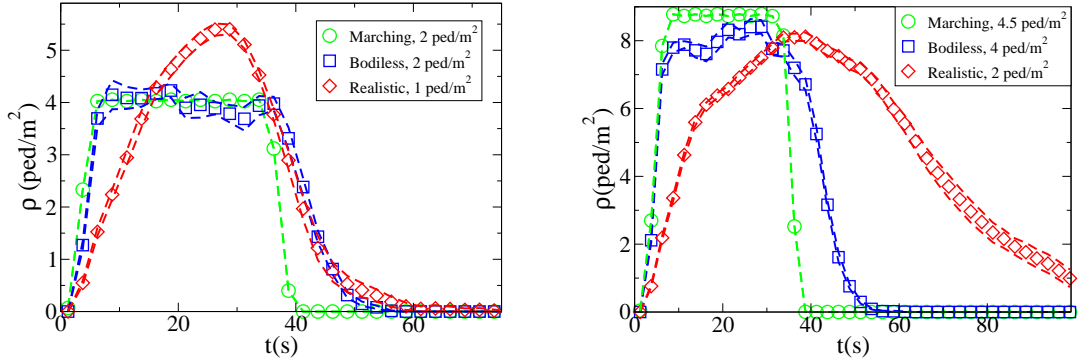

\begin{center}
\includegraphics[width=0.45\textwidth]{f2a.eps}\hspace{.07\textwidth}\includegraphics[width=0.45\textwidth]{f2b.eps}
\caption{Density in the crossing area as a function of time for all scenarios. Densities are averaged over 10 different randomly chosen initial conditions
  (excluding for the deterministic ``marching'' settings) and on time intervals of 2.5 s. Dashed lines provide standard error bars.}
\label{f2}
\end{center}
\end{figure}

\begin{figure}[t]
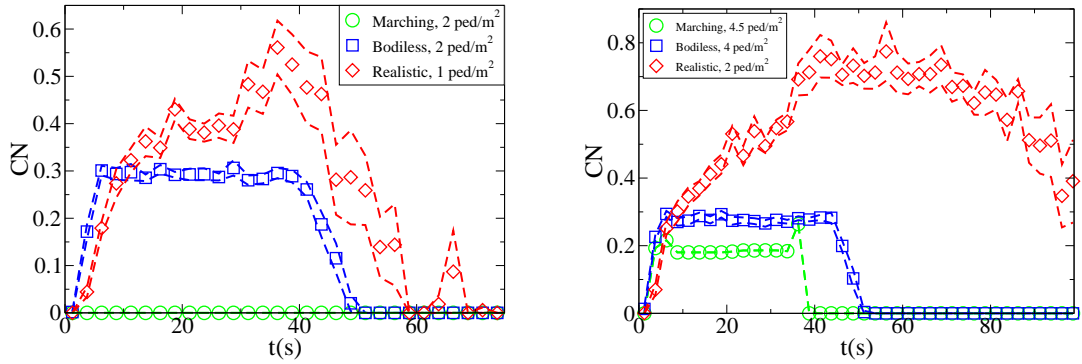

\begin{center}
\includegraphics[width=0.45\textwidth]{f3a.eps}\hspace{.07\textwidth}\includegraphics[width=0.45\textwidth]{f3b.eps}
\caption{Maximum $CN$ (maximum over all cells, averaged over 10 different randomly chosen initial conditions) as a function of time for all scenarios. Dashed lines provide standard error bars.}
\label{f3}
\end{center}
\end{figure}

\begin{figure}[t]
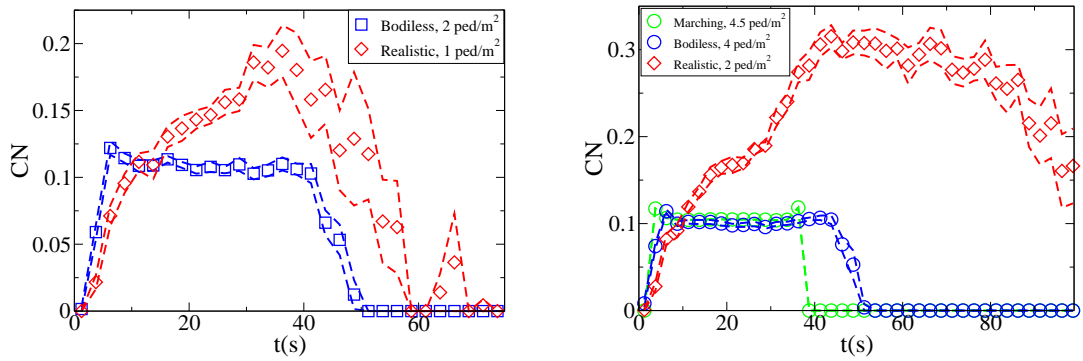

\begin{center}
\includegraphics[width=0.45\textwidth]{f4a.eps}\hspace{.07\textwidth}\includegraphics[width=0.45\textwidth]{f4b.eps}
\caption{Average $CN$ (computed as an average over non-zero $CN$ cells, and averaged over 10 different randomly chosen initial conditions) as a function of time for all scenarios. Dashed lines provide standard error bars.}
\label{f4}
\end{center}
\end{figure}

\begin{figure}[t]
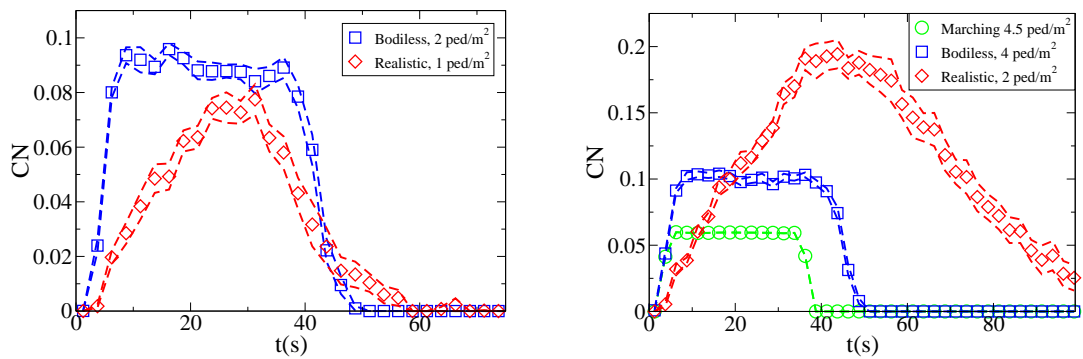

\begin{center}
\includegraphics[width=0.45\textwidth]{f4a_bis.eps}\hspace{.07\textwidth}\includegraphics[width=0.45\textwidth]{f4b_bis.eps}
\caption{Average $CN$ (computed as an average over all cells in the corridors, and averaged over 10 different randomly chosen initial conditions) as a function of time for all scenarios. Dashed lines provide standard error bars.}
\label{f4bis}
\end{center}
\end{figure}

The studied problem is the intersection of two corridors (width of 3 meters), each one with a high density/high flux uni-directional flow in it. We use three approaches for ``simulating'' this system
\begin{itemize}
\item ``Marching pedestrians'': the velocity and position of the pedestrians is decided in advance in a centralised way, such that even at a density as high as 9 pedestrians per squared meter the pedestrians do not need to slow down. Two uni-directional flow densities are proposed: $\approx$ 2 (Fig. \ref{fv1}) and $\approx$ 4.5 (Fig. \ref{fv2}) ped/m$^2$, while the walking velocity is fixed to 1 m/s, using respectively 408 and 816 pedestrians.
\item ``Bodiless pedestrians'': each pedestrian has a different velocity (from a Gaussian distribution centred on 1 m/s with $\sigma=0.1$ m/s) and performs a ``drifted random walk'' towards its goal. Nevertheless, physical dynamics is performed in the zero body-size limit (collisions happen only with the walls) and pedestrians do not perform collision avoidance. Two uni-directional flow densities are proposed: 2 and 4 (Fig. \ref{fv3}) ped/m$^2$, using respectively 408 and 816 pedestrians.
\item ``Realistic'': pedestrians have rigid bodies ($a=0.225$ and $b=0.1$ m ellipses) and perform collision avoidance.   Two uni-directional flow densities are proposed: 1 (Fig. \ref{fv4}) and 2 (Fig. \ref{fv5}) ped/m$^2$, using respectively 128 and 256 pedestrians. Velocity is Gaussian distributed around 1 m/s  with $\sigma=0.1$ m/s (here ``realistic'' refers to the fact that the pedestrians have a finite and realistic body size, and try to stop/avoid collisions; their behaviour may still differ from the one of actual pedestrians).
\end{itemize}
The time evolution of the average  density in the crossing area is shown for each setting and model in Fig. \ref{f2}. Unidirectional flows and conditions have been chosen in
such a way to have similar density peak values, and similar crossing time scales in each one of the ``lower density'' and ``higher density'' setting for all models.
Anyway, as 128 ``realistic'' pedestrians need roughly the same time to clear the crossing area than 408 ``bodiless'' one, and 256 ``realistic'' ones double the time than 816 ``bodiless'' ones, it is clear that the
flux (and velocity) is strongly decreased in the ``realistic'' case. Despite this, as shown in Fig. \ref{f3}, the maximum of $CN$ attains an higher value in the ``realistic case''. In this (recognising the crowd congestion
regardless of a slower velocity/lower flux) clearly the velocity scale independence of $CL$ plays an important role. Figs. \ref{f4} and \ref{f4bis} show results concerning average $CN$.


\begin{figure}[t]
\begin{center}
  \includegraphics[width=0.4\textwidth]{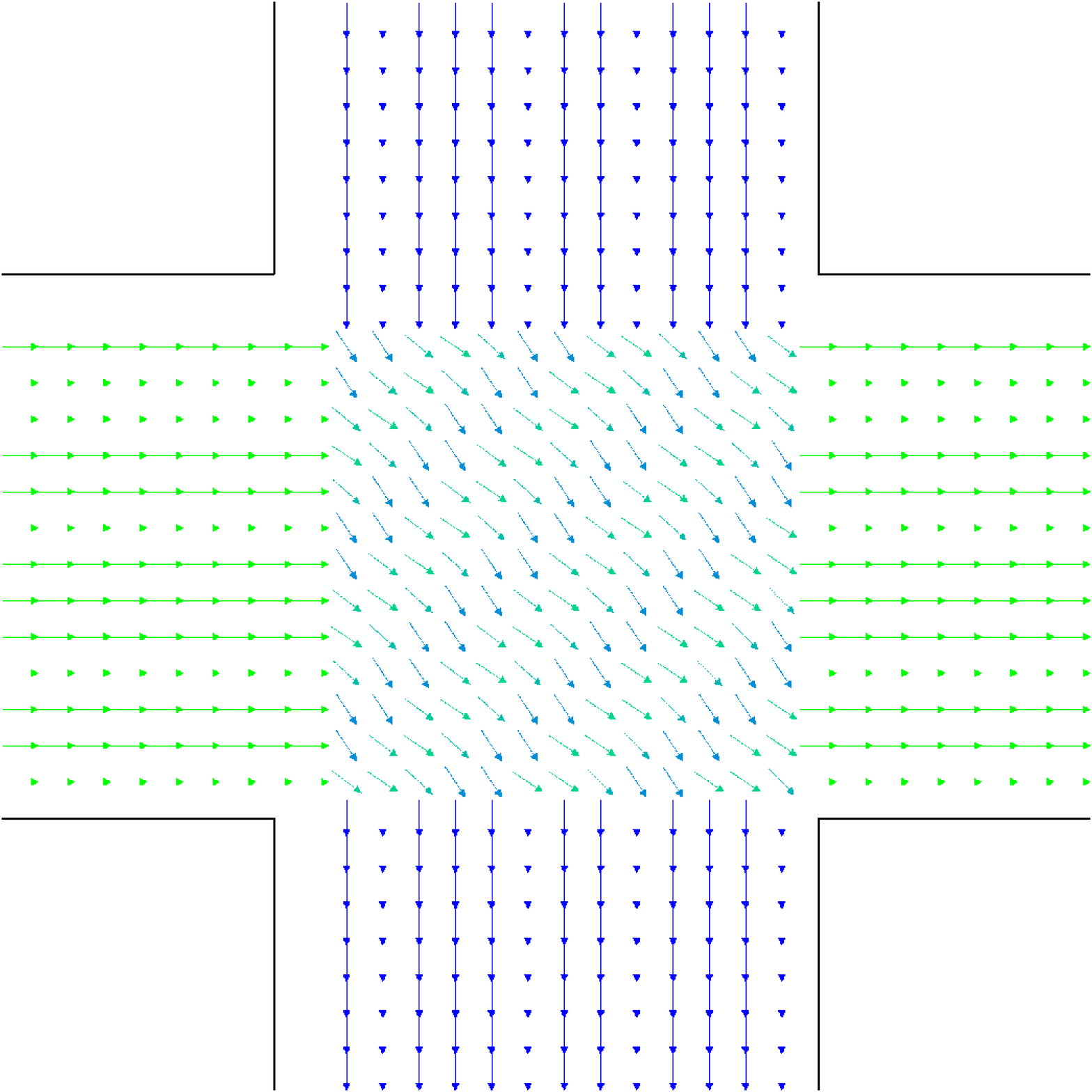}\includegraphics[width=0.1\textwidth]{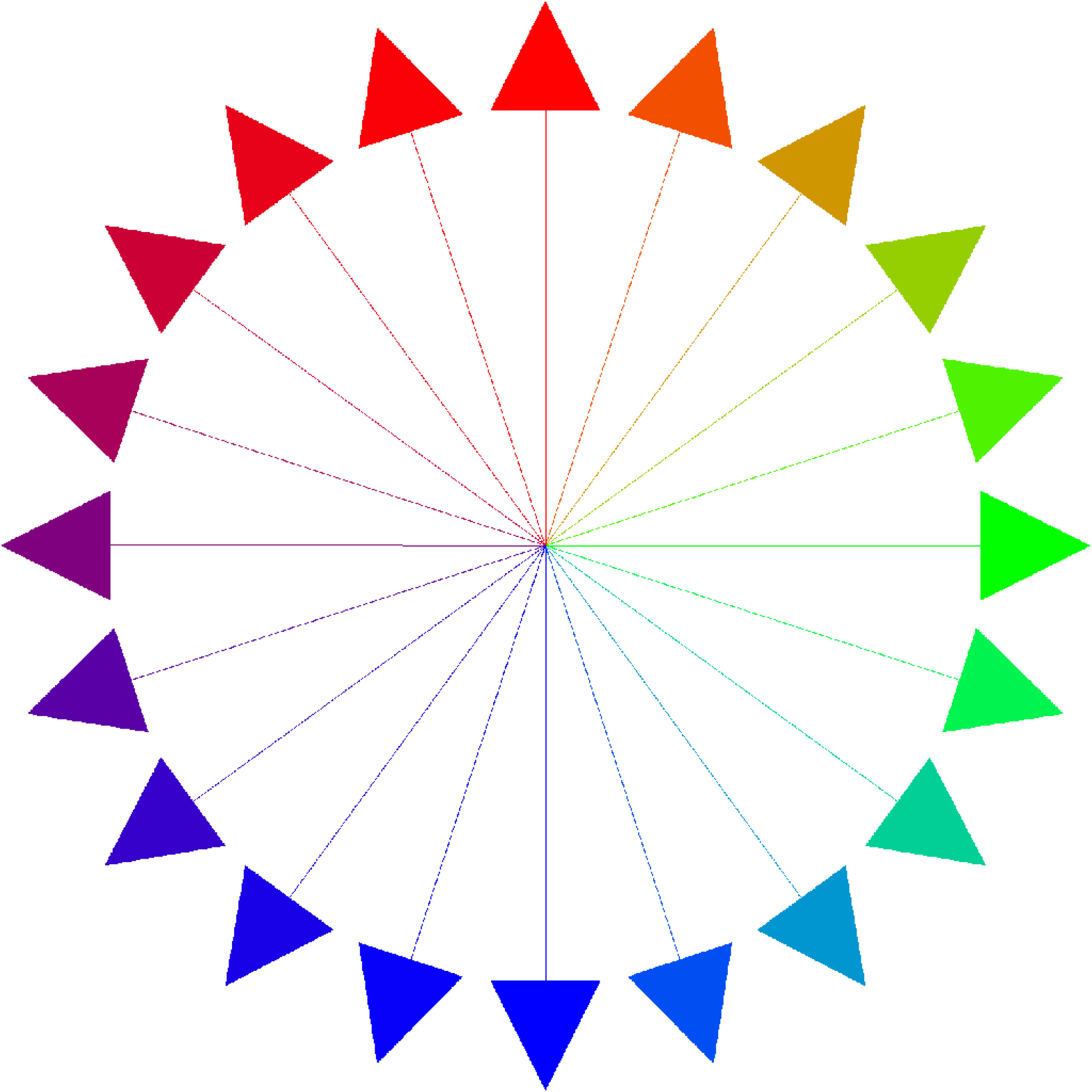}\hspace{0.02\textwidth}
  \includegraphics[width=0.4\textwidth]{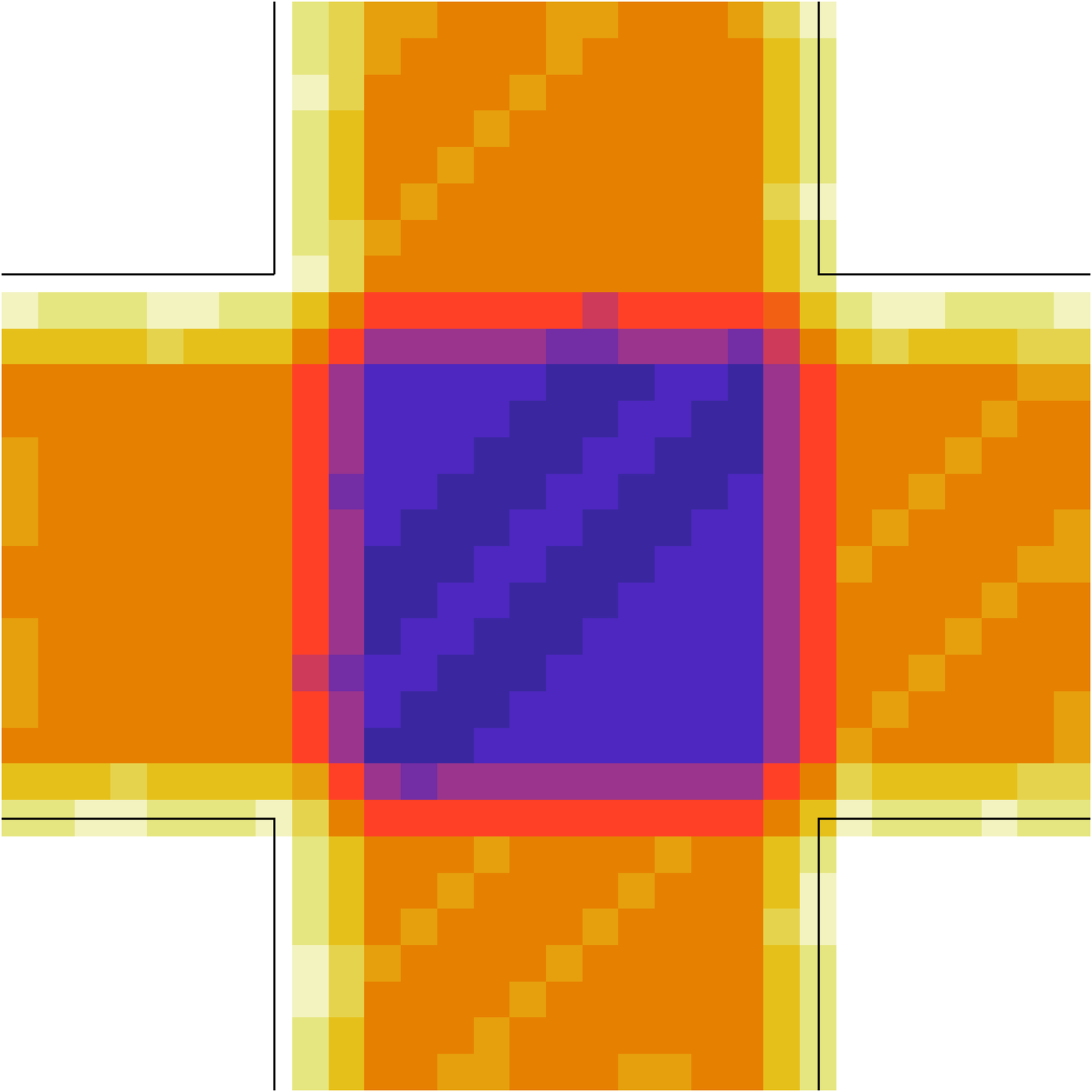}\hspace{0.02\textwidth}\includegraphics[width=0.05\textwidth]{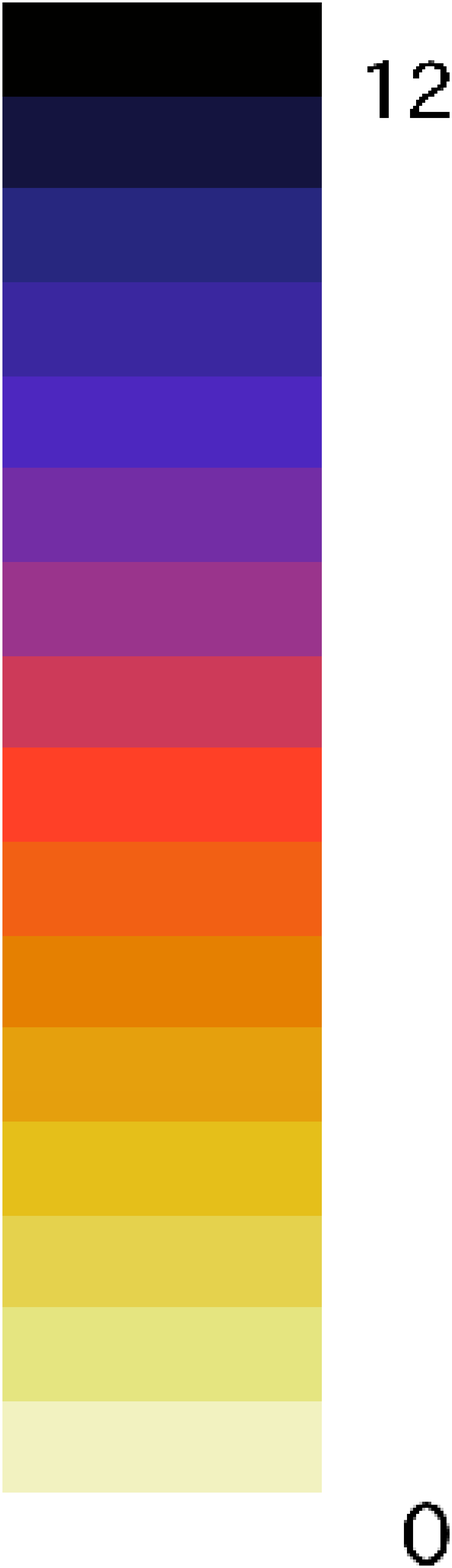}
  \includegraphics[width=0.4\textwidth]{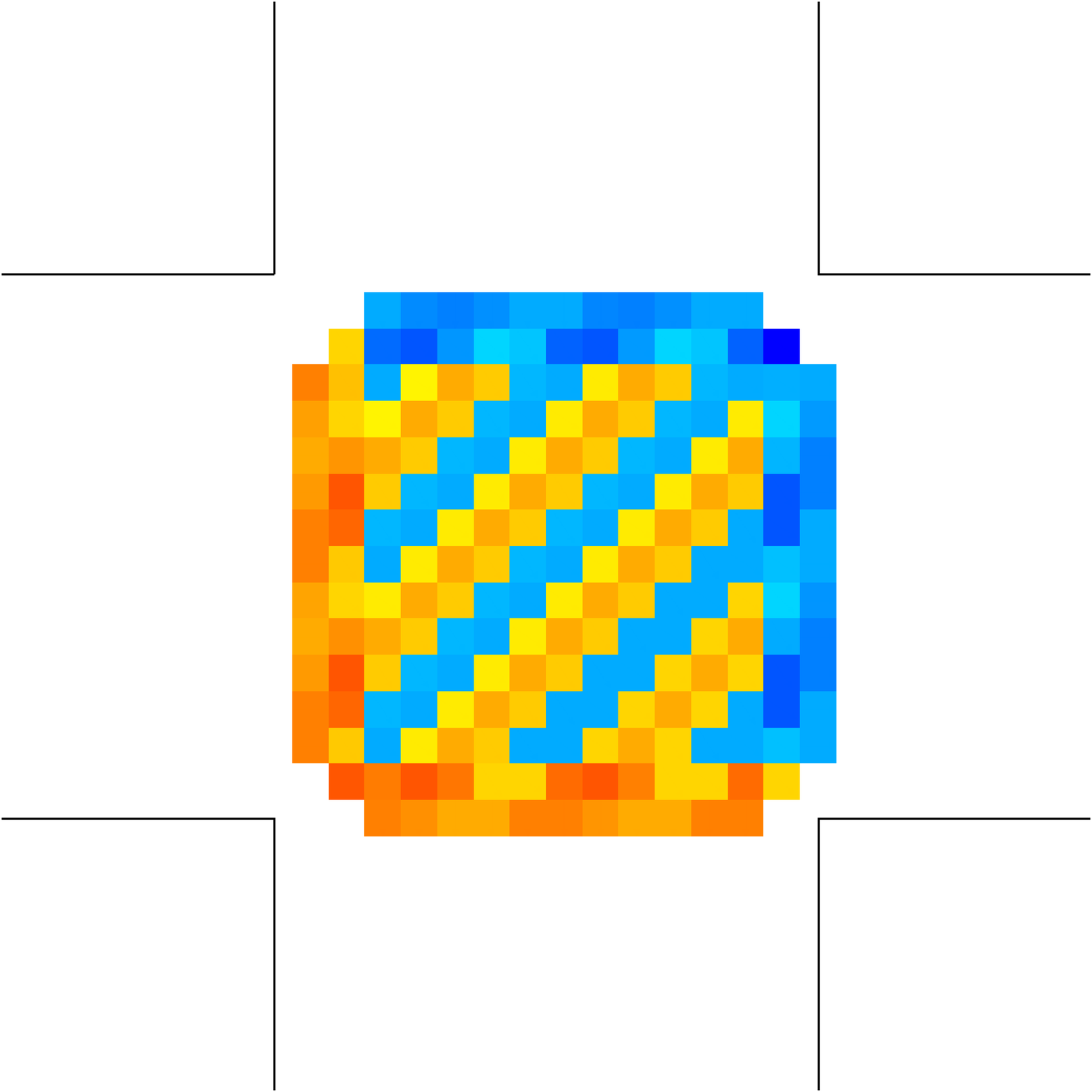}\hspace{0.02\textwidth}\includegraphics[width=0.05\textwidth]{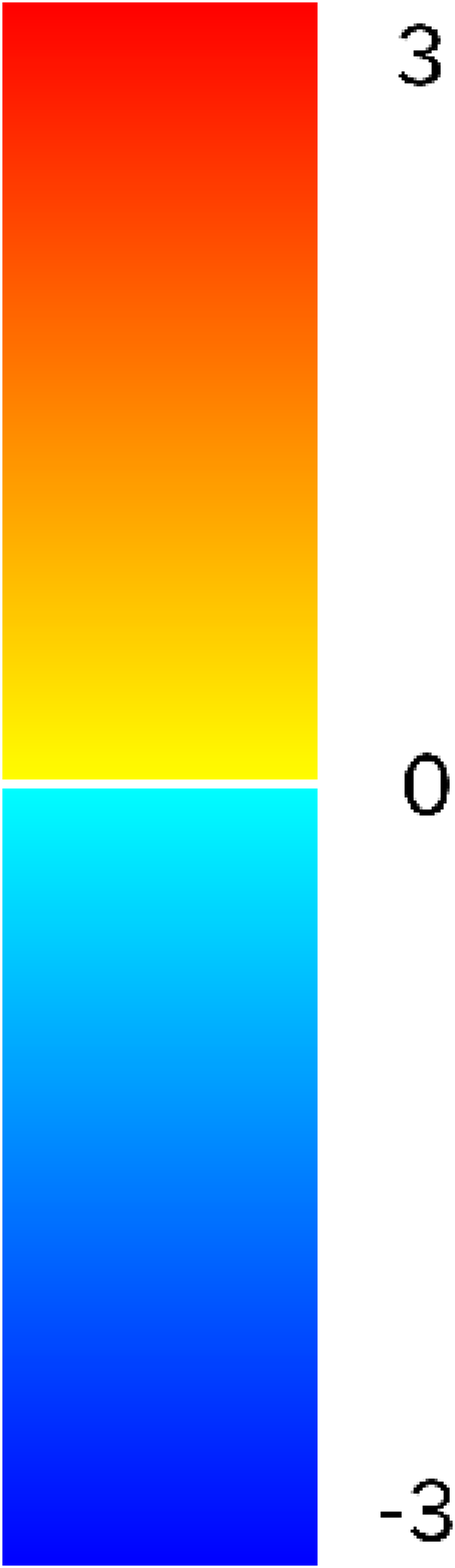}\hspace{0.05\textwidth}
  \includegraphics[width=0.4\textwidth]{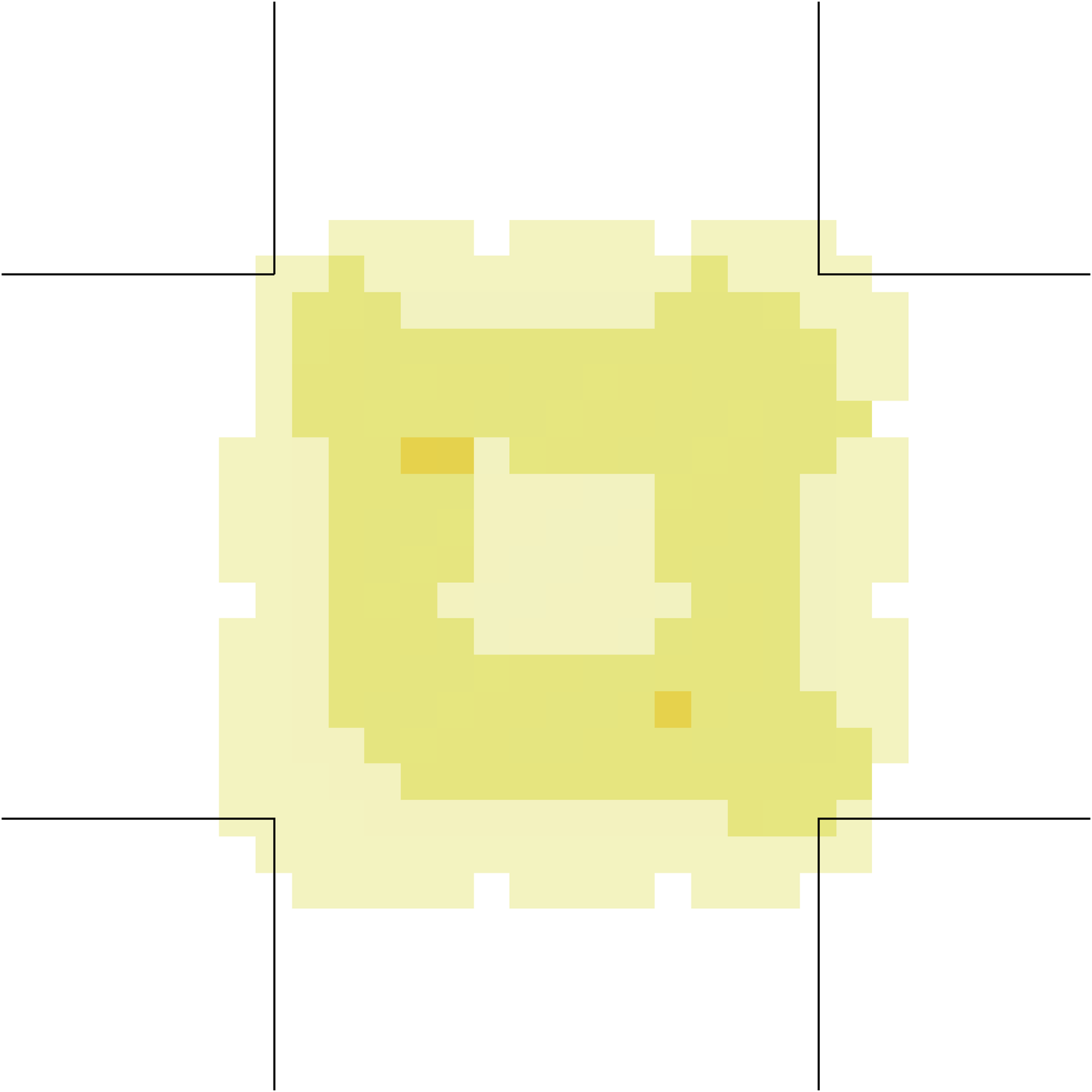}\hspace{0.02\textwidth}\includegraphics[width=0.05\textwidth]{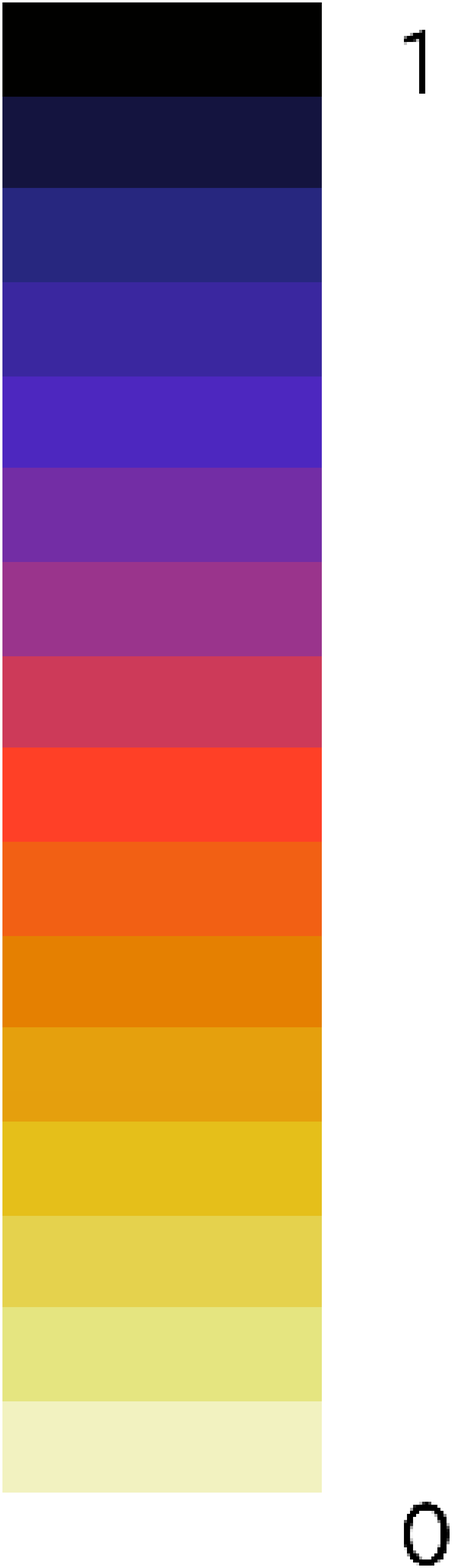}
  \caption{Higher density ``marching'' pedestrians at the time in which the maximum density is attained (8.78 ped/m$^2$ during the $[27.5,30)$ s interval). Top, left: $\mathbf{v}$ field; top, right: density field; bottom left: $(\nabla \wedge \mathbf{v})_z$ field; bottom, right: $CN$ field. In the velocity field, the length of the arrow is proportional to the magnitude (full length $v>0.5$), while the colour gives the orientation, as shown in the
      colour wheel legend. The density field is represented using a moving average over the Moore neighbourhood.}
\label{f5}
\end{center}
\end{figure}

\begin{figure}[t]
\begin{center}
  \includegraphics[width=0.4\textwidth]{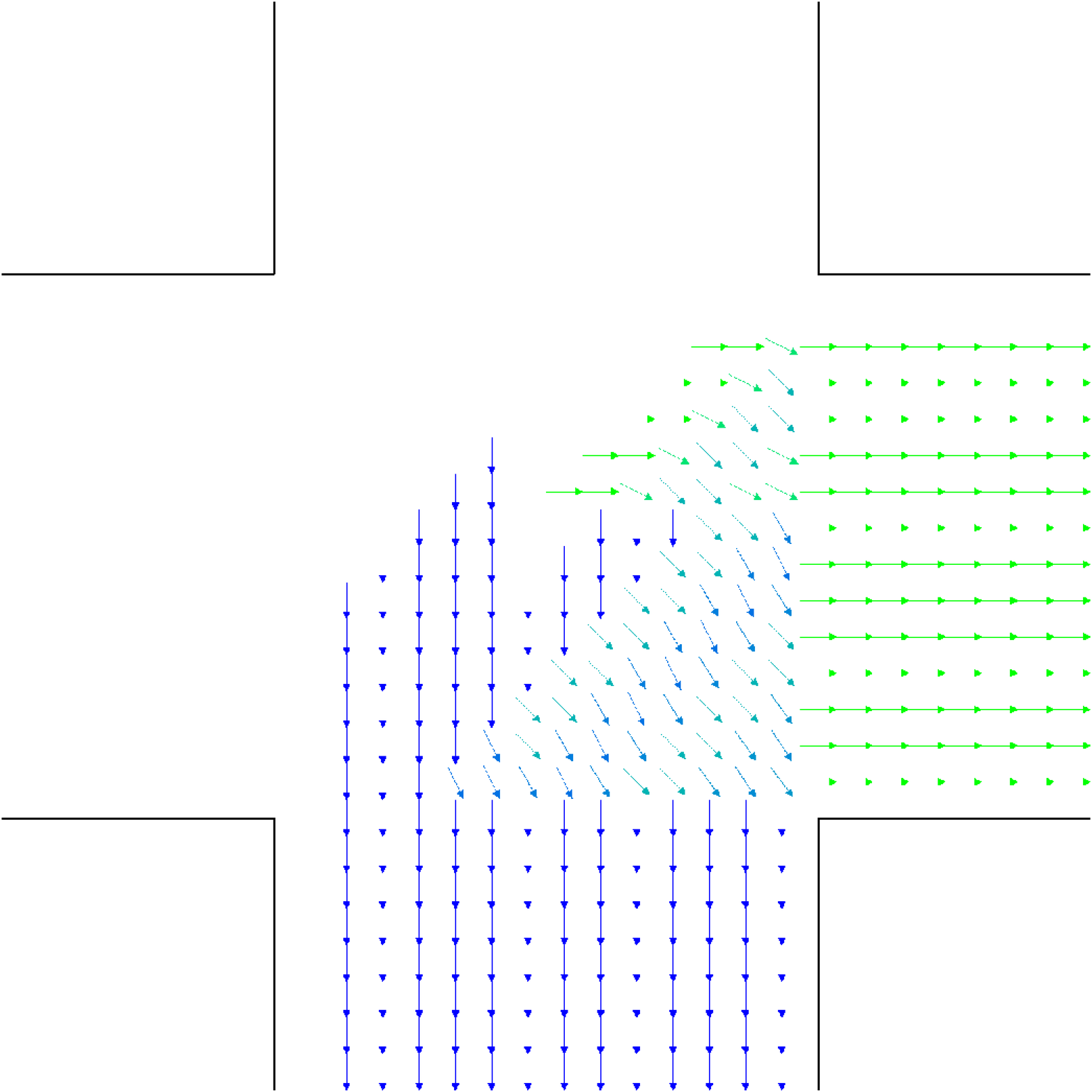}\includegraphics[width=0.1\textwidth]{frecce.eps}\hspace{0.02\textwidth}
  \includegraphics[width=0.4\textwidth]{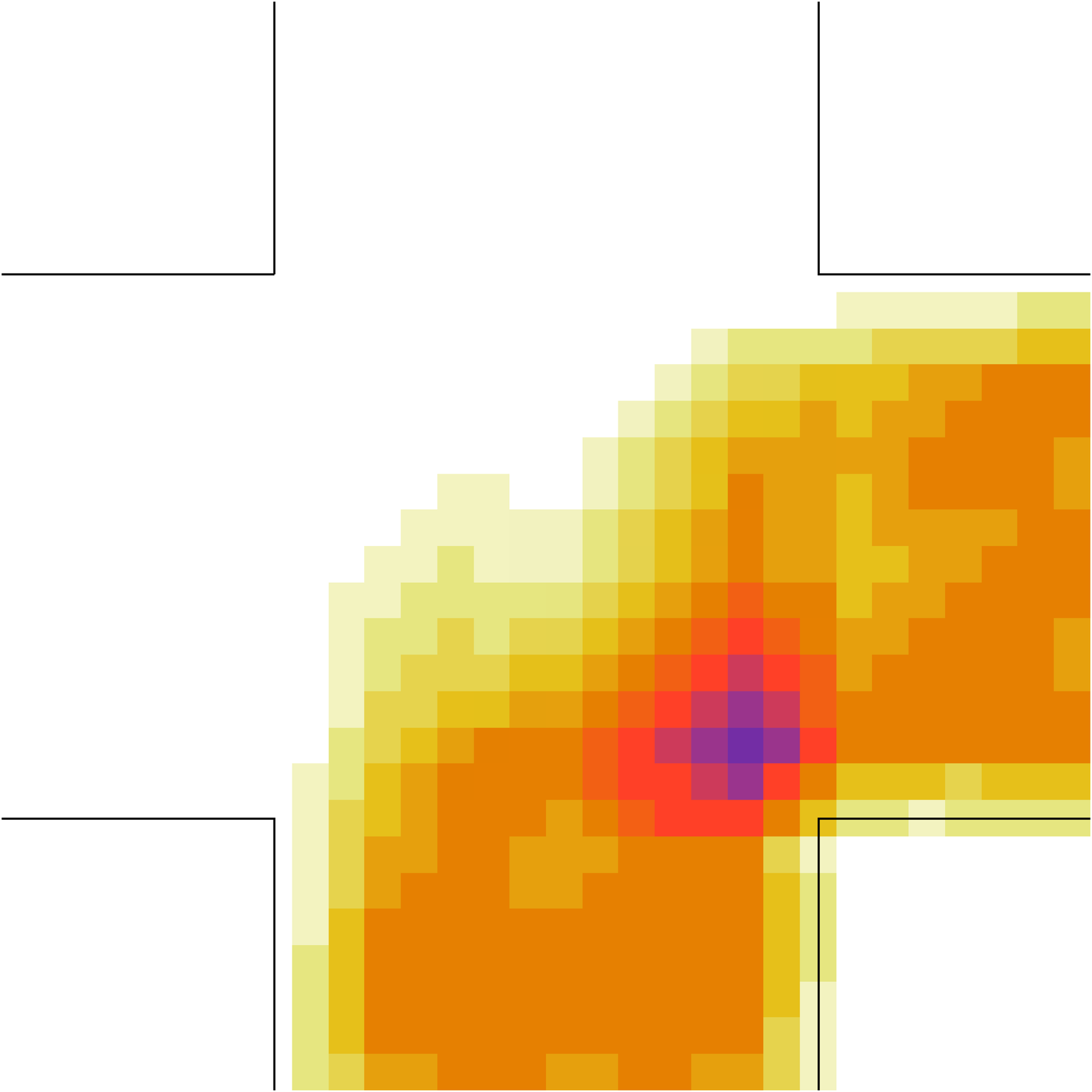}\hspace{0.02\textwidth}\includegraphics[width=0.05\textwidth]{scaledens.eps}
  \includegraphics[width=0.4\textwidth]{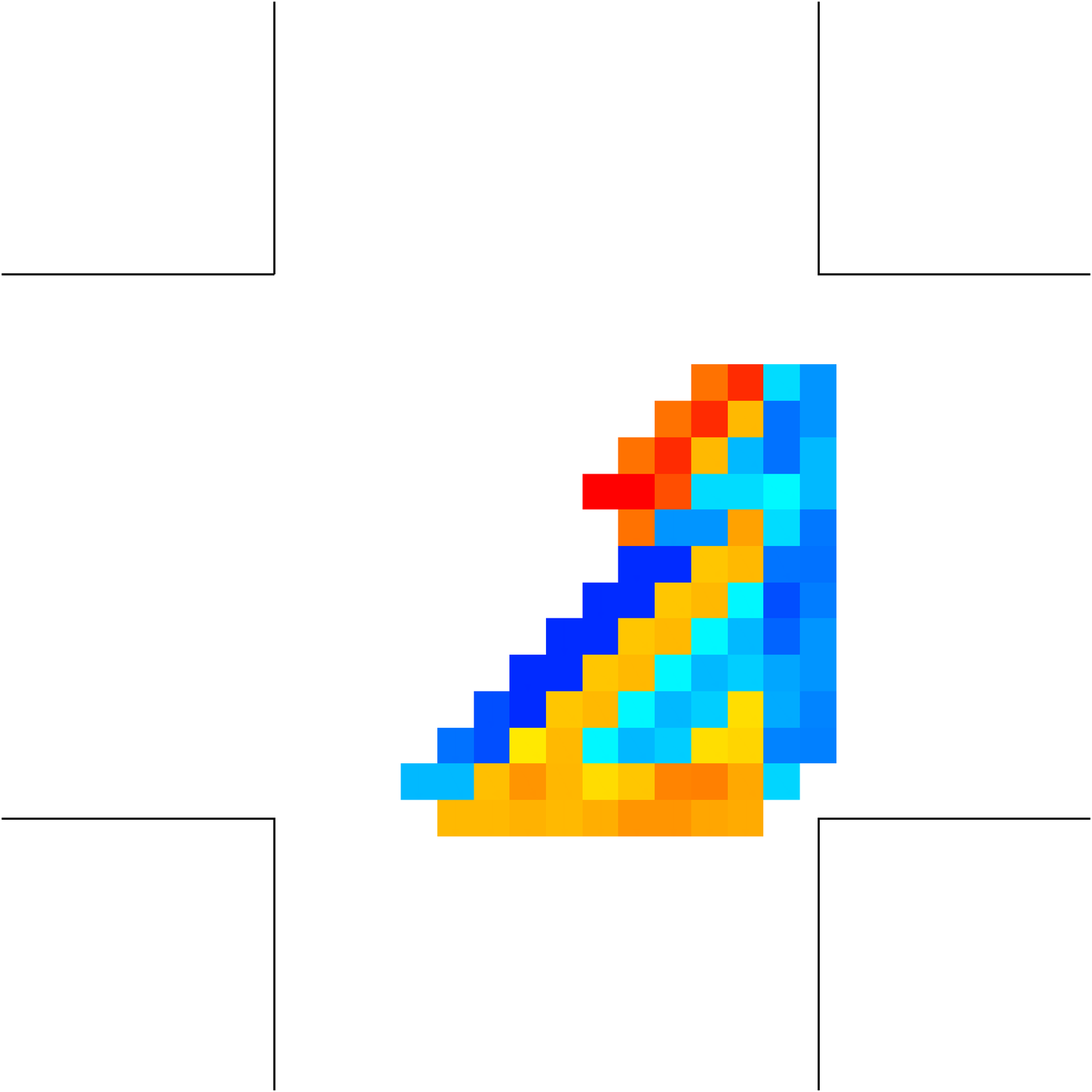}\hspace{0.02\textwidth}\includegraphics[width=0.05\textwidth]{scalepm0.eps}\hspace{0.05\textwidth}
  \includegraphics[width=0.4\textwidth]{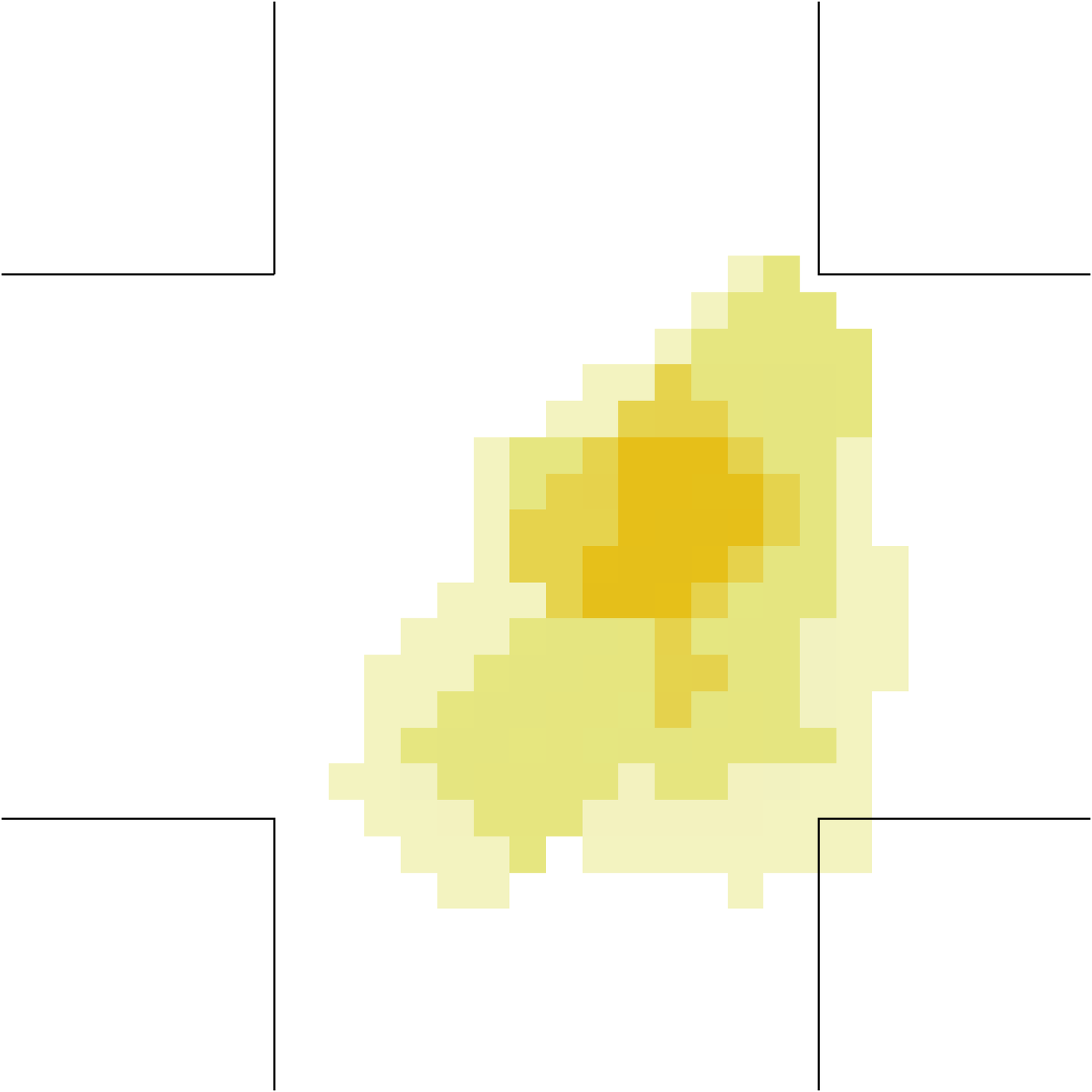}\hspace{0.02\textwidth}\includegraphics[width=0.05\textwidth]{scalenuovo.eps}
  \caption{Higher density ``marching'' pedestrians at the time in which the highest maximum $CN$ is attained (0.264 during the $[35,37.5)$ s interval). Top, left: $\mathbf{v}$ field; top, right: density field; bottom left: $(\nabla \wedge \mathbf{v})_z$ field; bottom, right: $CN$ field. In the velocity field, the length of the arrow is proportional to the magnitude (full length $v>0.5$), while the colour gives the orientation, as shown in the
      colour wheel legend. The density field is represented using a moving average over the Moore neighbourhood.}
\label{f6}
\end{center}
\end{figure}

Let us better analyse the results. The $CN$ for the lower density ``marching'' setting is constantly zero. This is due to the fact that the velocity grid has many empty spaces, and the few points in which the rotor may be computed present the same local structure, and thus the same rotor (furthermore, the rotor field is even zero everywhere). The high density case is more interesting. The maximum attained $CN$ value is constant at $\approx 0.2$,
reaching slightly higher values at the times in which the flows start and finish crossing. The $\mathbf{v}$, density, rotor and $CN$ fields at the time of maximum density are shown if Fig. \ref{f5}, while those at the time of maximum $CN$ are shown in Fig. \ref{f6}.
At maximum density, the velocity patterns are regular over the central crossing area and in the corridors, and as a result in this areas we have low $CN$. $CN$ reaches higher values in the ``corners'' were flows meet and separate. This effect is stronger when the flows are separating, as the velocity and rotor fields are less uniform.

\begin{figure}[t]
\begin{center}
  \includegraphics[width=0.4\textwidth]{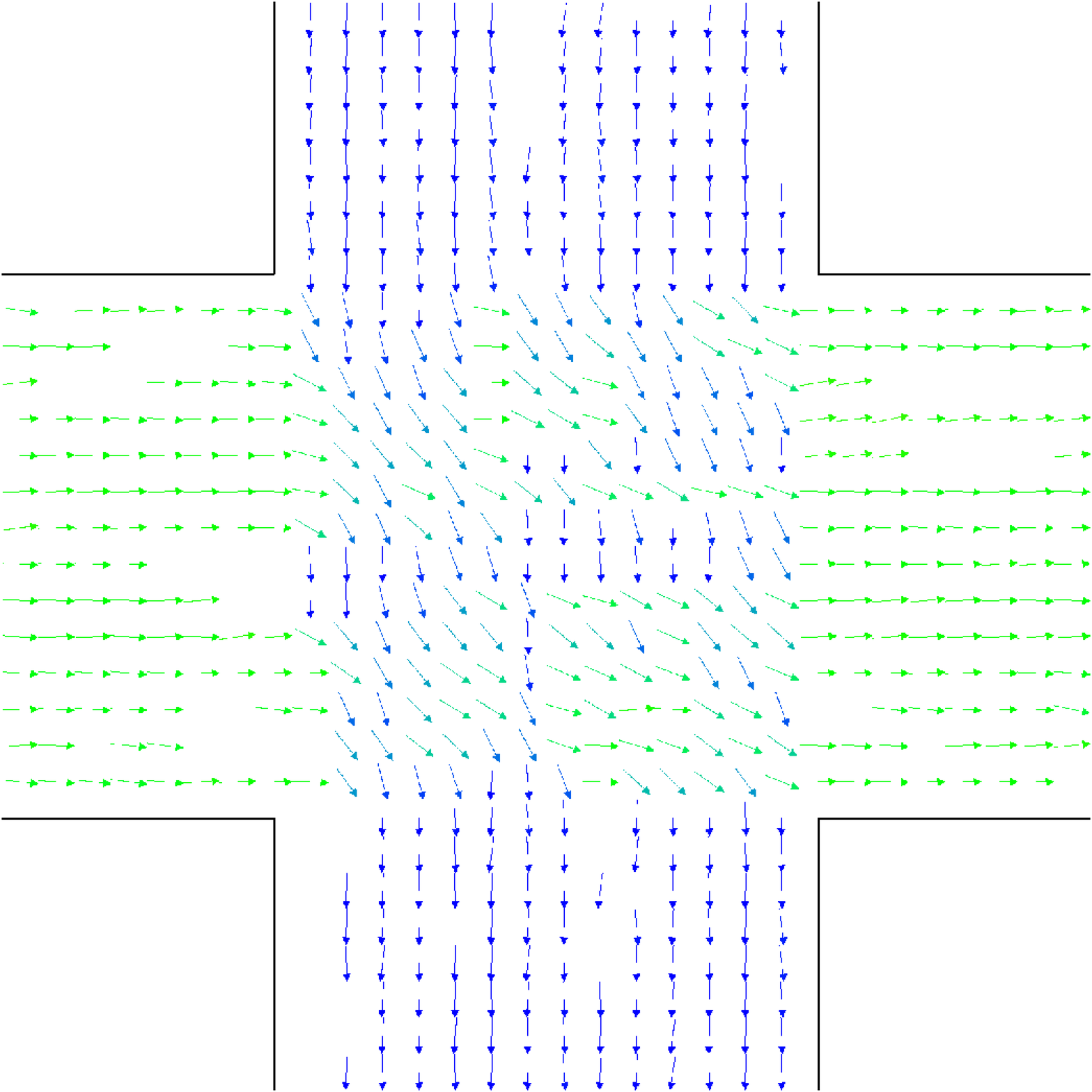}\includegraphics[width=0.1\textwidth]{frecce.eps}\hspace{0.02\textwidth}
  \includegraphics[width=0.4\textwidth]{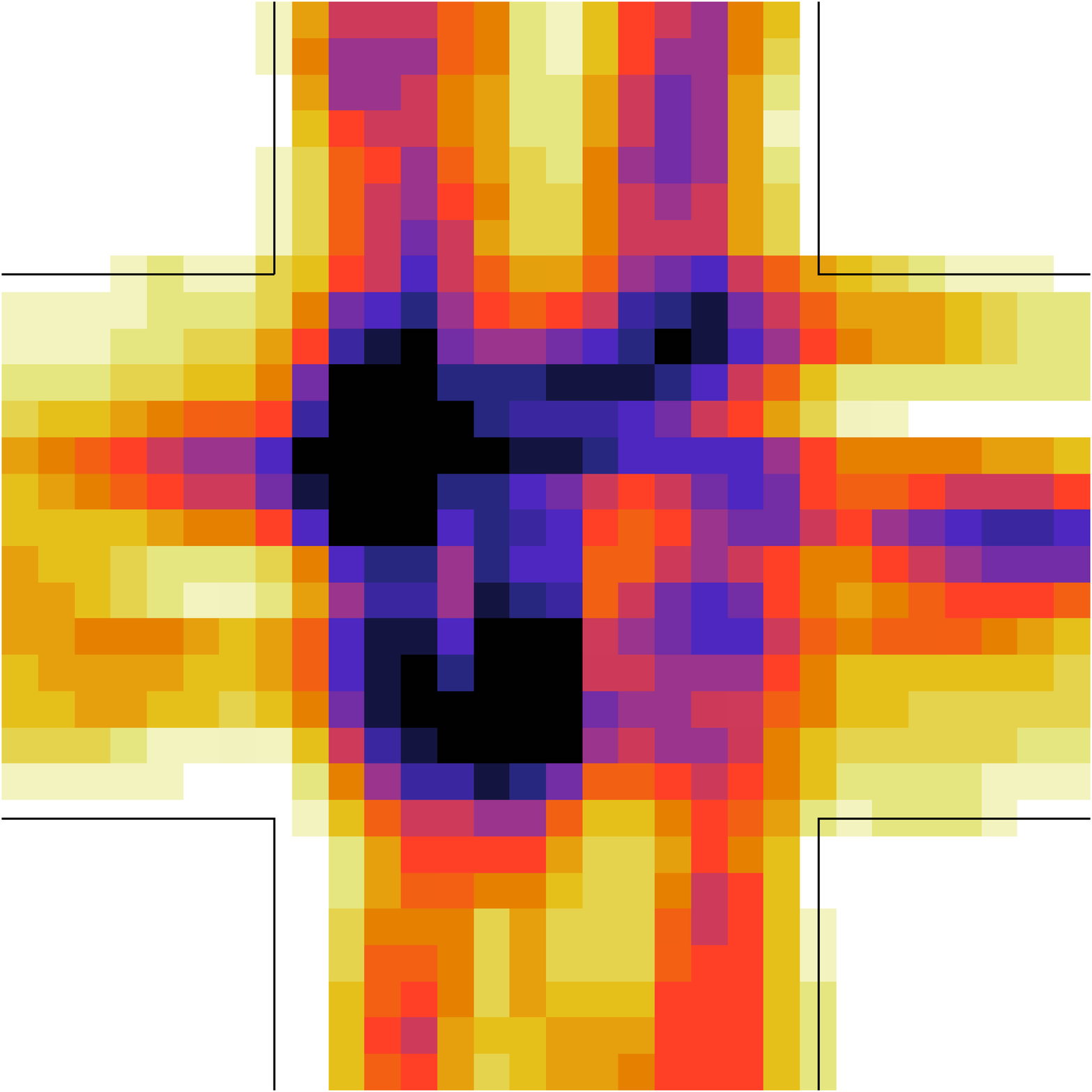}\hspace{0.02\textwidth}\includegraphics[width=0.05\textwidth]{scaledens.eps}
  \includegraphics[width=0.4\textwidth]{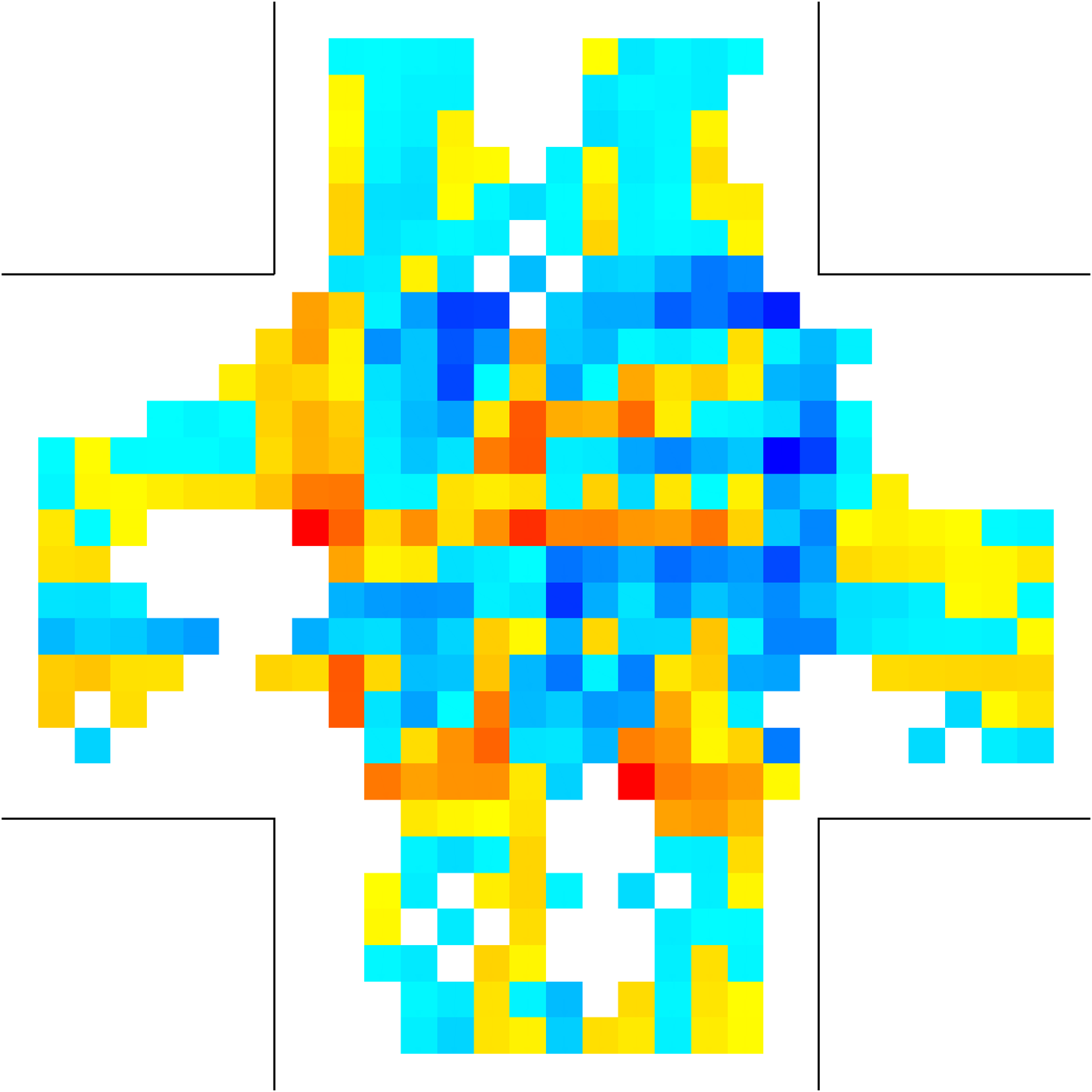}\hspace{0.02\textwidth}\includegraphics[width=0.05\textwidth]{scalepm0.eps}\hspace{0.05\textwidth}
  \includegraphics[width=0.4\textwidth]{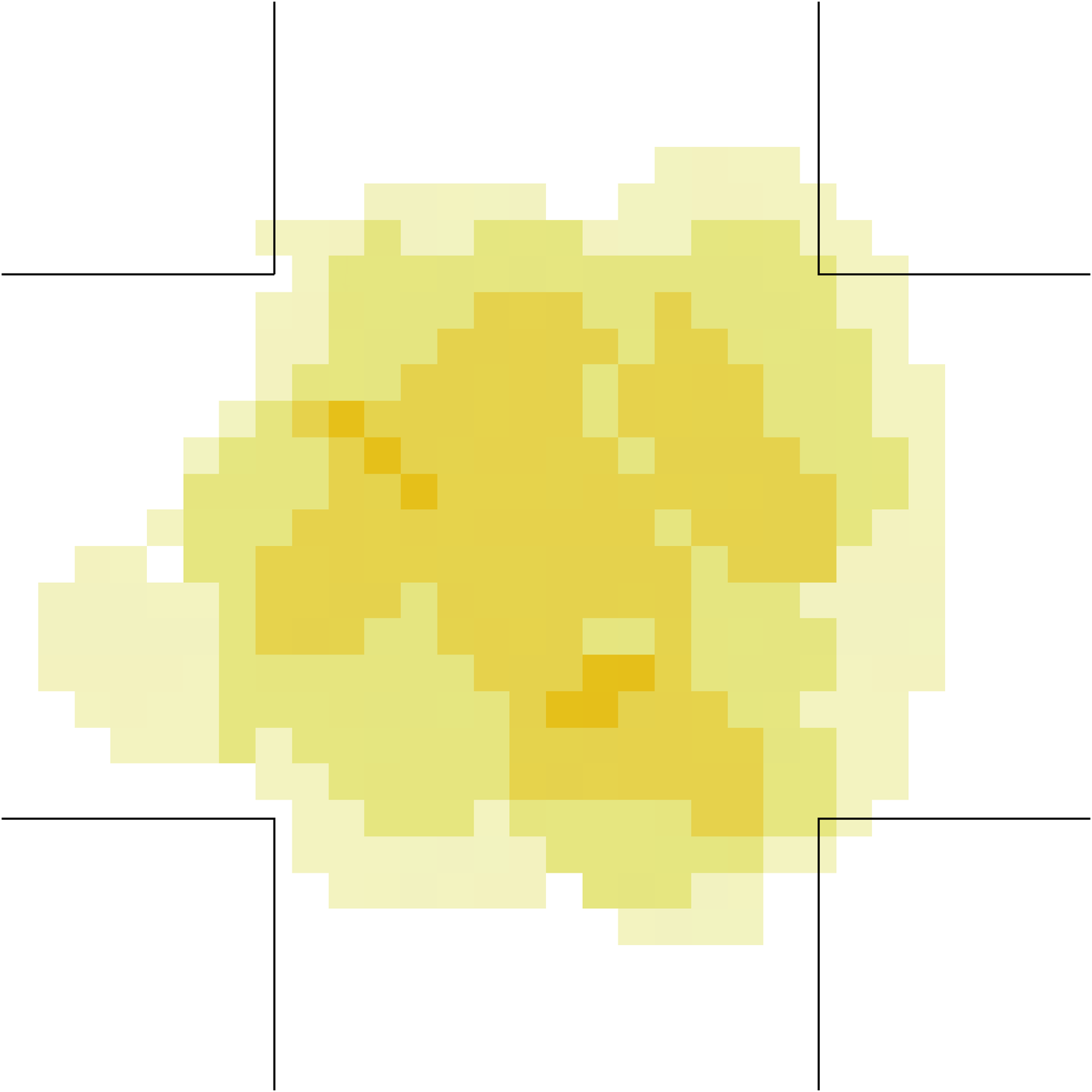}\hspace{0.02\textwidth}\includegraphics[width=0.05\textwidth]{scalenuovo.eps}
\caption{Higher density ``bodiless'' pedestrians at the time in which the highest density is attained (9.5 ped/m$^2$ during the $[25,27.5)$ s interval of the 6th repetition). Top, left: $\mathbf{v}$ field; top, right: density field; bottom left: $(\nabla \wedge \mathbf{v})_z$ field; bottom, right: $CN$ field. In the velocity field, the length of the arrow is proportional to the magnitude (full length $v>0.5$), while the colour gives the orientation, as shown in the
      colour wheel legend. The density field is represented using a moving average over the Moore neighbourhood.}
\label{f7}
\end{center}
\end{figure}

\begin{figure}[t]
\begin{center}
  \includegraphics[width=0.4\textwidth]{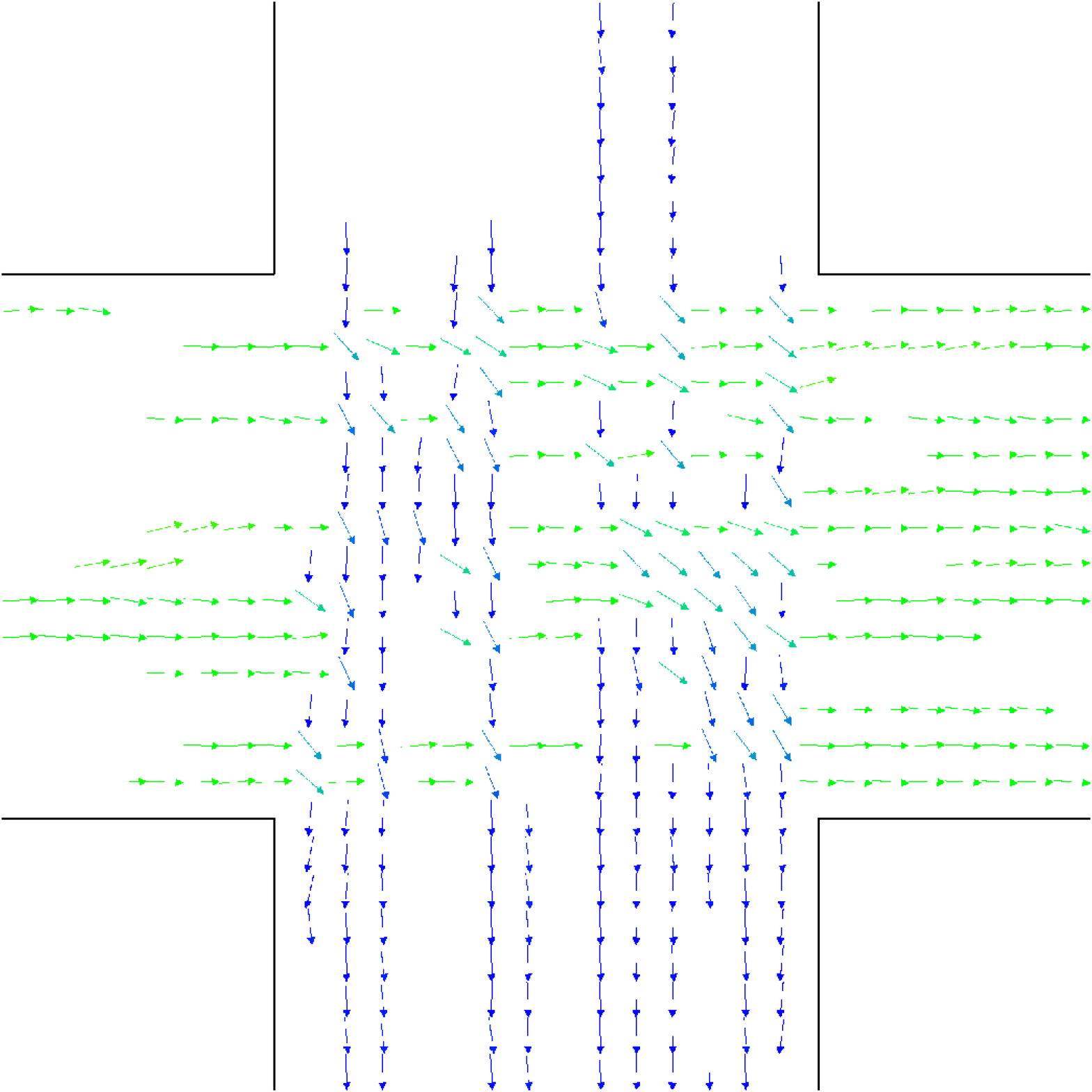}\includegraphics[width=0.1\textwidth]{frecce.eps}\hspace{0.02\textwidth}
  \includegraphics[width=0.4\textwidth]{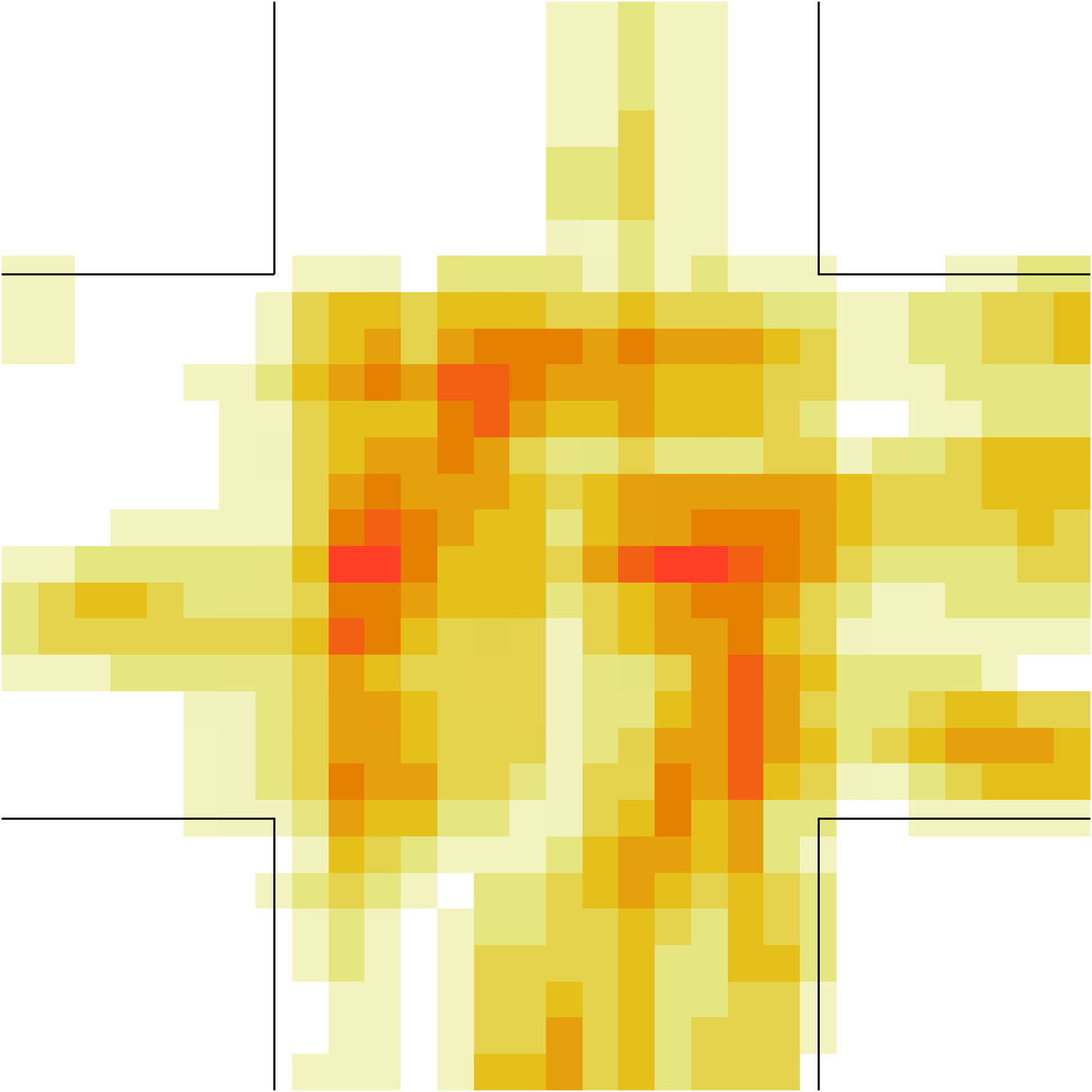}\hspace{0.02\textwidth}\includegraphics[width=0.05\textwidth]{scaledens.eps}
  \includegraphics[width=0.4\textwidth]{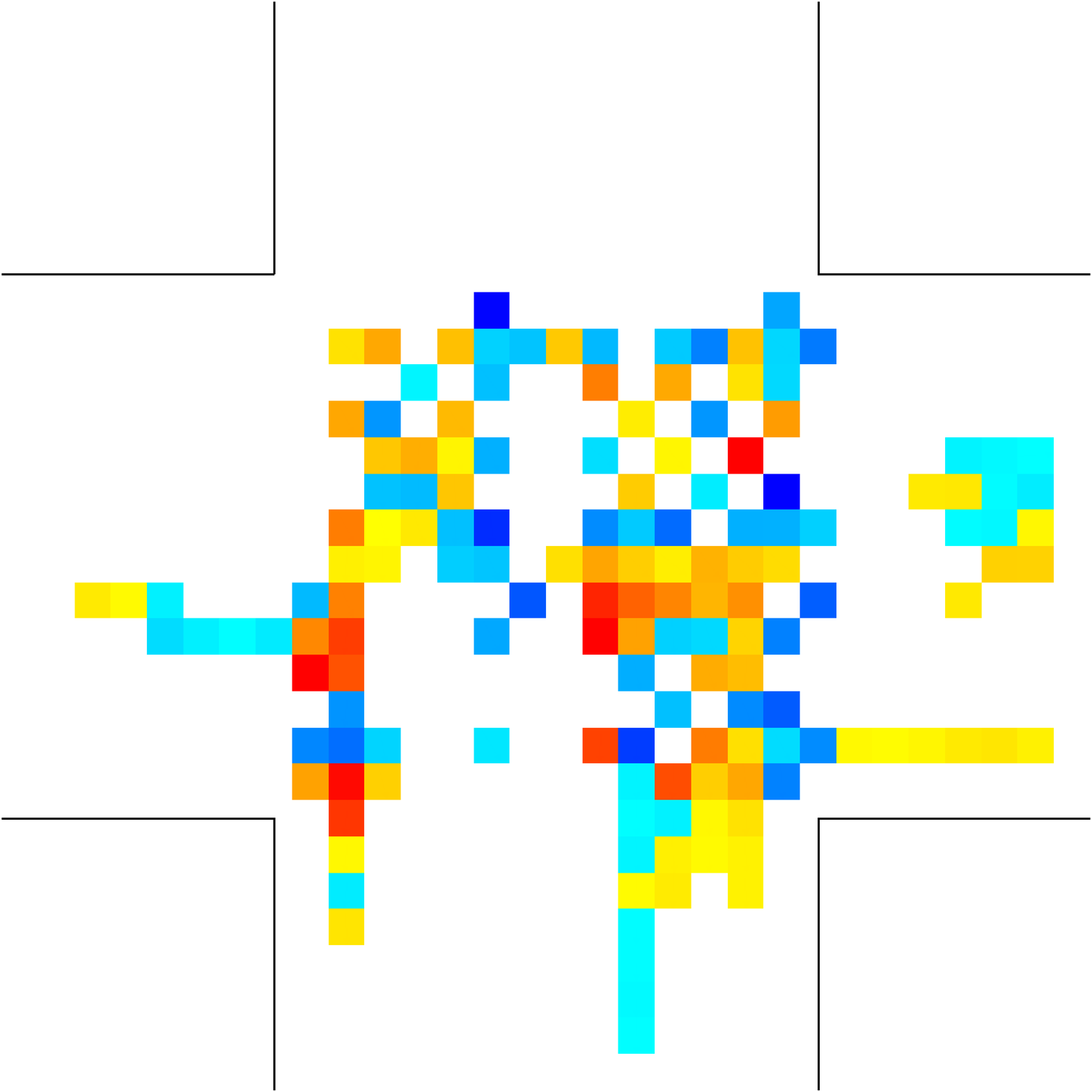}\hspace{0.02\textwidth}\includegraphics[width=0.05\textwidth]{scalepm0.eps}\hspace{0.05\textwidth}
  \includegraphics[width=0.4\textwidth]{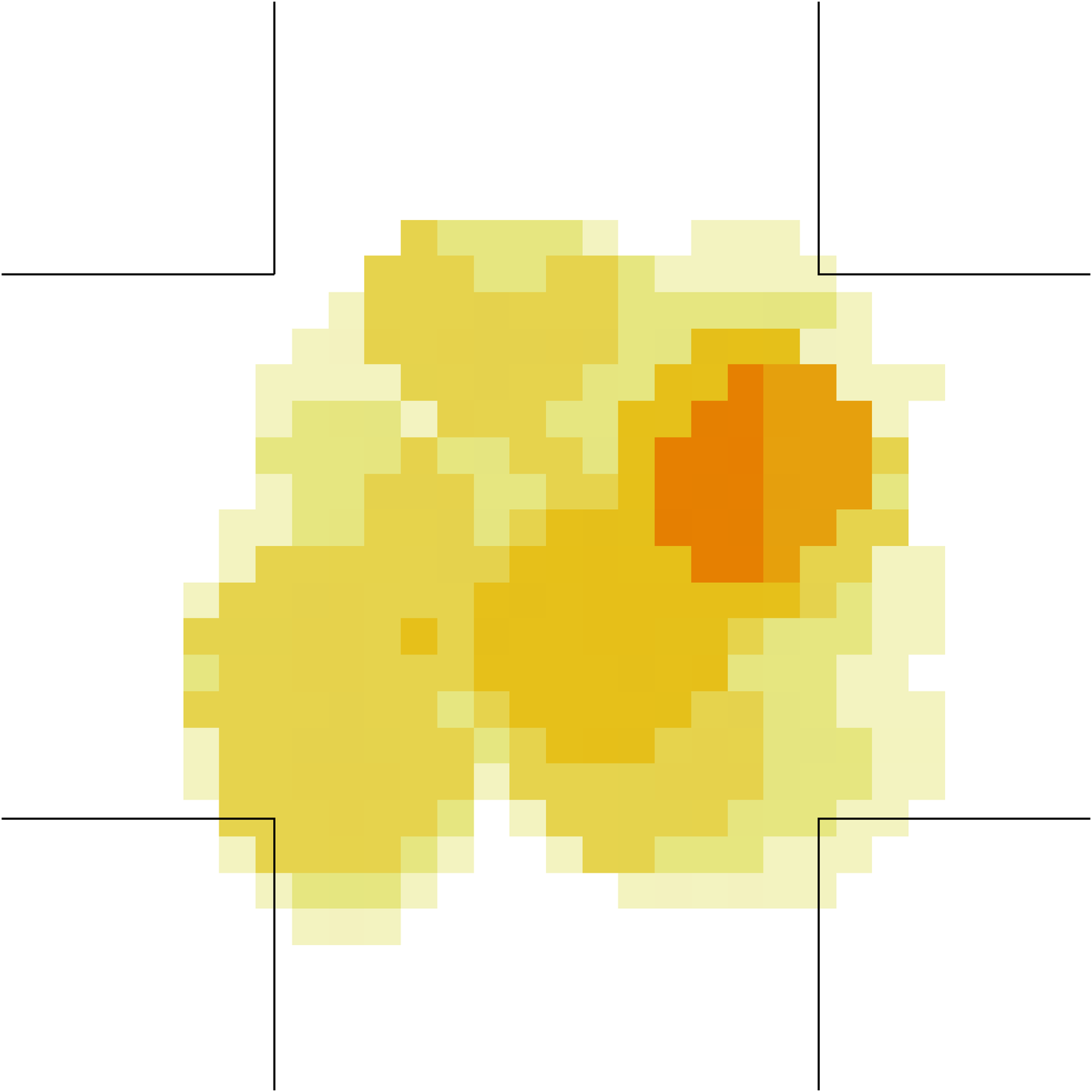}\hspace{0.02\textwidth}\includegraphics[width=0.05\textwidth]{scalenuovo.eps}
\caption{Higher density ``bodiless'' pedestrians at the time in which the highest $CN$ is attained (0.370 during the $[42.5,45)$ s interval of the 6th repetition). Top, left: $\mathbf{v}$ field; top, right: density field; bottom left: $(\nabla \wedge \mathbf{v})_z$ field; bottom, right: $CN$ field. In the velocity field, the length of the arrow is proportional to the magnitude (full length $v>0.5$), while the colour gives the orientation, as shown in the
      colour wheel legend. The density field is represented using a moving average over the Moore neighbourhood.}
\label{f8}
\end{center}
\end{figure}

An interesting result concerning the ``bodiless'' case is that the $CN$ evolution is basically independent of the flow intensity. This is to be expected, since for non-interacting ``particles'' the increase in flow only changes the statistical sample, with no other effect on the vector fields. It is nevertheless important to see that the $CN$ metric is not ``tricked'' by the increased flow.
As shown in particular in Figs. \ref{f7} (maximum density) and \ref{f8} (maximum $CN$), the ``bodiless'' fields are basically ``noisy'' versions of the ``marching'' ones.

\begin{figure}[t]
\begin{center}
  \includegraphics[width=0.4\textwidth]{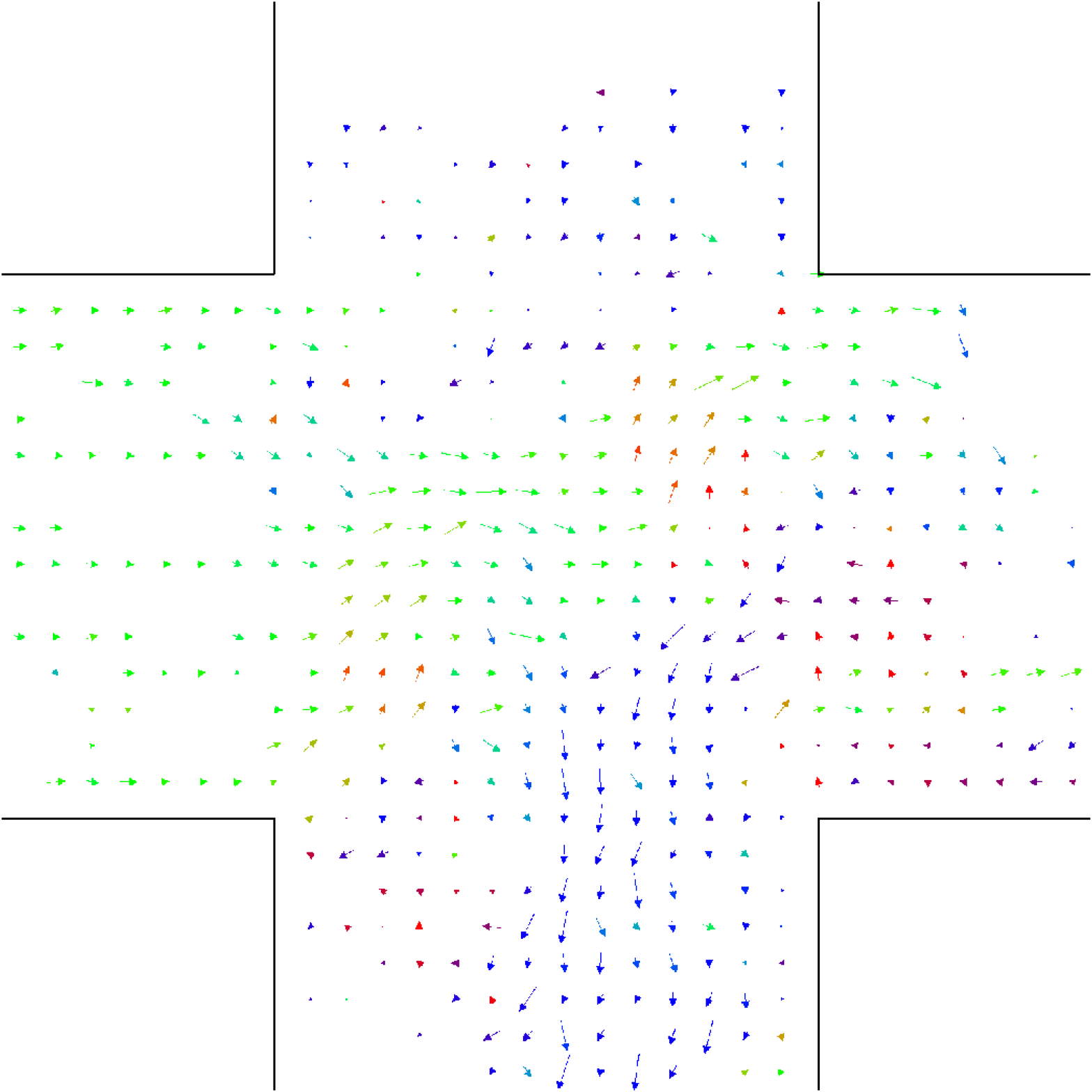}\includegraphics[width=0.1\textwidth]{frecce.eps}\hspace{0.02\textwidth}
  \includegraphics[width=0.4\textwidth]{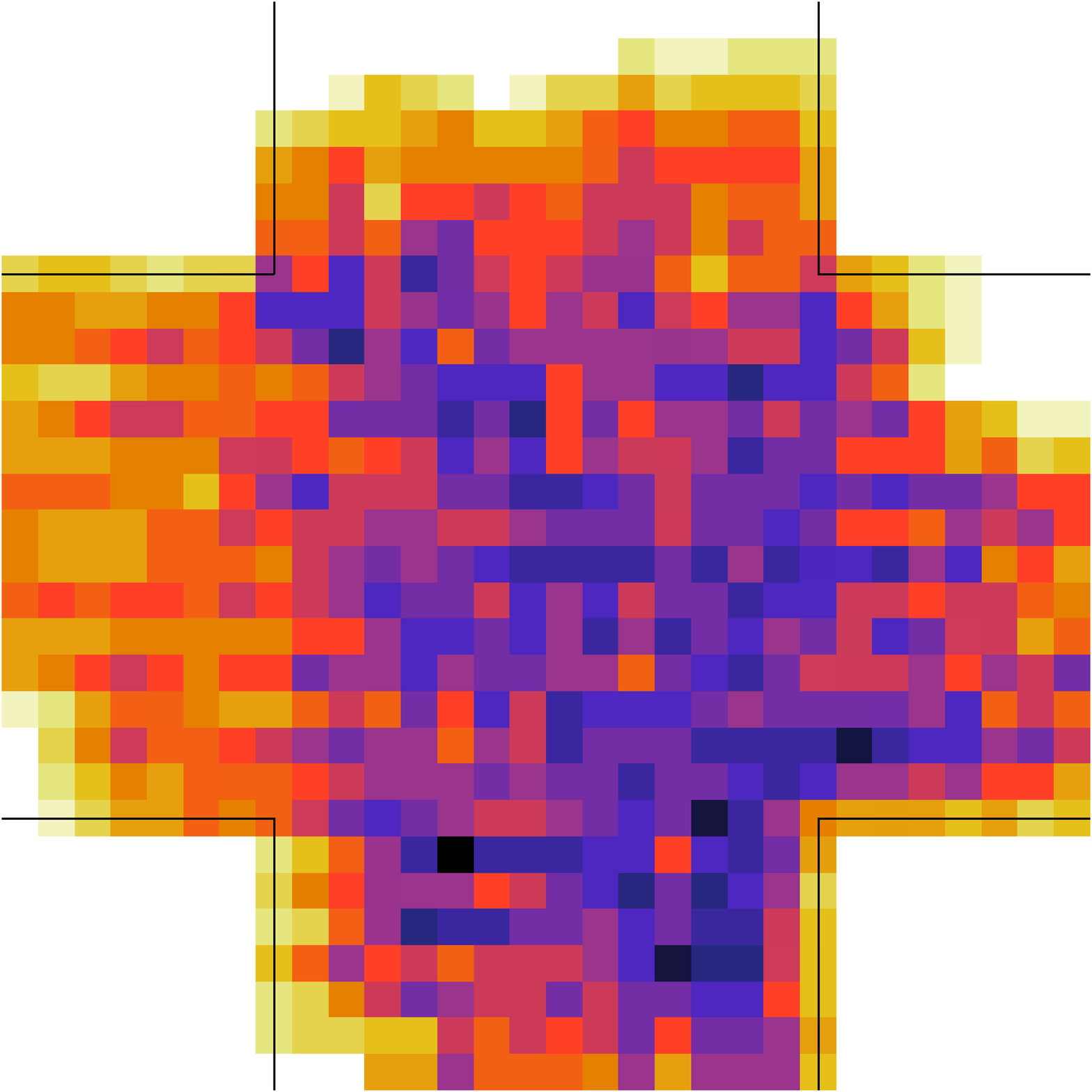}\hspace{0.02\textwidth}\includegraphics[width=0.05\textwidth]{scaledens.eps}
  \includegraphics[width=0.4\textwidth]{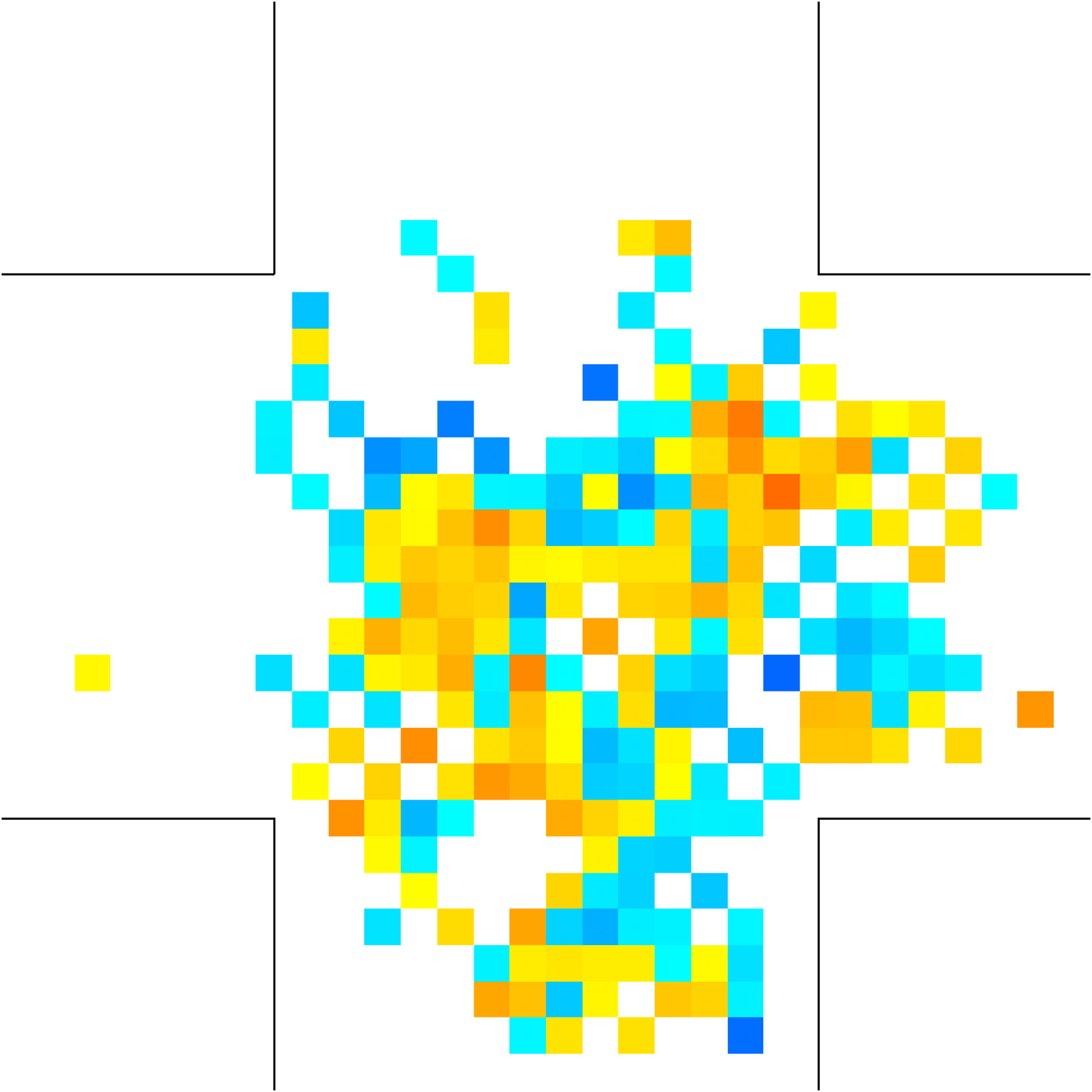}\hspace{0.02\textwidth}\includegraphics[width=0.05\textwidth]{scalepm0.eps}\hspace{0.05\textwidth}
  \includegraphics[width=0.4\textwidth]{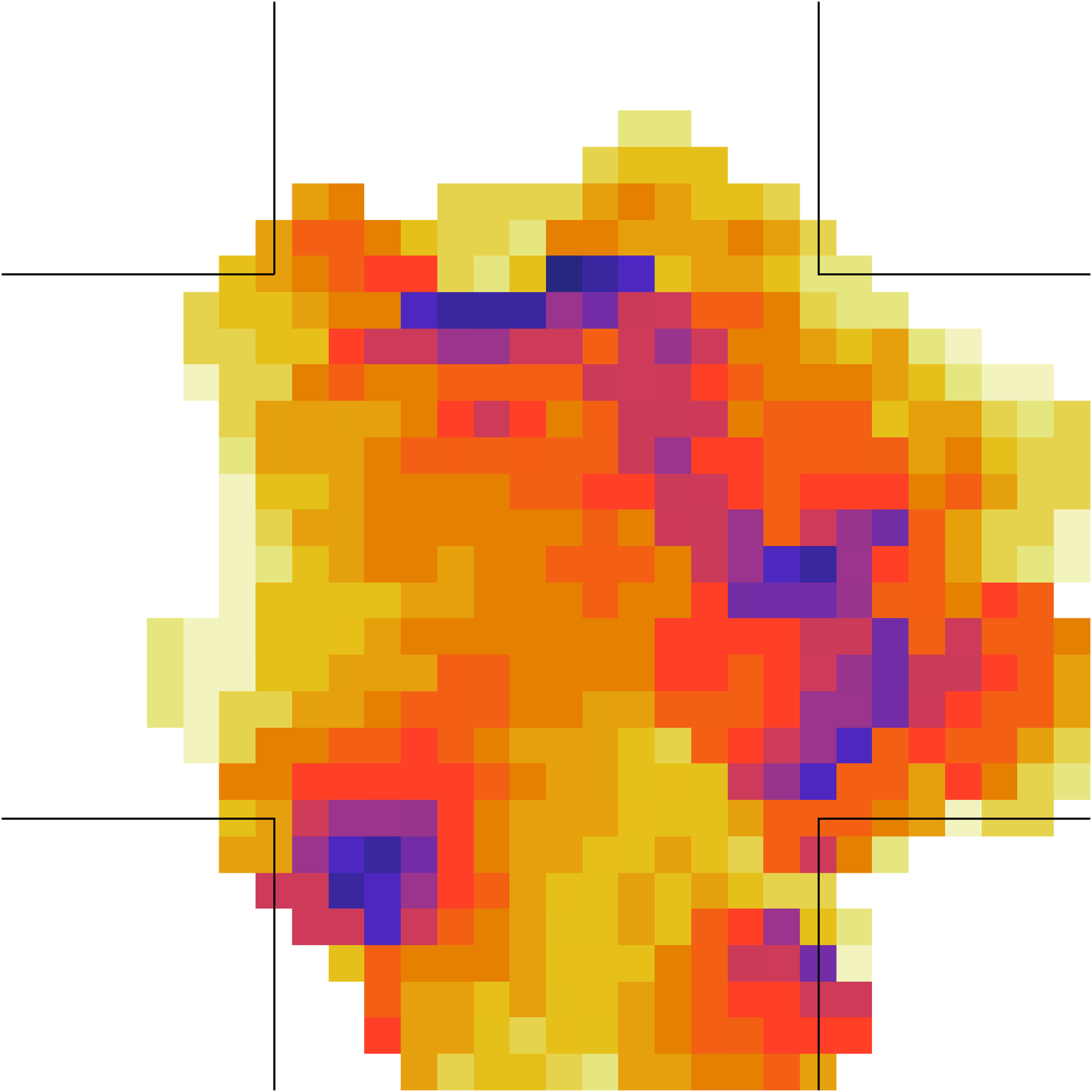}\hspace{0.02\textwidth}\includegraphics[width=0.05\textwidth]{scalenuovo.eps}
  \caption{Higher density ``realistic'' pedestrians at the time in which the maximum density is attained (8.79 ped/m$^2$ during the $[37.5,40)$ s interval of the 4th repetition).  Top, left: $\mathbf{v}$ field; top, right: density field; bottom left: $(\nabla \wedge \mathbf{v})_z$ field; bottom, right: $CN$ field. In the velocity field, the length of the arrow is proportional to the magnitude (full length $v>0.5$), while the colour gives the orientation, as shown in the
      colour wheel legend. The density field is represented using a moving average over the Moore neighbourhood.}
\label{f9}
\end{center}
\end{figure}

\begin{figure}[t]
\begin{center}
  \includegraphics[width=0.4\textwidth]{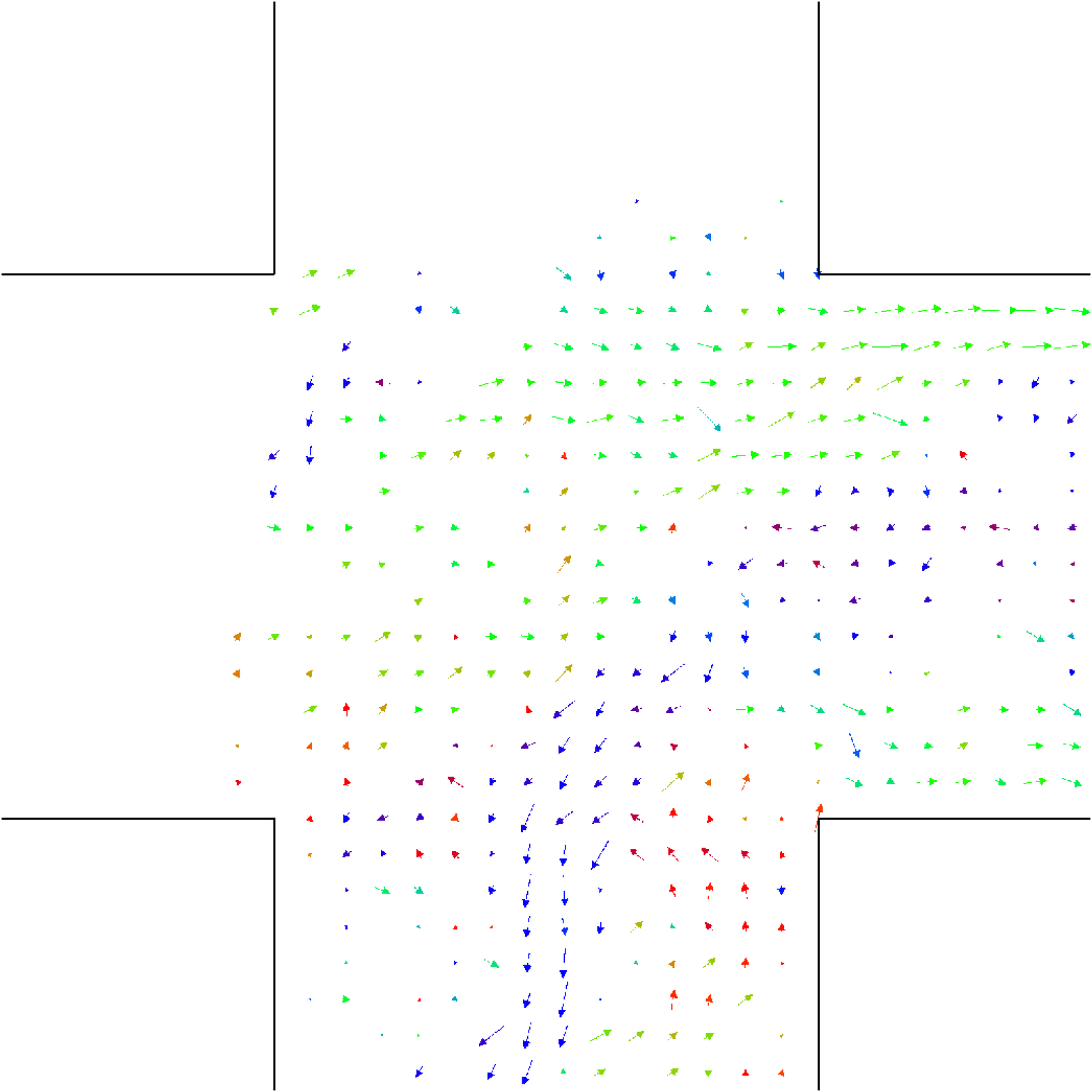}\includegraphics[width=0.1\textwidth]{frecce.eps}\hspace{0.02\textwidth}
  \includegraphics[width=0.4\textwidth]{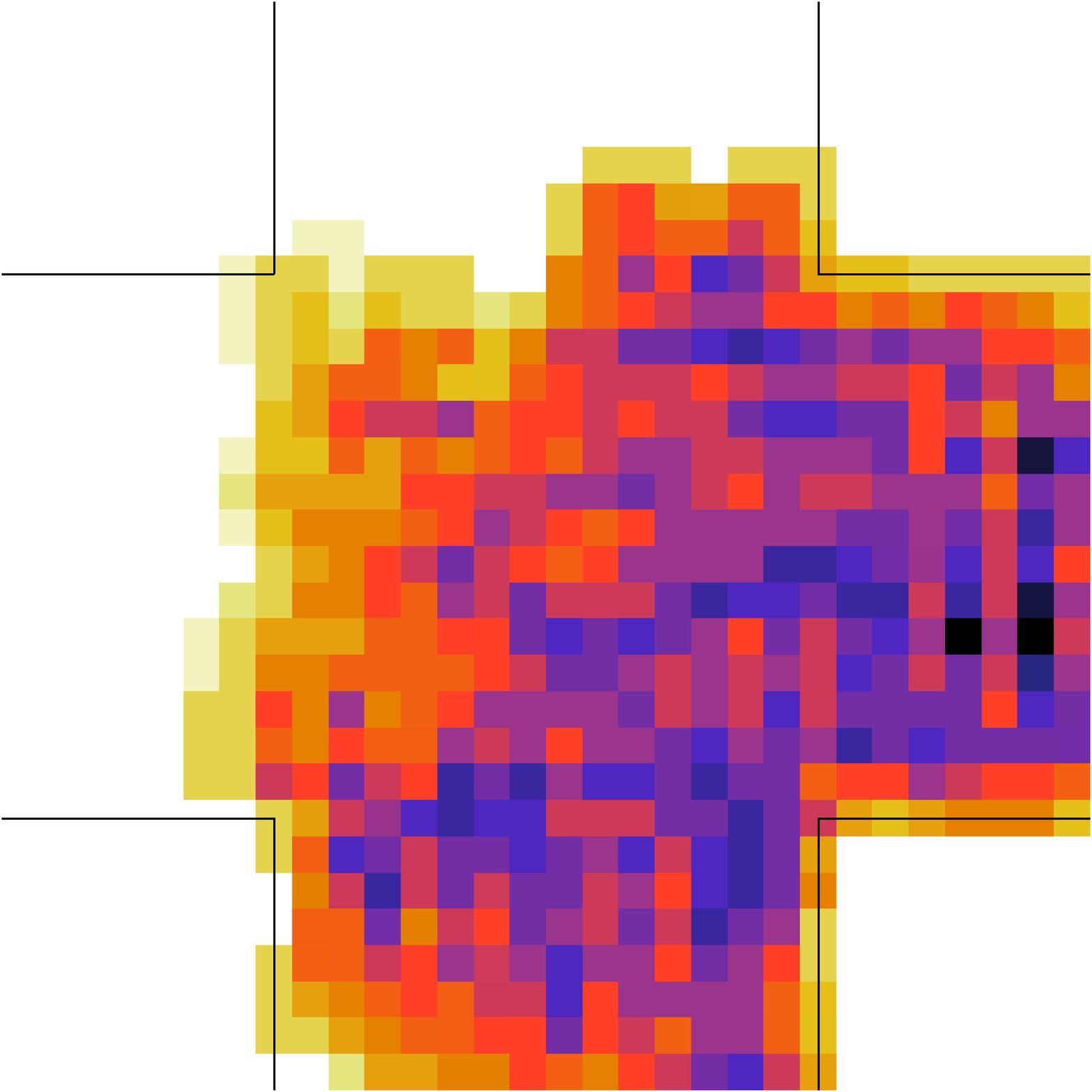}\hspace{0.02\textwidth}\includegraphics[width=0.05\textwidth]{scaledens.eps}
  \includegraphics[width=0.4\textwidth]{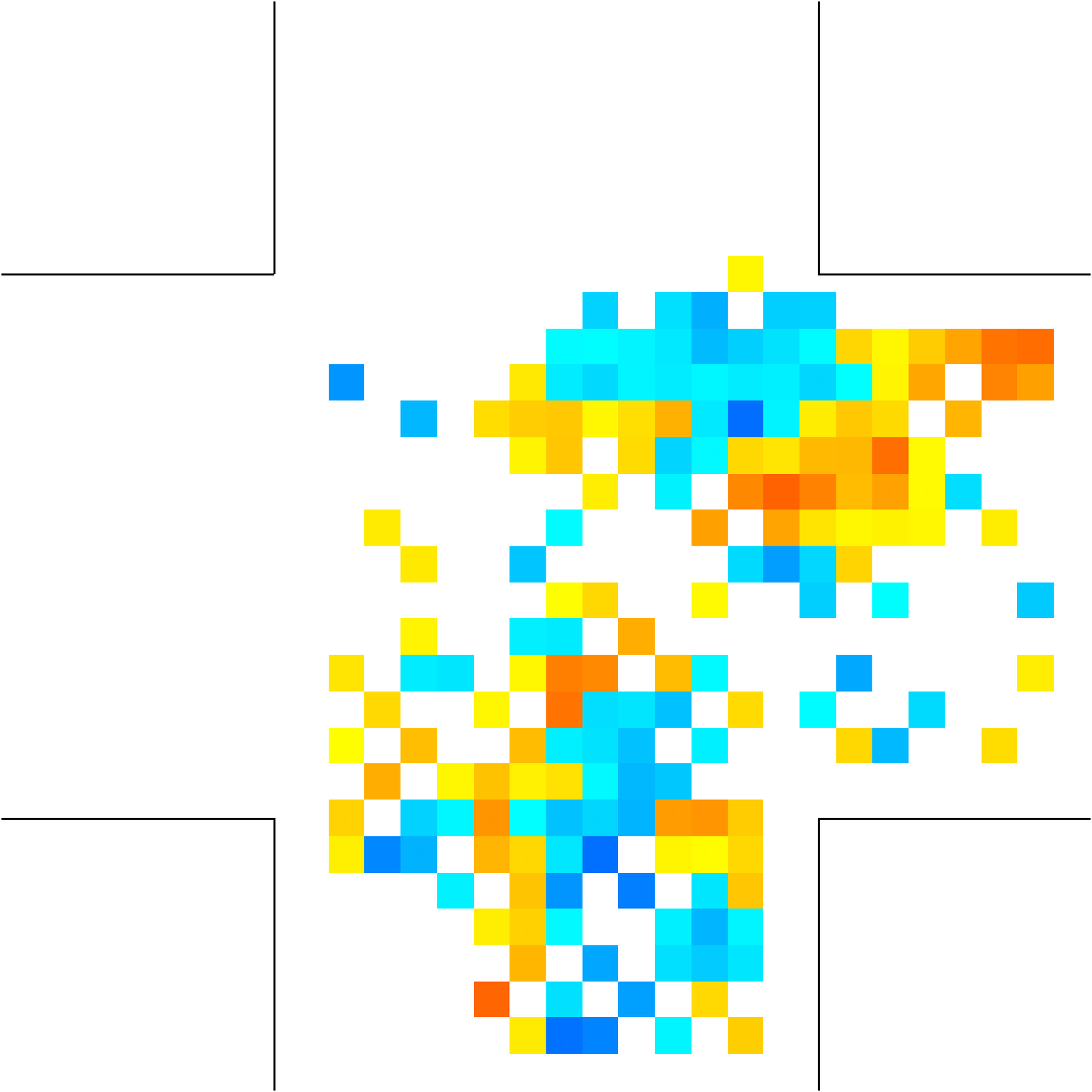}\hspace{0.02\textwidth}\includegraphics[width=0.05\textwidth]{scalepm0.eps}\hspace{0.05\textwidth}
  \includegraphics[width=0.4\textwidth]{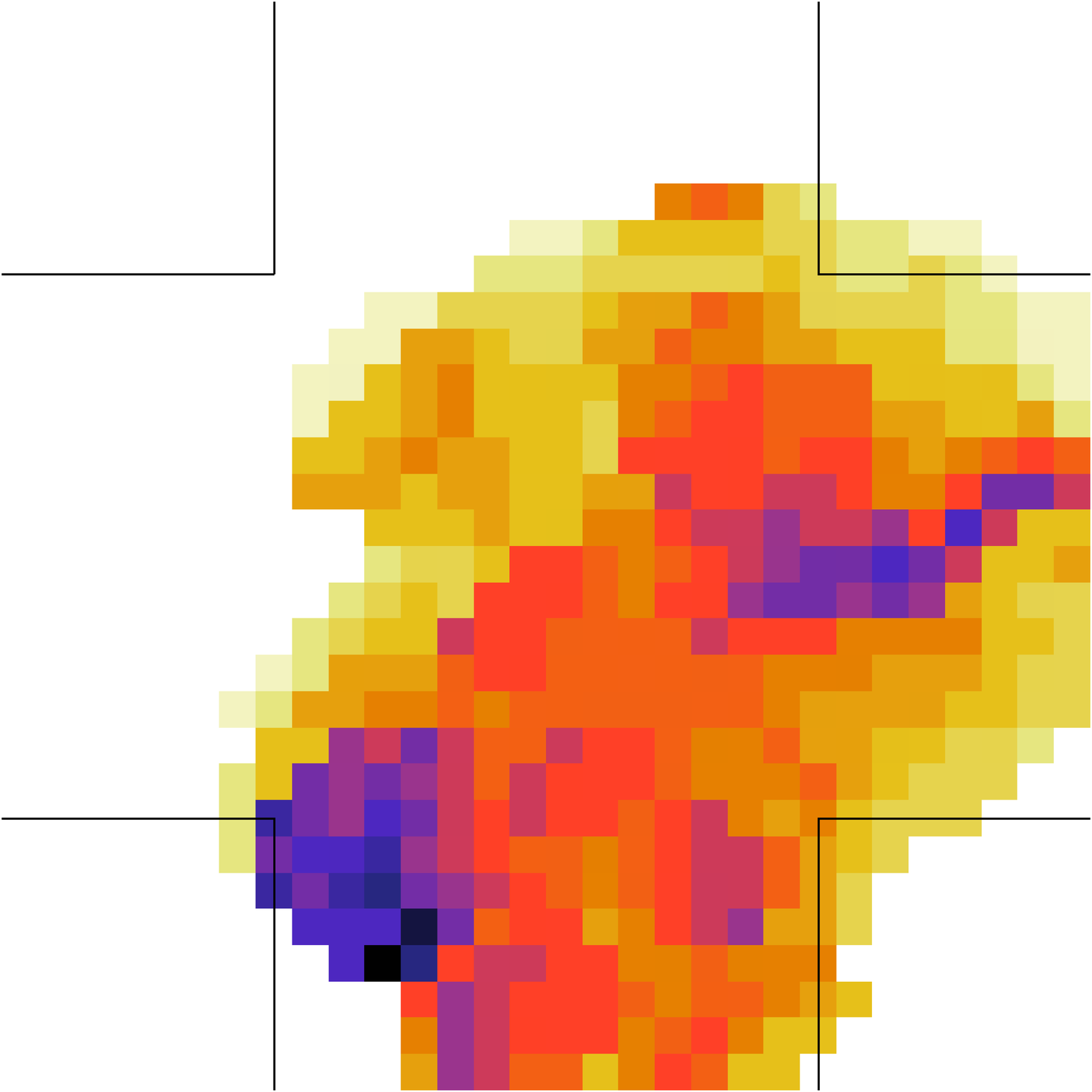}\hspace{0.02\textwidth}\includegraphics[width=0.05\textwidth]{scalenuovo.eps}
  \caption{Lower density ``realistic'' pedestrians at the time in which the highest maximum $CN$ is attained (1.258 during the $[55,57.5)$ s interval of the 9th repetition).  Top, left: $\mathbf{v}$ field; top, right: density field; bottom left: $(\nabla \wedge \mathbf{v})_z$ field; bottom, right: $CN$ field. In the velocity field, the length of the arrow is proportional to the magnitude (full length $v>0.5$), while the colour gives the orientation, as shown in the
      colour wheel legend. The density field is represented using a moving average over the Moore neighbourhood.}
\label{f10}
\end{center}
\end{figure}

\begin{figure}[t]
\begin{center}
  \includegraphics[width=0.4\textwidth]{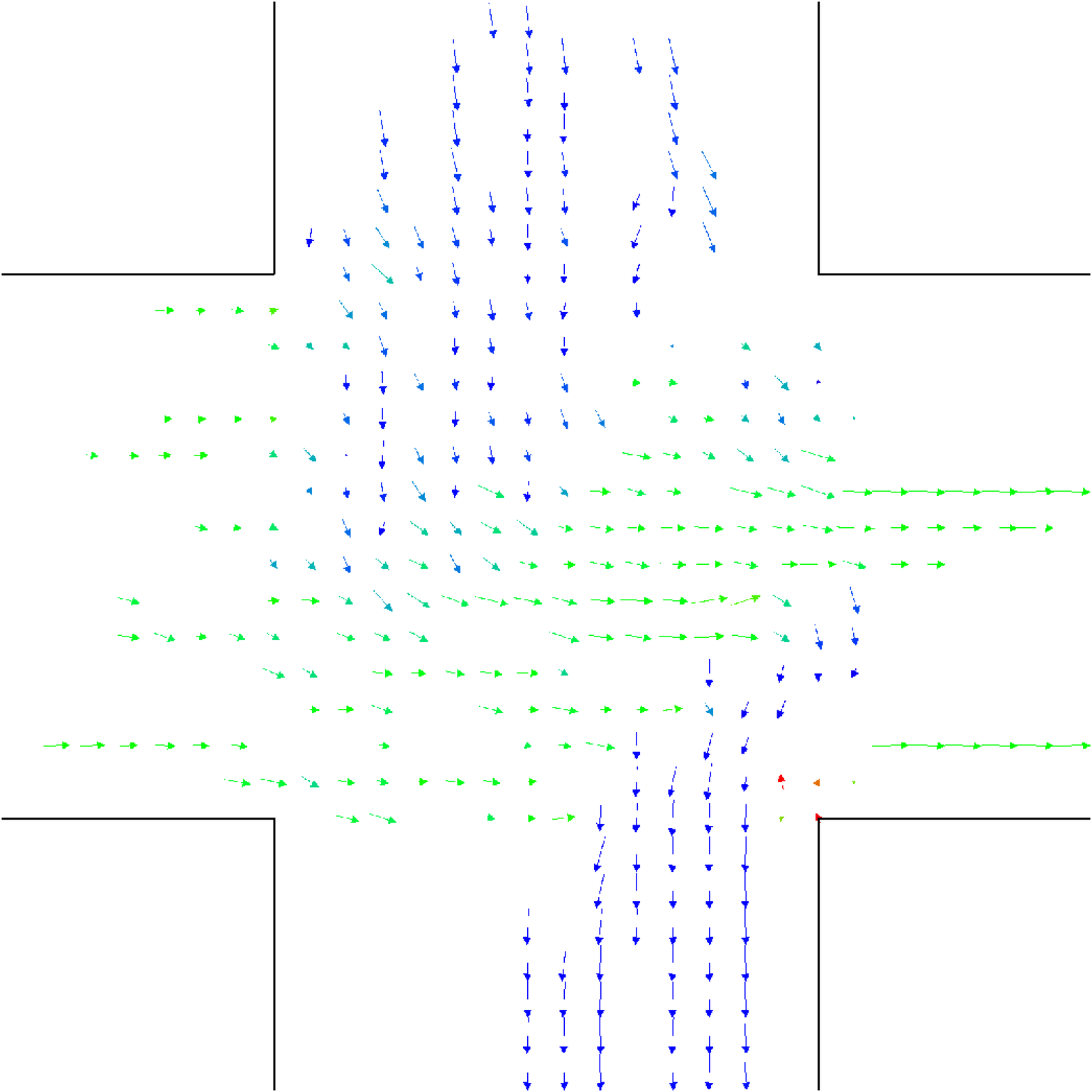}\includegraphics[width=0.1\textwidth]{frecce.eps}\hspace{0.02\textwidth}
  \includegraphics[width=0.4\textwidth]{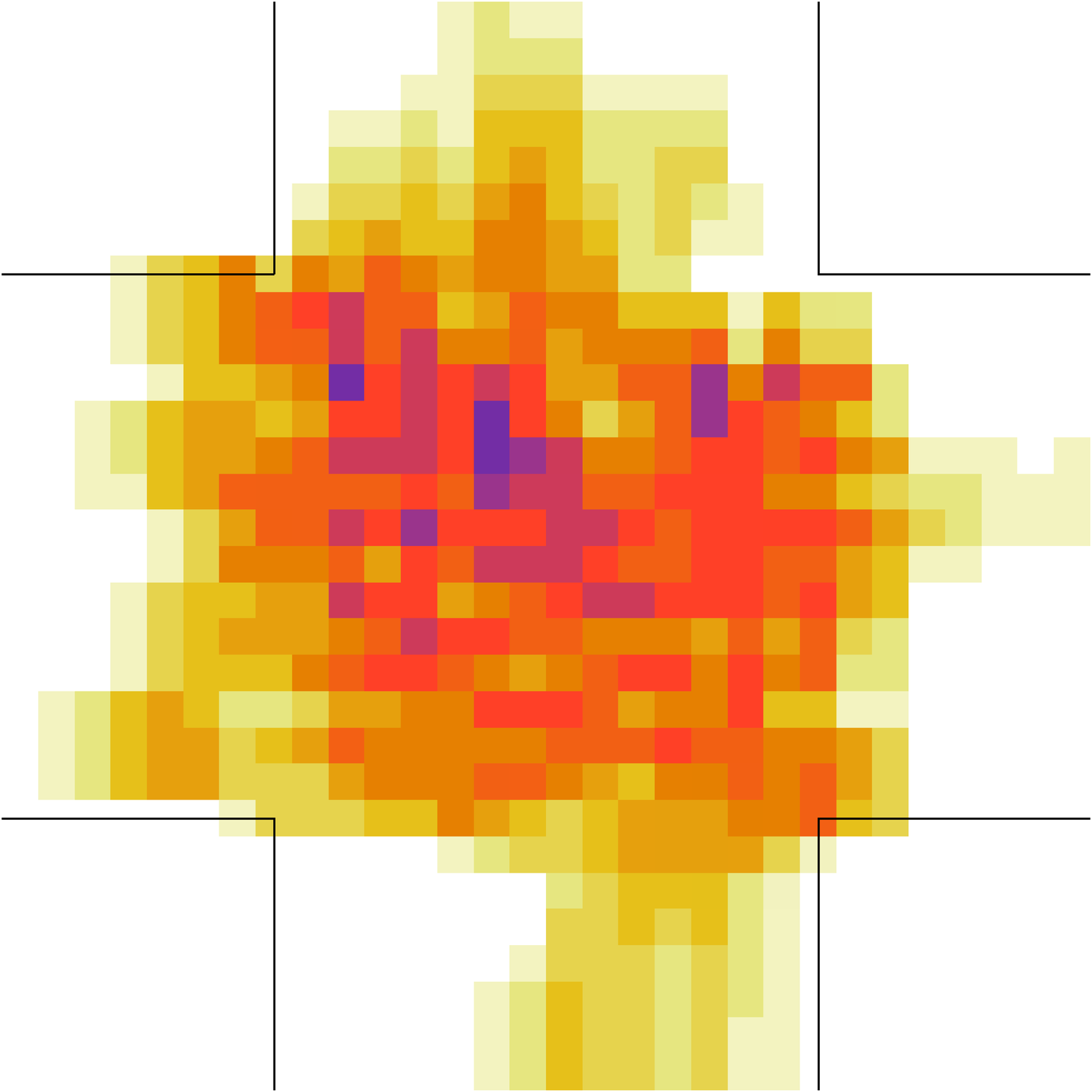}\hspace{0.02\textwidth}\includegraphics[width=0.05\textwidth]{scaledens.eps}
  \includegraphics[width=0.4\textwidth]{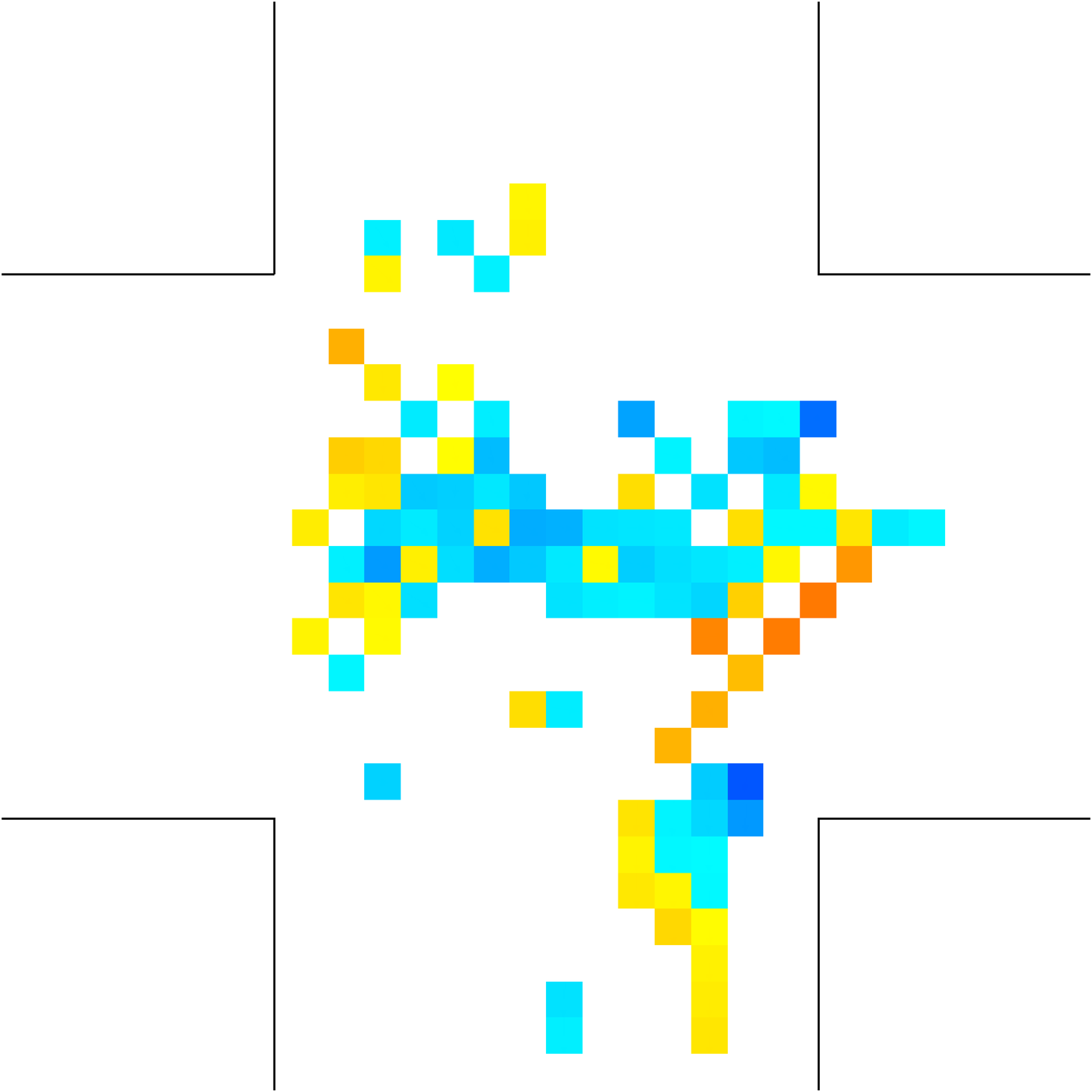}\hspace{0.02\textwidth}\includegraphics[width=0.05\textwidth]{scalepm0.eps}\hspace{0.05\textwidth}
  \includegraphics[width=0.4\textwidth]{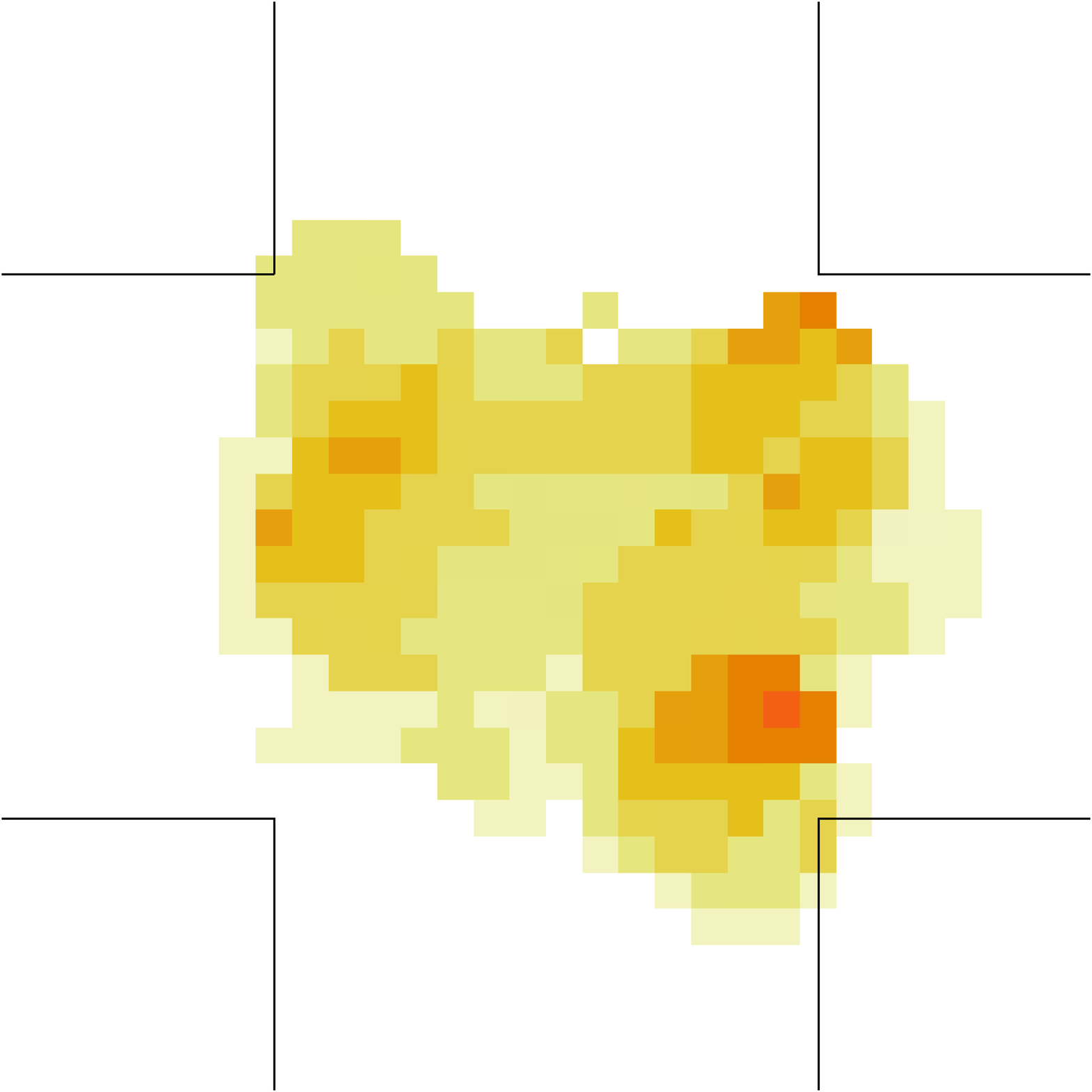}\hspace{0.02\textwidth}\includegraphics[width=0.05\textwidth]{scalenuovo.eps}
  \caption{Lower density ``realistic'' pedestrians at the time in which the maximum density is attained (5.80 ped/m$^2$ during the $[25,27.5)$ s interval of the 9th repetition).  Top, left: $\mathbf{v}$ field; top, right: density field; bottom left: $(\nabla \wedge \mathbf{v})_z$ field; bottom, right: $CN$ field. In the velocity field, the length of the arrow is proportional to the magnitude (full length $v>0.5$), while the colour gives the orientation, as shown in the
      colour wheel legend. The density field is represented using a moving average over the Moore neighbourhood.}
\label{f11}
\end{center}
\end{figure}

\begin{figure}[t]
\begin{center}
  \includegraphics[width=0.4\textwidth]{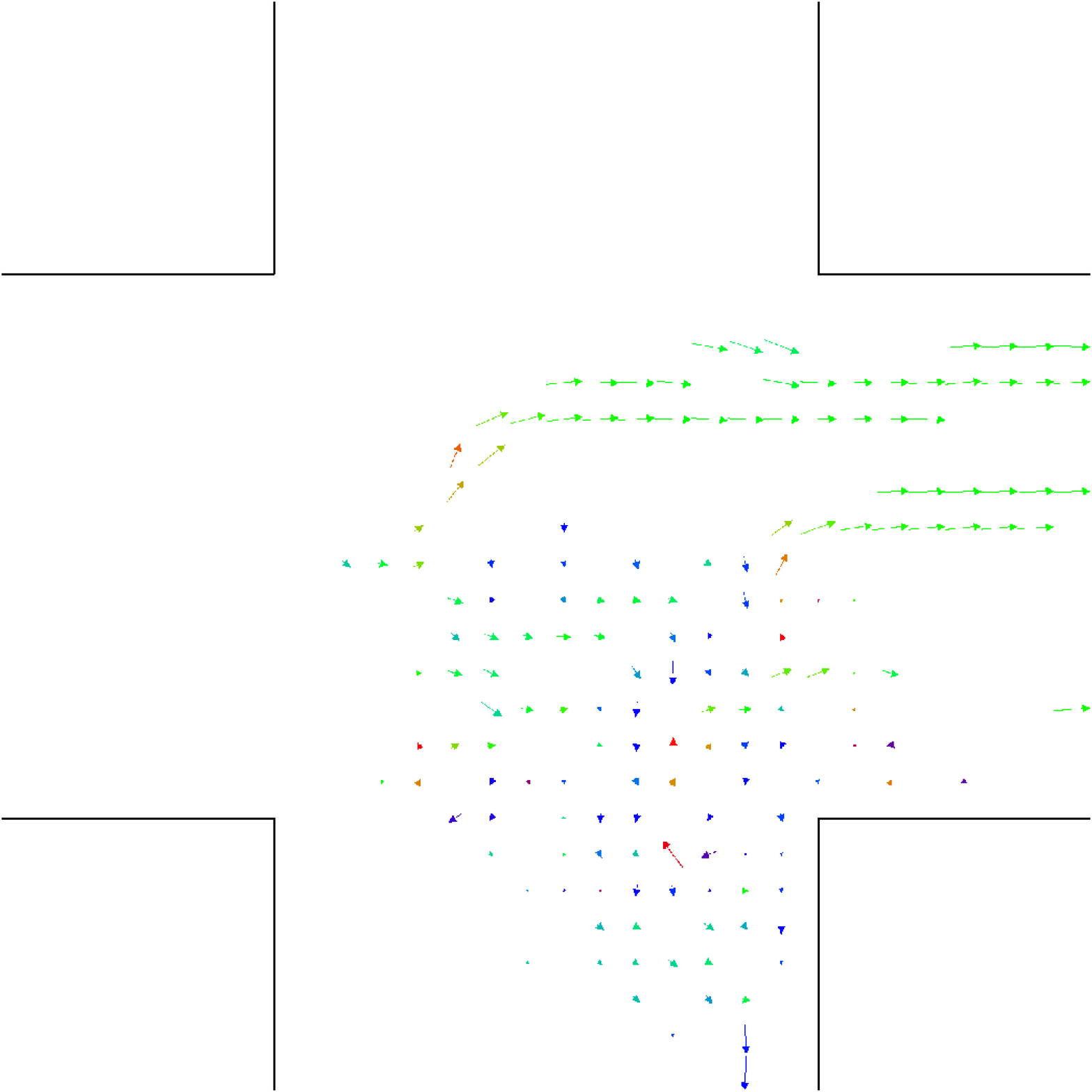}\includegraphics[width=0.1\textwidth]{frecce.eps}\hspace{0.02\textwidth}
  \includegraphics[width=0.4\textwidth]{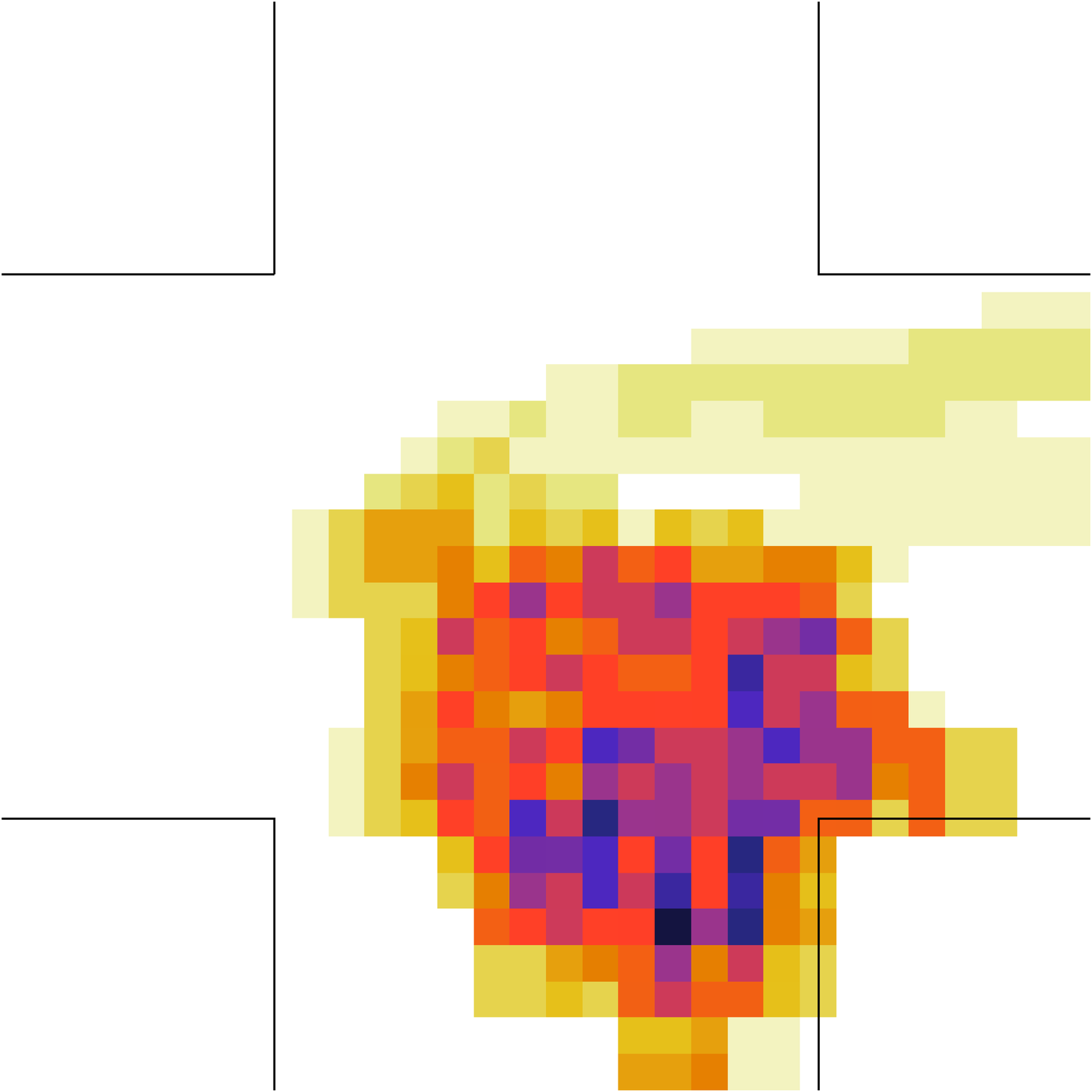}\hspace{0.02\textwidth}\includegraphics[width=0.05\textwidth]{scaledens.eps}
  \includegraphics[width=0.4\textwidth]{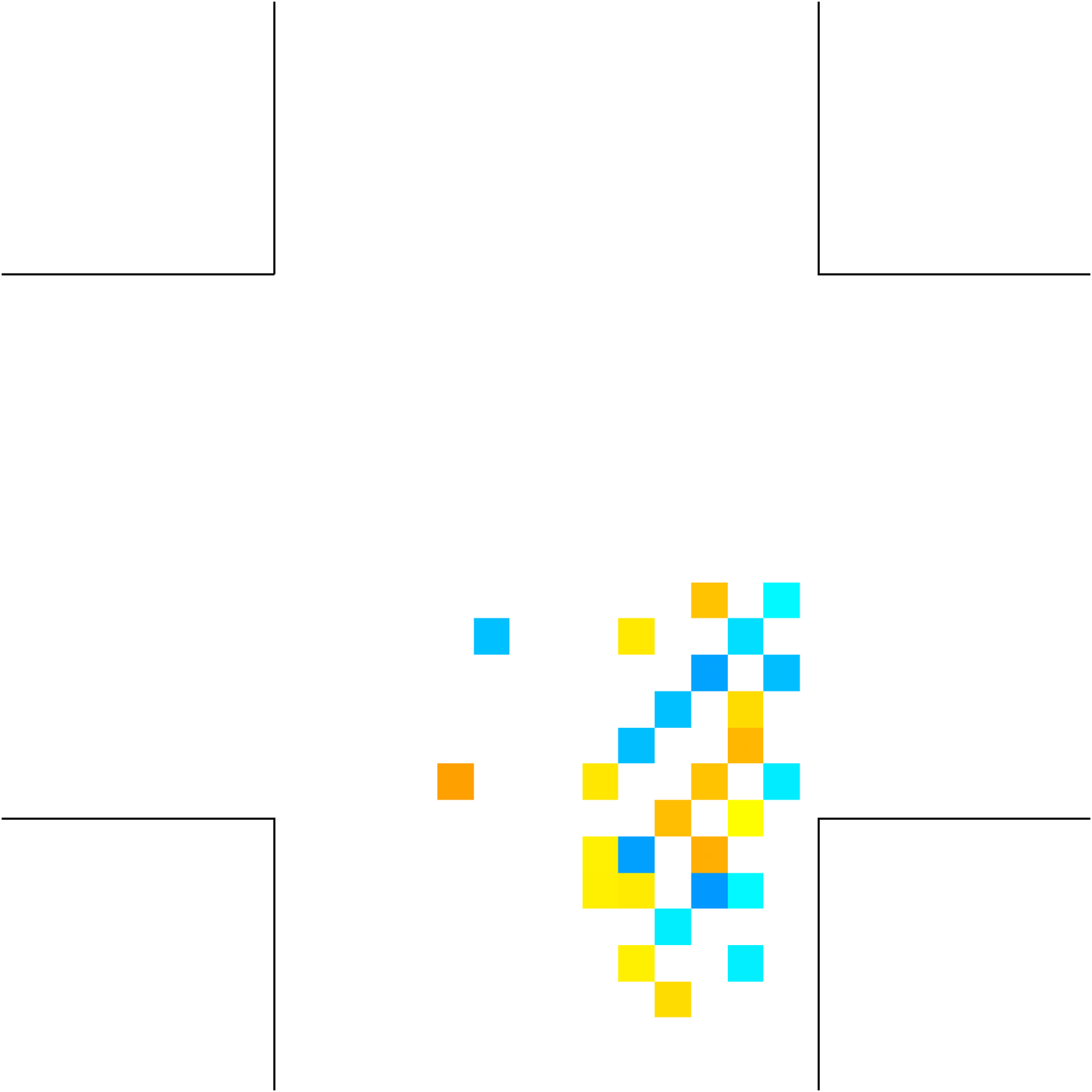}\hspace{0.02\textwidth}\includegraphics[width=0.05\textwidth]{scalepm0.eps}\hspace{0.05\textwidth}
  \includegraphics[width=0.4\textwidth]{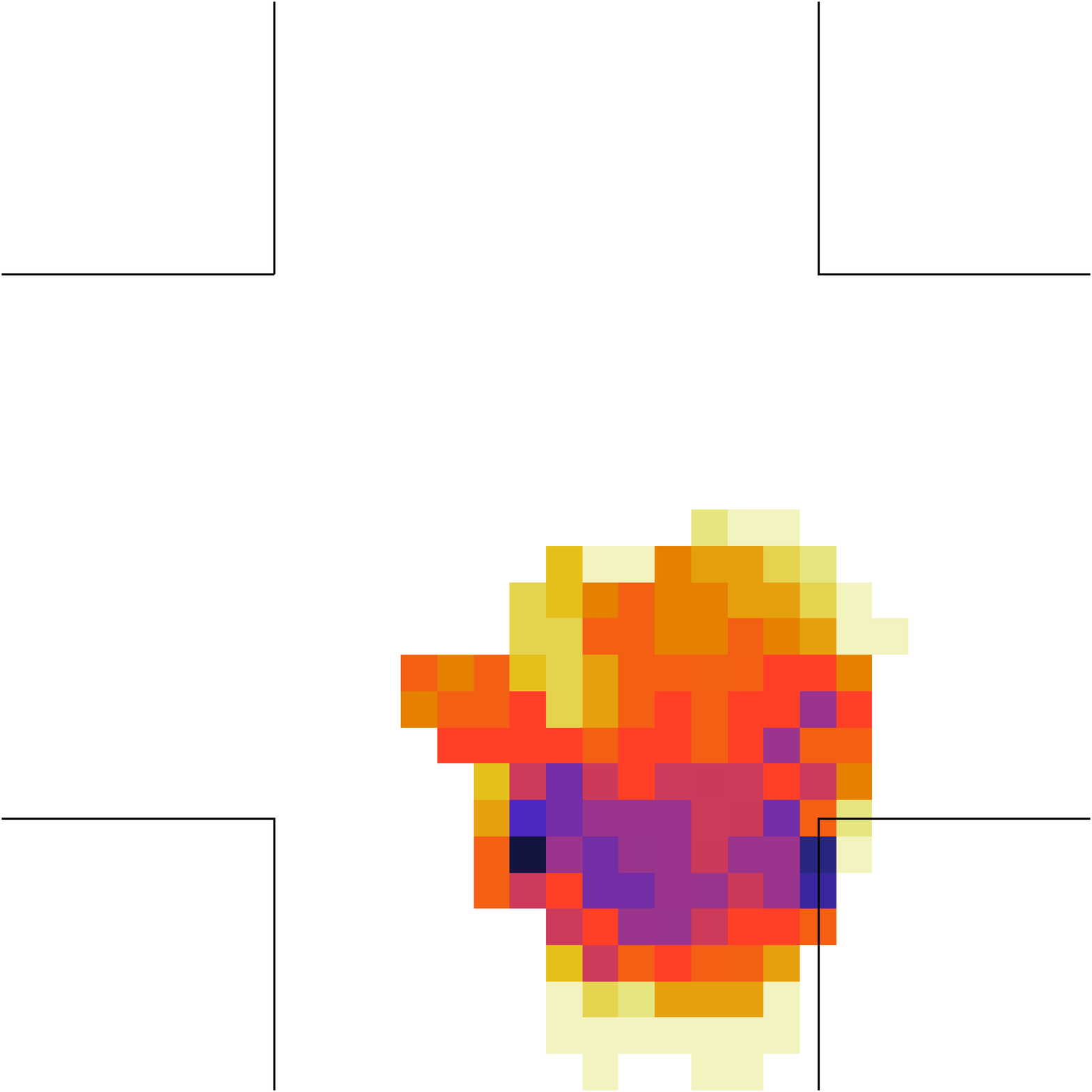}\hspace{0.02\textwidth}\includegraphics[width=0.05\textwidth]{scalenuovo.eps}
  \caption{Lower density ``realistic'' pedestrians at the time in which the highest maximum $CN$ is attained (0.892 during the $[37.5,40)$ s interval of the 5th repetition).  Top, left: $\mathbf{v}$ field; top, right: density field; bottom left: $(\nabla \wedge \mathbf{v})_z$ field; bottom, right: $CN$ field. In the velocity field, the length of the arrow is proportional to the magnitude (full length $v>0.5$), while the colour gives the orientation, as shown in the
      colour wheel legend. The density field is represented using a moving average over the Moore neighbourhood.}
\label{f12}
\end{center}
\end{figure}

For the ``realistic'' pedestrians, the crossing of the flows causes an actual increase in crowd congestion, due to the limited space and to the corresponding stopping/deviating behaviour.
This is properly indicated by the higher (with respect to the ``marching'' and ``bodiless'' cases) value of $CN$, and by the increase in $CN$ for higher flows. In the higher density setting, $CN$ attains values $\approx 1$. Fig. \ref{f9} shows the higher density setting $\mathbf{v}$, density, rotor and $CN$ fields at the time of maximum density, and Fig. \ref{f10} the same fields at the time of maximum $CN$ (which resulted to be
as high as 1.26). For reference, the lower density setting is shown in Figs. \ref{f11} and \ref{f12}.

An interesting feature is that in the higher density setting, $CN$ reaches (locally) extremely high values also when density is decreasing. This is due to
the situation depicted on the right in Fig. \ref{fv5}: some pedestrians are ``dragged'' in the wrong direction and high pressure/collision/congestion happens in the areas where they try to ``go back''.
Around the bottom-right corner, such behaviour continues even when the majority of pedestrians has crossed (the occurrence of such a dynamics may be related to the simplicity in the pedestrians' local behaviour, and could be absent or reduced in actual pedestrians or in a more realistic model; anyway here we are not judging the pedestrian model but the ability of the metric to recognise congestion and dangerous areas). The areas where such congestion is happening are correctly identified by $CN$ (Fig. \ref{f10}).

\section{$CN$ in a crossing scenario experiment}
We now present some data from a controlled experiments in which 54 participants (27 for each flow) were asked to move in two different flows, using the same geometry as in the simulations
settings\footnote{More precisely, the simulation settings where chosen to fit to those of this experiment, performed as part of a different research project by the Tokyo University team.}..
Results corresponding to two different initial conditions (1 and 2 ped/m$^2$) are shown (6 independent experiments were performed for each initial condition).

\begin{figure}[t]
\begin{center}
\includegraphics[width=0.8\textwidth]{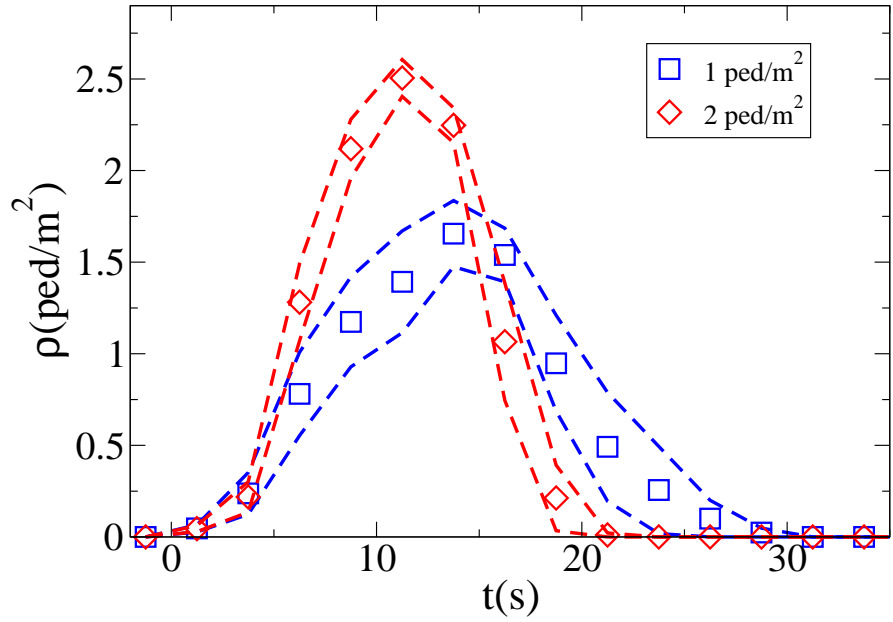}
\caption{Density in the crossing area as a function of time in the controlled experiments with subjects. Densities are averaged over 6 different initial conditions
 and on time intervals of 2.5 s. Dashed lines provide standard error bars.}
\label{f13}
\end{center}
\end{figure}

\begin{figure}[t]
\begin{center}
\includegraphics[width=0.8\textwidth]{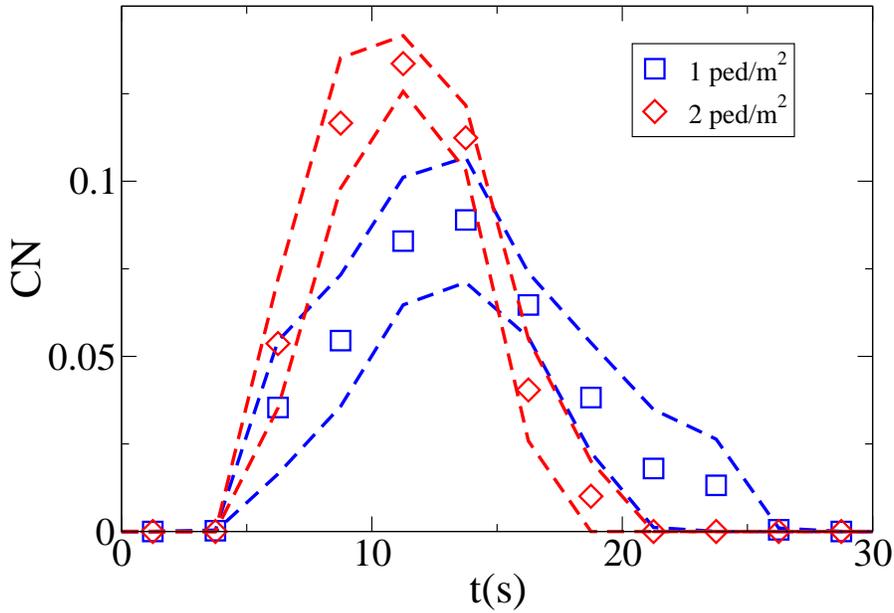}
\caption{Average $CN$ in the crossing area as a function of time in the controlled experiments with subjects. Dashed lines provide standard error bars, computed over the 6 different initial conditions.}
\label{f14}
\end{center}
\end{figure}

\begin{figure}[t]
\begin{center}
\includegraphics[width=0.8\textwidth]{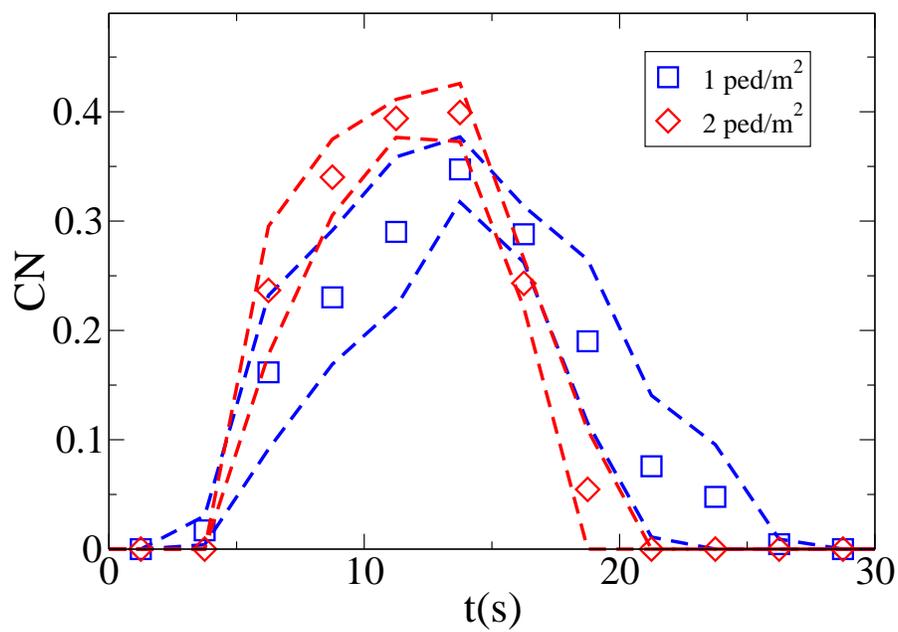}
\caption{Maximum $CN$ as a function of time in the controlled experiments with subjects. Dashed lines provide standard error bars, computed over the 6 different initial conditions.}
\label{f15}
\end{center}
\end{figure}

In Fig. \ref{f13} we report the density in the crossing area as a function of time for both settings, while in Fig. \ref{f14} we report the time evolution of
the average $CN$ in the central area, and in Fig. \ref{f15} the time evolution of the maximum $CN$.

\begin{figure}[t]
\begin{center}
  \includegraphics[width=0.4\textwidth]{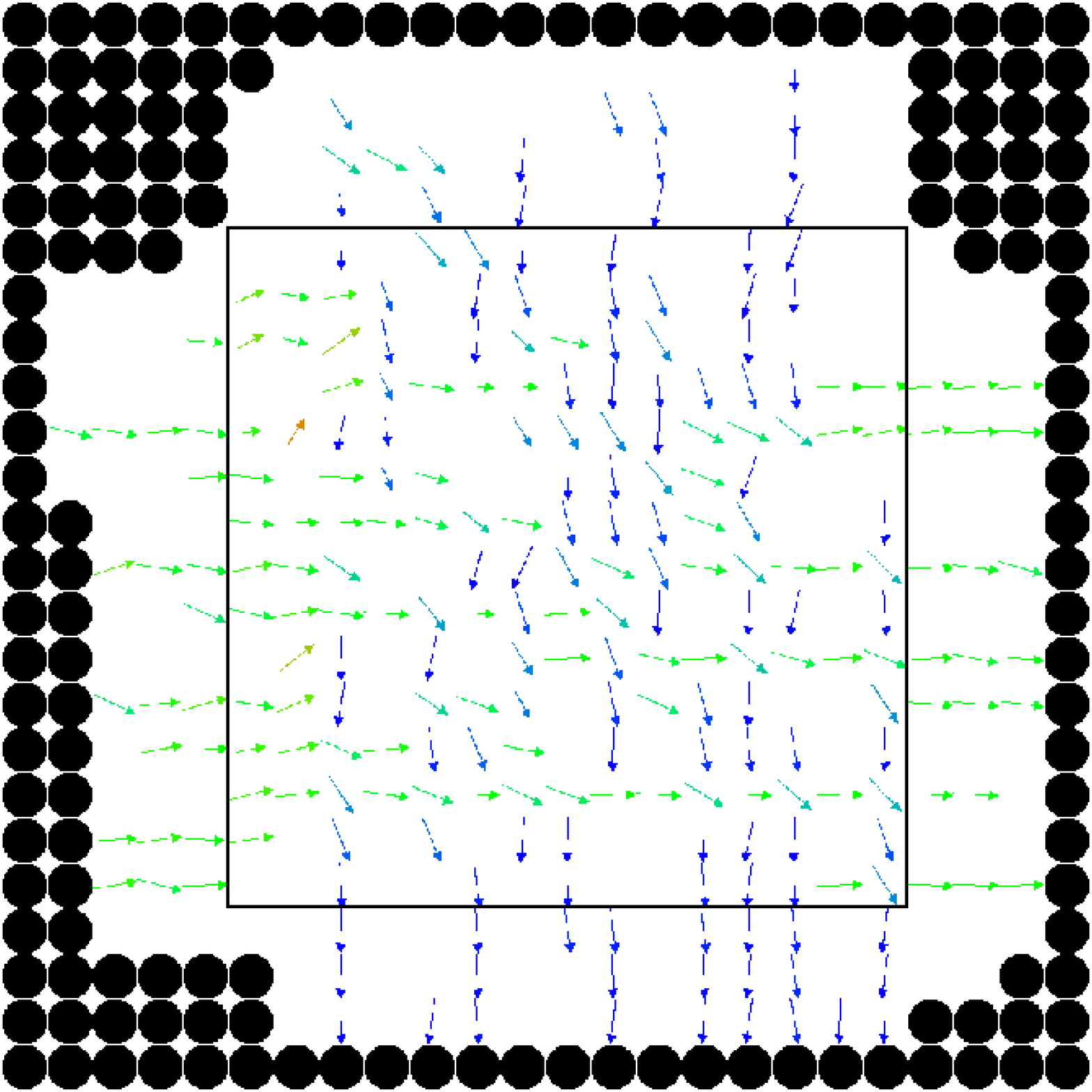}\includegraphics[width=0.1\textwidth]{frecce.eps}\hspace{0.02\textwidth}
  \includegraphics[width=0.4\textwidth]{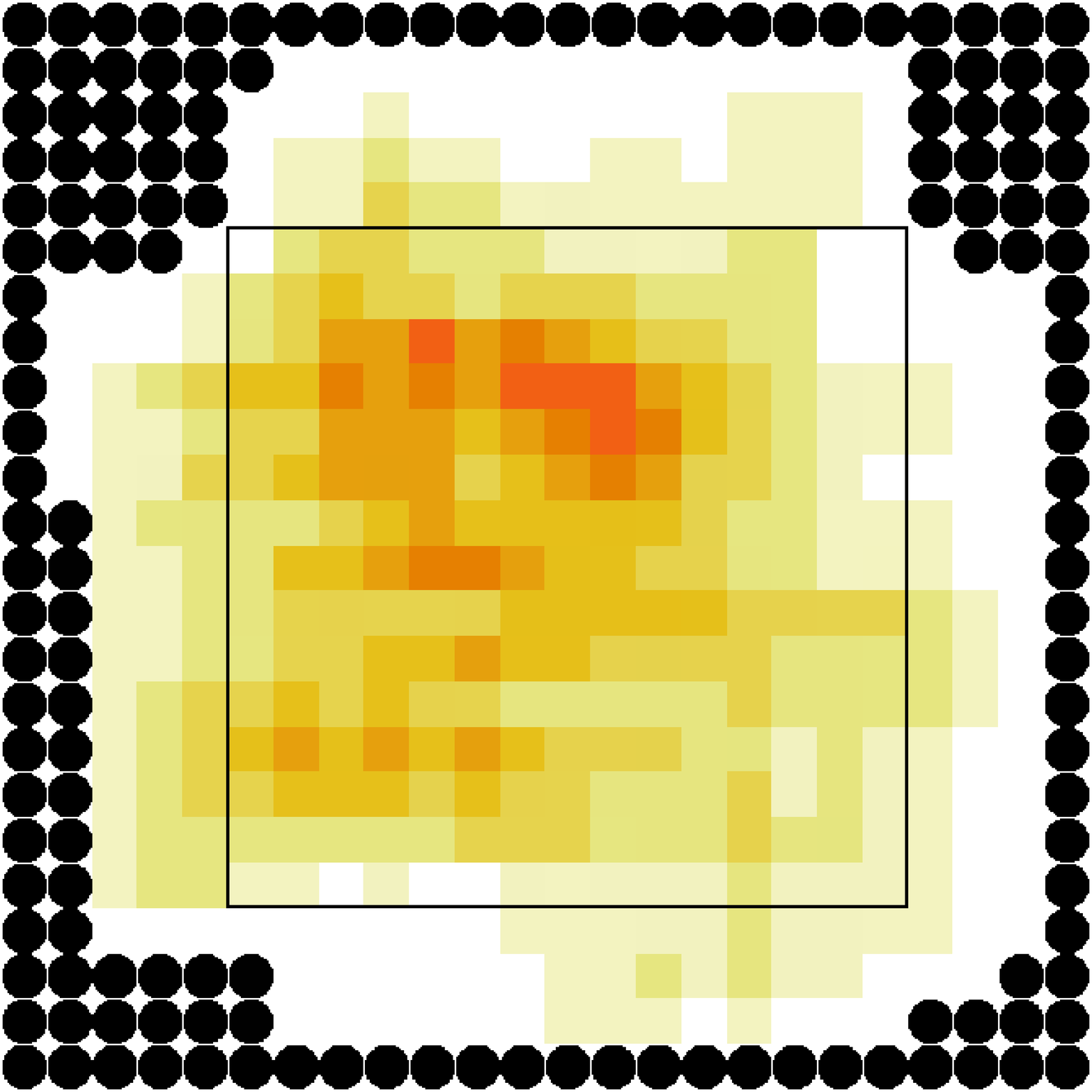}\hspace{0.02\textwidth}\includegraphics[width=0.05\textwidth]{scaledens.eps}
  \includegraphics[width=0.4\textwidth]{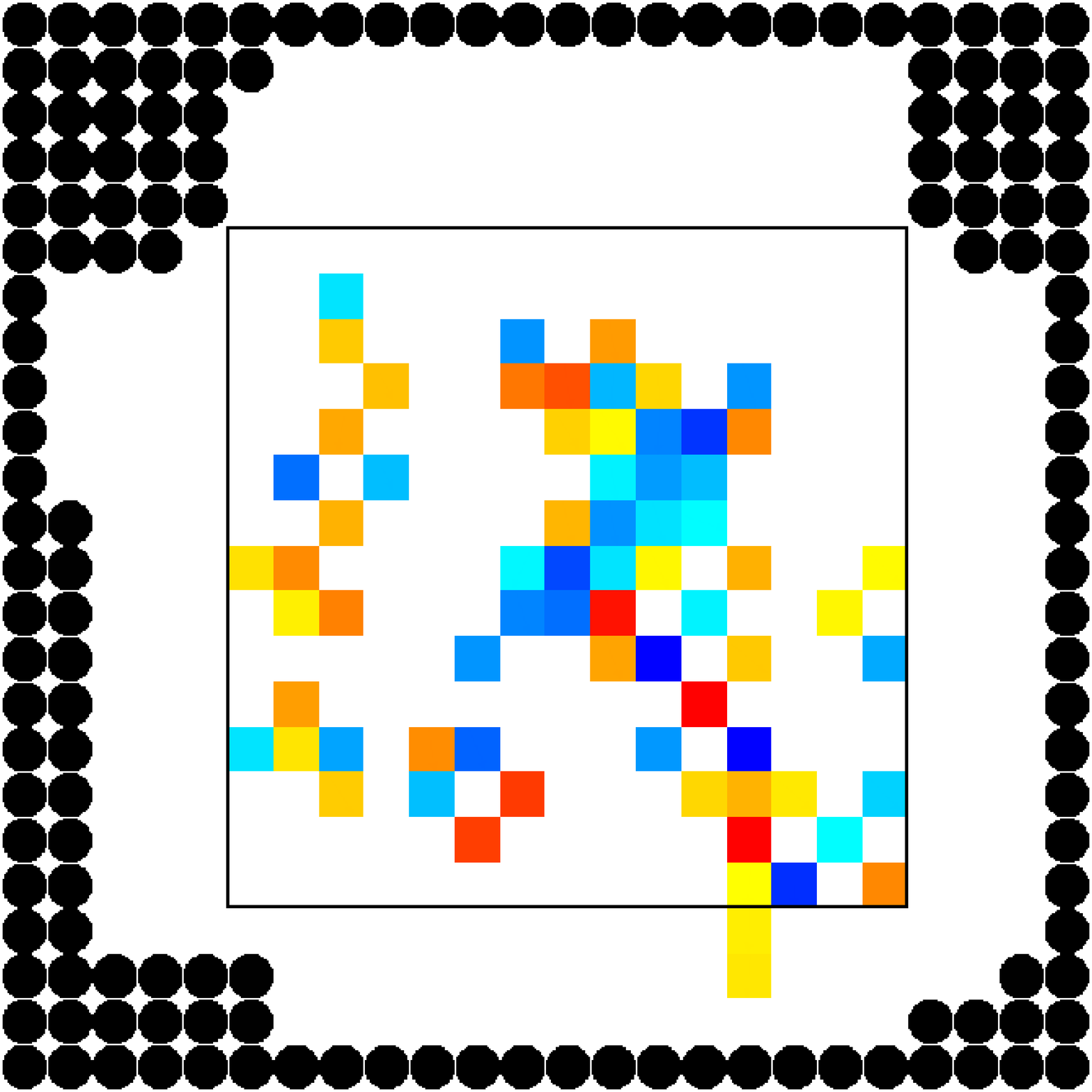}\hspace{0.02\textwidth}\includegraphics[width=0.05\textwidth]{scalepm0.eps}\hspace{0.05\textwidth}
  \includegraphics[width=0.4\textwidth]{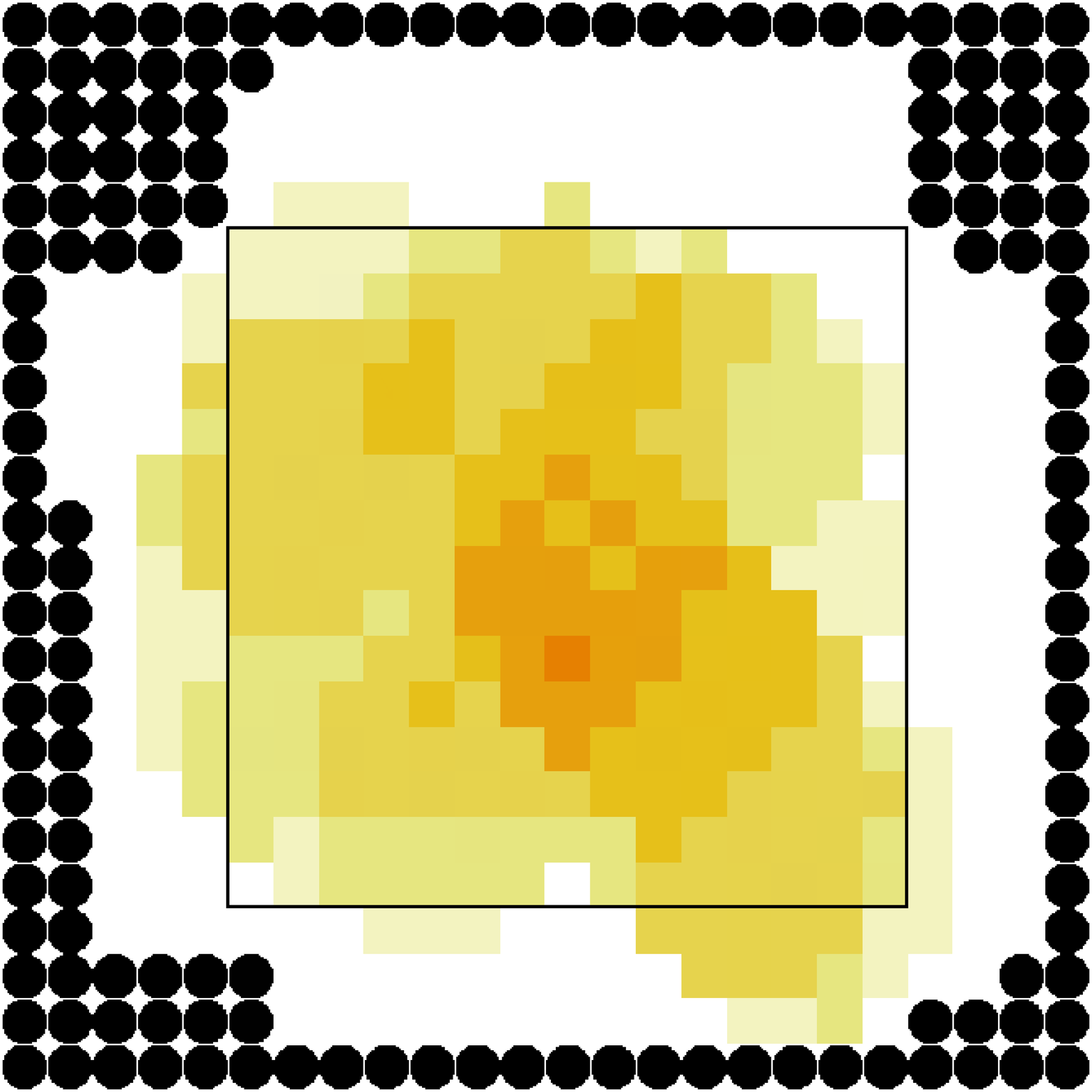}\hspace{0.02\textwidth}\includegraphics[width=0.05\textwidth]{scalenuovo.eps}
  \caption{Pedestrians in the controlled experiments at the time in which the maximum density is attained (2.72 ped/m$^2$ during the $[10,12.5)$ s interval of the 1st repetition). Top, left: $\mathbf{v}$ field; top, right: density field; bottom left: $(\nabla \wedge \mathbf{v})_z$ field; bottom, right: $CN$ field. In the velocity field, the length of the arrow is proportional to the magnitude (full length $v>0.5$), while the colour gives the orientation, as shown in the
      colour wheel legend. The density field is represented using a moving average over the Moore neighbourhood. Black discs correspond to cells without tracking data.}
\label{f16}
\end{center}
\end{figure}

\begin{figure}[t]
\begin{center}
  \includegraphics[width=0.4\textwidth]{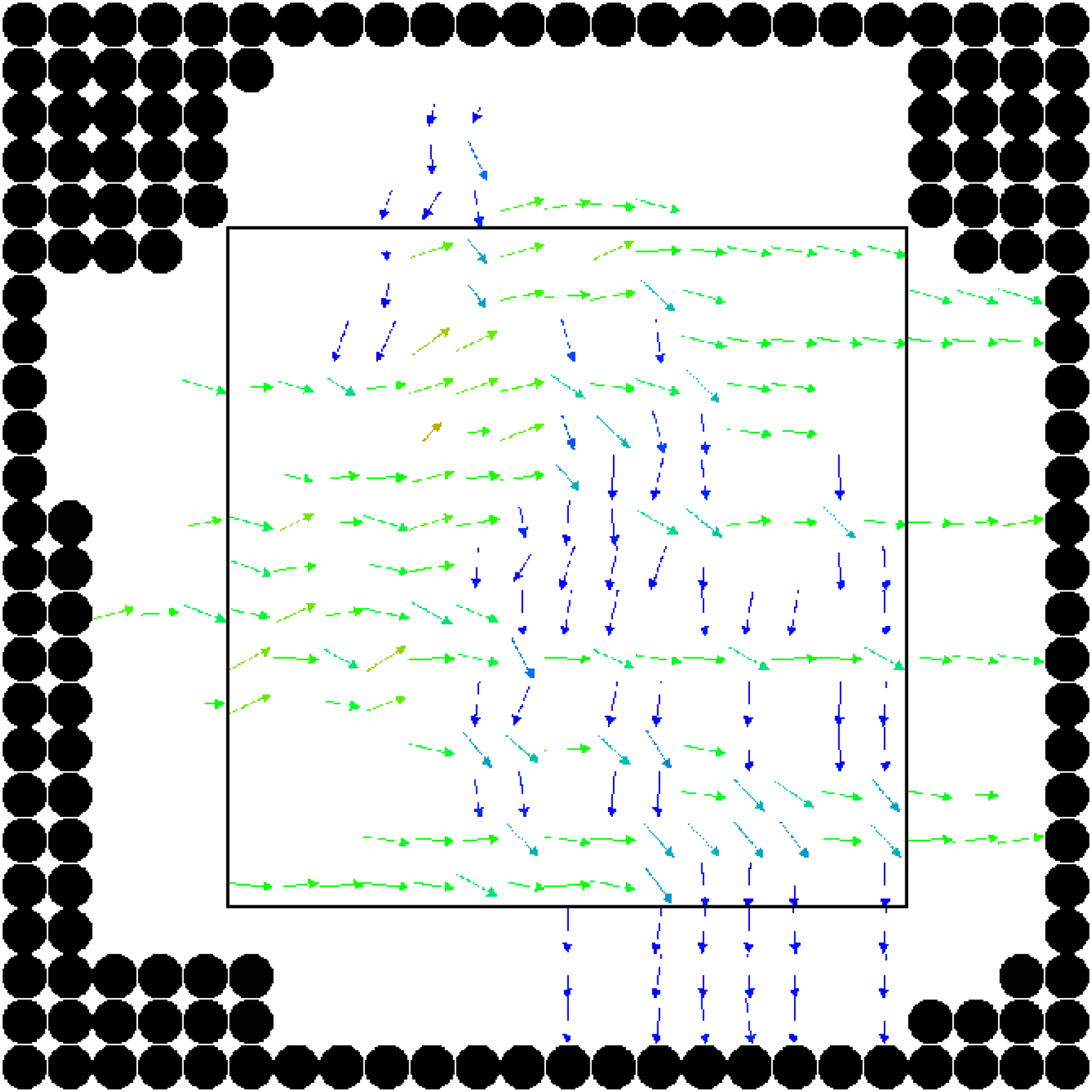}\includegraphics[width=0.1\textwidth]{frecce.eps}\hspace{0.02\textwidth}
  \includegraphics[width=0.4\textwidth]{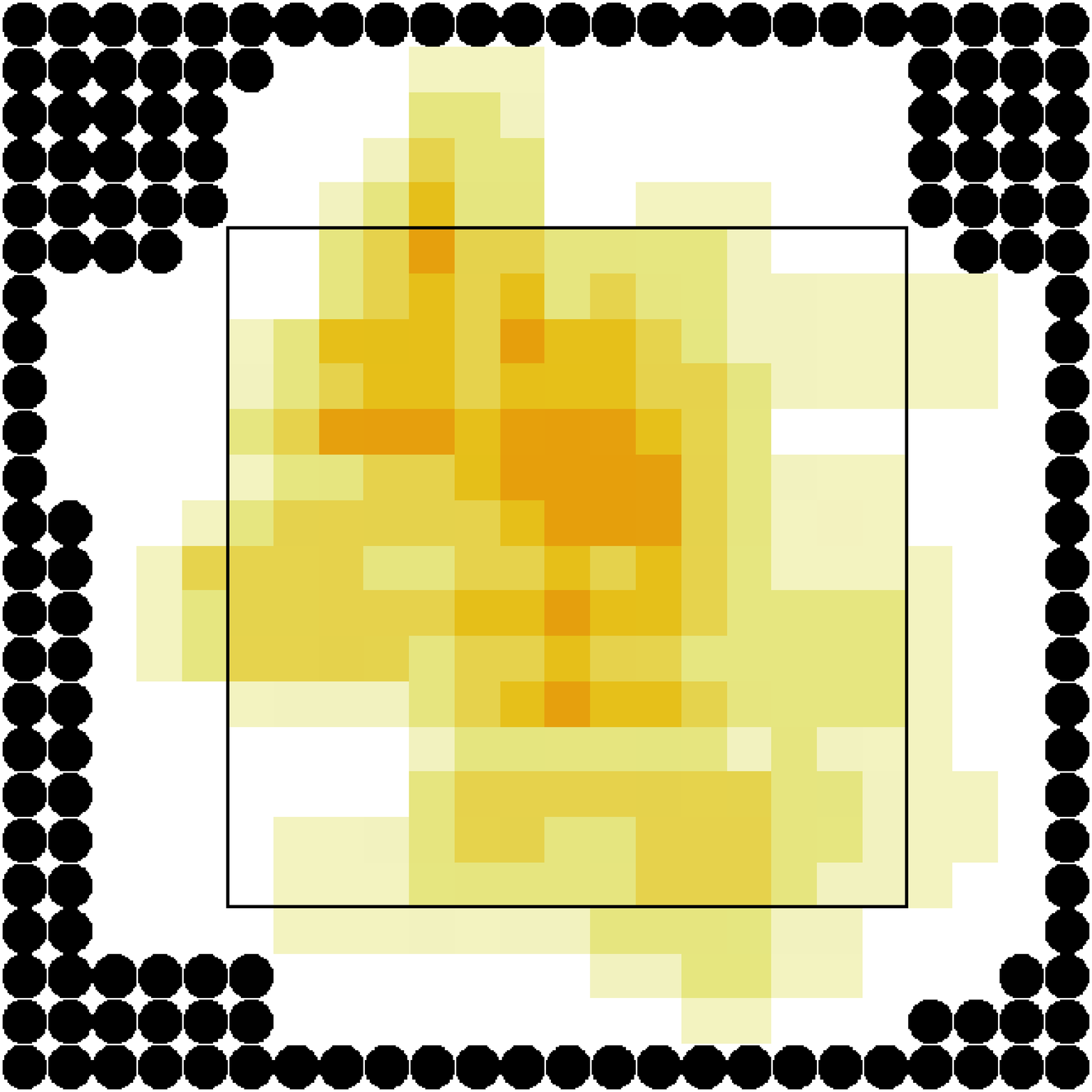}\hspace{0.02\textwidth}\includegraphics[width=0.05\textwidth]{scaledens.eps}
  \includegraphics[width=0.4\textwidth]{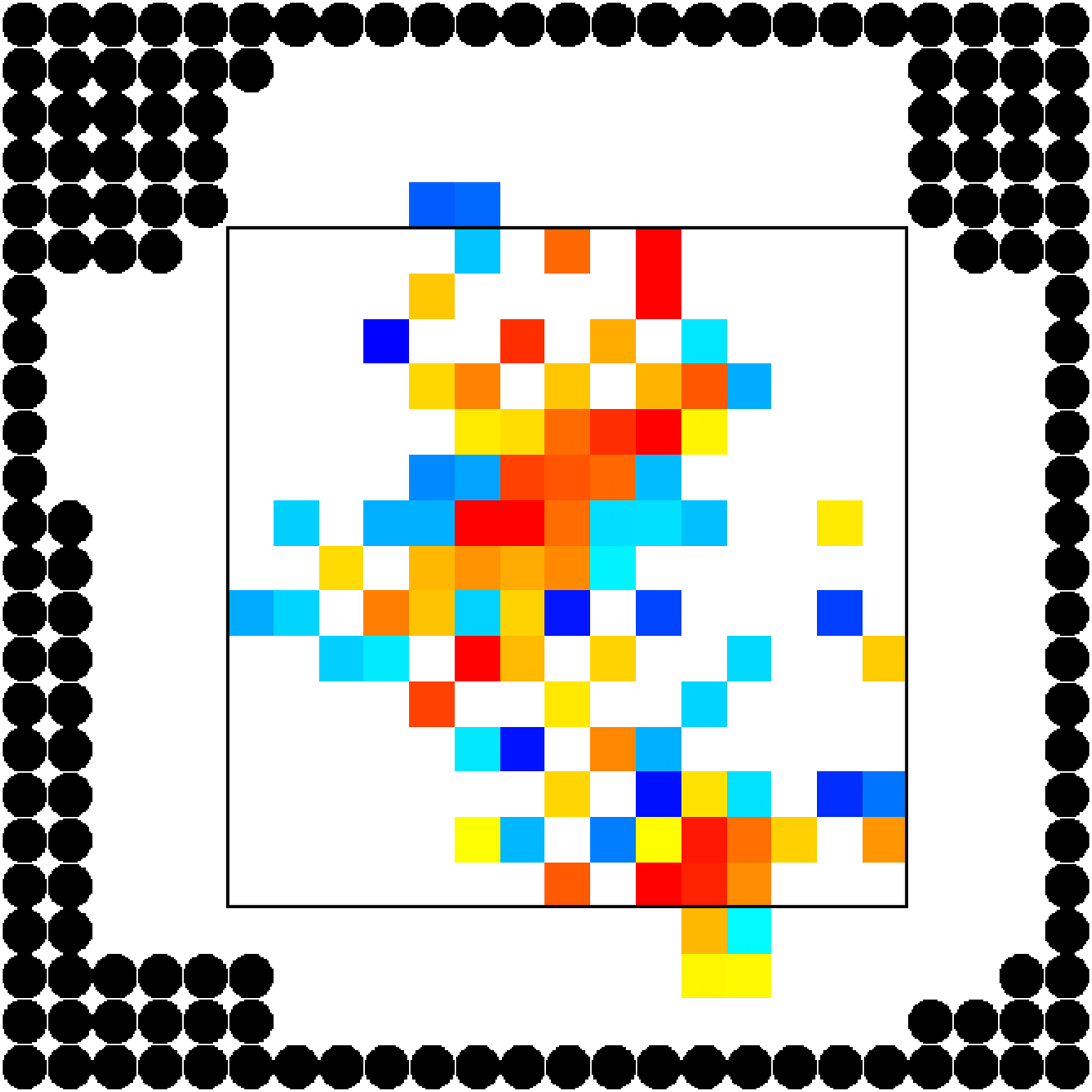}\hspace{0.02\textwidth}\includegraphics[width=0.05\textwidth]{scalepm0.eps}\hspace{0.05\textwidth}
  \includegraphics[width=0.4\textwidth]{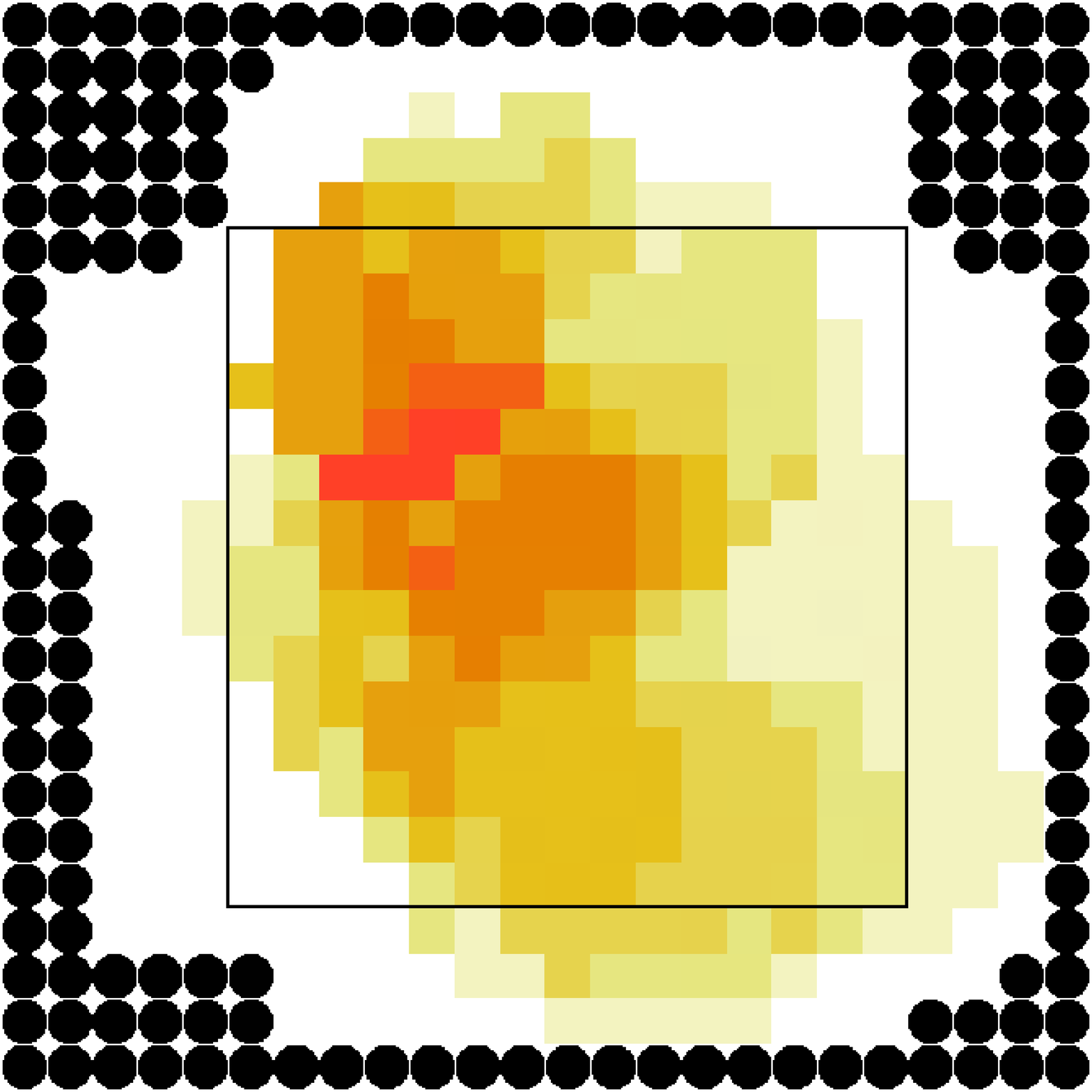}\hspace{0.02\textwidth}\includegraphics[width=0.05\textwidth]{scalenuovo.eps}
  \caption{Pedestrians in the controlled experiments at the time the maximum $CN$ is attained (0.513 during the $[12.5,15)$ s interval of the 2nd repetition). Top, left: $\mathbf{v}$ field; top, right: density field; bottom left: $(\nabla \wedge \mathbf{v})_z$ field; bottom, right: $CN$ field. In the velocity field, the length of the arrow is proportional to the magnitude (full length $v>0.5$), while the colour gives the orientation, as shown in the
      colour wheel legend. The density field is represented using a moving average over the Moore neighbourhood. Black discs correspond to cells without tracking data.}
\label{f17}
\end{center}
\end{figure}

Finally, for the 2 ped/m$^2$ condition, we show in figure \ref{f16} the $\mathbf{v}$, density, rotor and $CN$ fields at the time of maximum density, while those at the time of maximum $CN$ are shown in Fig. \ref{f17}.

\clearpage

\section*{Acknowledgements}

This research work was in part supported by: JSPS KAKENHI Grant Number 18H04121JSPS, JSPS KAKENHI Grant Number 20K14992, JST-Mirai Program Grant Number JPMJMI17D4.


\begin{thebibliography}{9}
\bibitem{CL1} C. Feliciani and K. Nishinari, {\it Measurement of congestion and intrinsic risk in pedestrian crowds}, Transportation Research part C: Emerging Technologies 91 (2018): 124-155
\bibitem{CL2} C. Feliciani and K. Nishinari, {\it Investigation of pedestrian evacuation scenarios through congestion level and crowd danger}, Collective Dynamics 5 (2020): 150-157.  
\bibitem{Tu} L. Tu, {\it An Introduction to Manifolds}, 2nd edition, Springer, 2011 (pag. 142-143).
\bibitem{tgf} F. Zanlungo, L. Crociani, Z. Y\"ucel and T. Kanda {\it The effect of social groups on the dynamics of bi-directional pedestrian flow: a numerical study} arXiv preprint arXiv:1910.04337 (2019).  
\end{thebibliography}
\end{document}